\documentclass[a4paper,12pt]{article}

\usepackage{amsmath,amssymb}   
\usepackage{epsfig}                    
\usepackage[matrix,arrow]{xy}          
\usepackage{xspace}                    
\usepackage{stmaryrd}                  
\usepackage{slashed}
\usepackage{appendix,graphicx}
\usepackage{enumitem}
\usepackage{cite}
\usepackage{hyperref}
\usepackage{mathtools}  

\usepackage{units} 
\usepackage{tikz}
\usepackage{color}

\usepackage[vcentermath]{youngtab}

\usepackage{multirow}

\DeclareSymbolFont{bbold}{U}{bbold}{m}{n}
\DeclareSymbolFontAlphabet{\mathbbold}{bbold}

\numberwithin{equation}{section}       

\allowdisplaybreaks                    

\newcommand{\ii}{\mathrm{i}}

\newcommand{\der}{\partial}

\newcommand{\bbZ}{\mathbb{Z}}
\newcommand{\bbR}{\mathbb{R}}
\newcommand{\bbC}{\mathbb{C}}

\DeclareMathOperator{\SU}{\mathit{SU}}
\DeclareMathOperator{\SO}{\mathit{SO}}
\DeclareMathOperator{\SL}{\mathit{SL}}
\DeclareMathOperator{\GL}{\mathit{GL}}
\DeclareMathOperator{\Symp}{\mathit{Sp}}
\DeclareMathOperator{\Spin}{\mathit{Spin}}

\DeclareMathOperator{\spin}{\mathit{spin}}

\DeclareMathOperator{\Cliff}{Cliff}

\newcommand{\rep}[1]{\mathbf{#1}}

\newcommand{\repp}[2]{(\rep{#1}, \rep{#2})}
\newcommand{\reppp}[3]{(\rep{#1}, \rep{#2}, \rep{#3})}
\newcommand{\id}{\mathbbold{1}}

\DeclareMathOperator{\tr}{tr}

\DeclareMathOperator{\ch}{ch}

\DeclareMathOperator{\adj}{ad}

\newcommand{\LC}{\nabla}

\newcommand{\proj}[1]{\times_{#1}}

\DeclareMathOperator{\Edd}{\mathit{E_{d(d)}}}

\DeclareMathOperator{\dHd}{\mathit{\tilde{H}_d}}

\newcommand{\tA}{{\tilde{A}}}

\newcommand{\tLambda}{{\tilde{\Lambda}}}

\newcommand{\um}{\underline{m}}
\newcommand{\un}{\underline{n}}

\newcommand{\hs}[1]{\hspace{#1}}

\newcommand{\ra}{\rightarrow}

\newcommand{\cN}{\mathcal{N}}

\newcommand{\maxalg}{\mathcal{A}}
\newcommand{\ctab}[1]{{\tiny\yng(#1)}}

\newcommand{\vac}{|0\rangle}

\newcommand{\cC}{\mathcal{C}}
\newcommand{\cG}{\mathcal{G}}

\newcommand{\suone}{1}
\newcommand{\sutwo}{2}

\newcommand{\lc}{\bbR}

\DeclareMathOperator{\Tr}{Tr}

\begin{document}

\setcounter{page}{0}
\begin{titlepage}
\titlepage

\begin{flushright}
IPhT-t20/047 
\end{flushright}

\vspace{30pt}
\begin{center}

{\Large {\bf   On symmetries and dynamics \\
\vspace{10pt}  of exotic supermultiplets}}

\vspace{25pt}

 Ruben Minasian$^a$, Charles Strickland-Constable$^b$ and Yi Zhang$^a$

\vspace{20pt}

\vspace{10pt}

{${}^a$\it Institut de Physique Th\'eorique, Universit\'e Paris Saclay, CNRS, CEA\\
91191 Gif-sur-Yvette Cedex, France}

\vspace{10pt}

{${}^b$\it School of Physics, Astronomy and Mathematics, University of Hertfordshire \\
College Lane, Hatfield, AL10 9AB, UK}

\vspace{40pt}

\underline{ABSTRACT}
\end{center}

Among the allowed representations of extended supersymmetry in six dimensions there are exotic chiral multiplets that, instead of a graviton, contain mixed-symmetry spin-$2$ tensor fields. 
Notably, an $\mathcal{N}=(4,0)$ multiplet has a four index exotic graviton and it was conjectured that an interacting theory based on this multiplet could arise as a strong coupling limit of M theory compactified on $T^6$. 
We present an algebraic study of these multiplets and their possible embedding into the framework of exceptional field theory, finding in particular that the six-dimensional momenta do not correspond to a conventional spacetime section. 
When compactified on a circle,
the six-dimensional multiplets give rise to the same degrees of freedom as five-dimensional supergravity theories with the same number of supersymmetries.  
However, by considering anomalies (computed using the product multiplets construction) and the generation of Chern-Simons couplings, 
we find reason to doubt that their dynamics will agree with the five-dimensional gravity theories. 
We propose an alternative picture, similar to F-theory, in which particular fixed-volume $T^3$-fibered spacetimes play a central role, suggesting that only on compactification to three-dimensions will one make contact with the dynamics of supergravity.

\vfill
\begin{flushleft}
{ \today }\\
\vspace{.5cm}
\end{flushleft}
\end{titlepage}

\newpage

\begin{center}
\tableofcontents
\end{center}
\thispagestyle{empty}
\section{Introduction}

It has been proposed that  a strong coupling limit of five-dimensional  quantum $\cN=8$ supergravity in which the Planck length becomes infinite could give a six-dimensional superconformal phase of M-theory~\cite{Hull1,Hull2,Hull3}. Moreover for the free theory this limit has been argued to be given by a six-dimensional theory with maximal $(4,0)$ supersymmetry. This theory is conformal and hence has no length scales. When put on a circle, the compactification scale $R$ becomes the five-dimensional Planck scale. Clearly understanding such a limit would require radically new ideas and these would be important for our overall understanding of the gravitational physics of M-theory. 
In recent years, there has been a revival of interest in this area, producing many interesting developments and new approaches~\cite{Action,magicsquare,magicpyramid,Anastasiou:2014qba,YangMillssquared,Tensorsquared,Cachazo:2018hqa,Chiodaroli:2011pp}. 

However, regardless of the implications for M theory, at the level of supermultiplets, the (free) multiplet with $N =(4,0)$ supersymmetry certainly exists~\cite{Strathdee} and has 32 supersymmetries and 32 conformal supersymmetries. Its dimensional reduction has the same degrees of freedom and the same field content as  the maximal supergravity in five dimensions. The latter theory has $E_{6(6)}$ global symmetry, and in addition to the graviton has 27 vector and 42 scalar fields, as well as eight gravitini and 48 spin 1/2 fermions. 
It has been suggested that the former has the same $E_{6(6)}$ symmetry, such that the fields appear in similar representations. Instead of gravity (rank two symmetric field) it has a rank 
four tensor gauge field with the symmetries of the Riemann tensor. Due to self-duality contraints on its double field strength this field has five degrees of freedom (just like the five-dimensional graviton) and its dimensional reduction gives conventional linearised gravity in five dimensions~\cite{Hull1}. 
Similarly, instead of 27 five-dimensional vectors, the $(4,0)$ multiplet has 27 self-dual tensors.\footnote{In our conventions, the six-dimensional $(2,0)$ gravity multiplet has five anti-self-dual tensor fields, while the $(2,0)$ tensor multiplets have  self-dual tensors.} In either case there are $27 \times 3$ degrees of freedom. The 48 spin 1/2 fermions simply become chiral fermions in six dimensions. Finally the eight gravitini (vector-spinor fields) are replaced by eight\footnote{We count the four quaternionic fields as eight complex fields and will use similar counting throughout.} ``exotic gravitini" $\psi_{\mu\nu}$ - spinor-valued two-forms with self-duality constraint on their field strength.\footnote{Like in much of the literature, the fields in $(4,0)$ and $(3,1)$ multiplets that do not appear in ordinary gravity or matter multiplets, but have direct counterparts, i.e. like eight spinor-valued two forms in $(4,0)$ vs eight gravitini in $(2,2)$, will be labeled as ``exotic". Due to its properties, for the exotic graviton in $(4,0)$ multiplet the self-dual Weyl (SDW) label will also be used.}

 In fact, the $(4,0)$ multiplet is not the only exotic six-dimensional theory. There exists also a $(3,1)$
multiplet, where the self-duality constraints are  partial, and from examining the scalar degrees of freedom one might guess that the symmetry governing the theory is $F_{4(4)}$. The multiplet has a rank 3 tensor field, with a partial self-duality, and 28 scalars which lie in the tangent space to the 
symmetric space $F_{4(4)}/ \Symp(2)\times\Symp(6)$. 
However, the 14 vector fields and 12 self-dual tensors only form the $\rep{26}$ representation of $F_{4(4)}$ when combined together. This suggests that in fact only the R-symmetry group $\Symp(2)\times\Symp(6)$ (and not the full $F_{4(4)}$) would be a true symmetry. 
This could make one suspicious as to whether $E_{6(6)}$ would be a true symmetry of the $N=(4,0)$ theory, and we will see some indications that it may indeed not be. As these symmetries do not follow directly from the supermultiplets, but appear only in the construction of the associated theories, the absence of a complete construction of the $(4,0)$ theory means that one cannot be sure. 
However, a simple argument in favour of the $E_{6(6)}$ symmetry is that the scalars of the 5d maximal supergravity are all lifted to scalars in 6d. Thus naively one would expect the 5d transformations of them also to lift to 6d. 
The fermionic fields of the $(3,1)$ multiplet comprise two exotic gravitini, six standard gravitini of negative chirality, 28 spin 1/2 fermions of positive chirality and 14 spin 1/2 fermions of negative chirality. The exotic and conventional gravitini reduce to give the eight standard gravitini in five dimensions, while the spin 1/2 fermions of either chirality simply reduce to five dimensional spin 1/2 fields. 
 
Finally, the exotic fields can appear in multiplets with less supersymmetry. These can be constructed via the usual representation-theoretic arguments. An alternative is to consider the decomposition of the maximally supersymmetric multiplets. For example, as we shall discuss, the  $(4,0)$ multiplet decomposes into an exotic $(2,0)$ gravity multiplet as well as 4 exotic $(2,0)$ gravitino multiplets and 5 $(2,0)$ tensor multiplets. This decomposition is very similar to the decomposition of the maximal $(2,2)$ six-dimensional supergravity. This can be decomposed into $(2,0)$ multiplets: one gravity, 4 gravitino and 5 tensors.\footnote{It is not hard to verify that even if individual multiplets are chiral the whole combination is not - for every chiral fermion or self-dual field there is another with the opposite chirality or anti-self-duality.}

One useful perspective on these multiplets is given by the fact that they can be seen as square or product theories~\cite{Chiodaroli:2011pp,YangMillssquared,Tensorsquared}, in analogy to the linearised maximal supergravity in six dimensions, i.e. the $(2,2)$ theory being the square of the six-dimensional super Yang-Mills. In the same vein, the $(4,0)$ multiplet can be seen as a square of $(2,0)$ tensor multiplets, while the $(3,1)$ theory - as a product of a $(2,0)$ multiplet with a $(1,1)$ vector one. Similar product structures appear in the exotic theories with less supersymmetry. 
While much of the interest in double copy constructions comes from the computation of amplitudes in perturbation theory~\cite{CK1,CK2,CK3} (see~\cite{Bern:2019prr} for a review) there have also been developments in off-shell field theoretical realisations~\cite{magicsquare,magicpyramid,Anastasiou:2014qba,Nagy:2014jza,YangMillssquared,Borsten:2015pla,Anastasiou:2017nsz} and the construction of classical solutions~\cite{ClassicalDCopy1,ClassicalDCopy2,Cardoso:2016ngt,Cardoso:2016amd,ClassicalDCopy3}. Unfortunately in our case of interest, the strongly coupled theory has no perturbative expansion and there may also be no classical limit with interactions, limiting the direct usefulness of these constructions. 

\subsection{Algebraic aspects}

Two questions that preoccupy us in this paper concern the algebraic symmetry-based reasons behind the existence of the  the exotic multiplets and the possibility of probing the existence of interacting forms of these exotic theories (as well as their existence on non-flat spaces). Some of the arguments here can be made for both $(4,0)$ and  $(3,1)$ multiplets, and some are specific only to $(4,0)$. 
Much of the algebraic discussion takes place in the context of the U-duality groups and their relation to the corresponding superalgebras. In particular, we will use the language of generalised geometry~\cite{CSW2,CSW3,CSC} and exceptional field theory~\cite{HS,E77EFT,HS-E88}, discussing the charges appearing in the supersymmetry algebra as generalised vectors in a generalised tangent space which transforms as a linear representation under the relevant U-duality group. 
In order to avoid encountering infinite dimensional duality algebras, we will work with dimensional splits of the theories considering three external dimensions separately from the rest.

As we will discuss in section~\ref{sec:superalgebras}, all supersymmetry algebras with 32 supercharges arise from a particular superalgebra $\mathcal{A}$ (with bosonic subalgebra $\mathfrak{sl}(32,\bbR)\ltimes \bbR^{528}$) by restricting $\mathfrak{sl}(32,\bbR)$ to different $\mathfrak{spin}(1,d-1)$ subalgebras. For example, one can obtain the superalgebras of 11d, type IIA and type IIB supergravities from this prescription. On performing a dimensional split, decomposing say $\spin(9,1) \ra \spin(3,1) \times \spin(6)$ in type IIA or IIB, one can see how the resulting $\spin(6)$ group would act on the charges appearing in the generalised tangent space of the supergravity theory on the internal Euclidean signature part. In this way, merely requiring the chiralities of the fermions present in type IIA and type IIB implies that one requires $E_{d(d)}$-inequivalent ``sections" (in the language of exceptional field theory) of the generalised tangent space to correspond to the physical momenta in spacetime for the two theories. For the particular case of type IIA vs type IIB, these inequivalent sections (or inequivalent embeddings of the general linear group into the U -duality group) have been discussed extensively in the literature~\cite{Schnakenburg:2001he,CSW2,CSC,HS}. 
A similar discussion of sections for half-maximal supersymmetry can be found in~\cite{Ciceri:2016hup}, where it was concluded that inequivalent sections gave the $\cN=(1,1)$ and $\cN=(2,0)$ supergravities in six dimensions (the former section extending to type I in ten dimensions).

Similarly, one can explore what happens if one instead requires $\cN =(4,0)$ supersymmetry in six-dimensions from the decomposition. 
We examine the intersection of the relevant $\Spin(5,1)$ group with the generalised spin group $\Spin(2,1)\times\SO(16)$. Under the common subgroup $\Spin(2,1)\times\Spin(3)$ we observe how the charges in the generalised tangent space are grouped into irreducible representations of the $\Spin(3)$ factor and of the $\SL(3,\bbR) \subset \SL(9,\bbR) \subset E_{8(8)}$  which contains it. 
This reveals a very different behaviour to the normal situation in generalised geometry or exceptional field theory. 

The root of this difference lies partly in the fact that in $E_{8(8)}$, the charges appearing in the supersymmetry algebra do not span the full $\rep{248}$ representation in which the generalised vector transforms, but rather only the $\rep{120}$ part under its $\SO(16)$ subgroup. 
Under the direct embedding into the $\rep{248}$, the momentum charges do not satisfy the section condition, even in standard supergravity. 

The $\Spin(3)$ triplet of momentum charges of the $(4,0)$ supersymmetry algebra thus embed into the generalised vector as a triplet of $\SO(3)$, which consists of two of the momenta that would be present in the conventional reduction of five-dimensional supergravity to three dimensions, plus part of the dual graviton charge, much as expected from~\cite{Hull1}. 
However, under the $\SL(3,\bbR)$ subgroup containing this $\SO(3)$, these three charges are combined with five others to form an octuplet. 
Ordinarily in supergravity one would expect them rather to be contained in a subspace of the sum of two triplets, a space in which one could identify an $\SL(3,\bbR)$ triplet solving the section condition. 
Here, this is not the case, and there is no such section. 
Further, this $\SL(3,\bbR)$ subgroup is related to that of $\cN=(2,2)$ supergravity by a transformation in $\SL(9,\bbR) \subset E_{8(8)}$, so any such section would be equivalent to the standard one anyway. 

Nonetheless, we go on to examine the decomposition of the generalised vector and the adjoint of $E_{8(8)}$ under $\SL(3,\bbR)\times E_{6(6)}$, noting that if we had enhanced $\SL(3,\bbR)$ to $\GL(3,\bbR)$ as one would usually in standard supergravity, this would break the $E_{6(6)}$ commutant to $\SO(5,5)$. We then look at these decompositions and attempt to apply the naive algebraic prescription (usually imagined only in the context of supergravity -- see e.g.~\cite{CSC} for a discussion) to extract the field content of a parent six-dimensional theory. We find that, with suitable identifications, this matches exactly what one would expect from the $\cN=(4,0)$ multiplet, though questions remain over whether one must decompose under $\SO(3) \subset \SL(3,\bbR)$ and $\Symp(8) \subset E_{6(6)}$ in order to make these identifications. Indeed, the algebraic construction of the generalised Lie derivative in flat space appears to reproduce a formula for the gauge transformation of the exotic graviton, which reassures us that our identification of the spacetime directions inside the generalised tangent space, together with the fields and charges, is somewhat correct.

\subsection{h-theories}

Of course, one can wonder if there is more to these multiplets than simply their algebraic properties. 
They stand out as multiplets with highest-spin $\leq 2$ which do not appear in standard supergravity theories, their decompositions under sub-superalgebras and compactifications or their matter multiplets. 
We shall present arguments that the fact that the conjectured $(4,0)$ symmetry group $E_{6(6)}$ has an $\SL(3, \bbR)$ commutant inside the three-dimensional symmetry group $E_{8(8)}$ serves not only as a helpful technical tool, 
but is closely connected to the very existence of the six-dimensional theory with $E_{6(6)}$ symmetry. 
Correspondingly, the symmetry groups for exotic $(2,0)$ and $(1,0)$ symmetry groups have $\SL(3, \bbR)$ commutants inside the symmetry groups of three-dimensional theories with 16 and 8 supercharges respectively.

In general, the exceptional $E_{d(d)}$ groups have $GL(n, \bbR)$ commutants inside bigger $E_{d+n(d+n)}$ groups. This is essentially by construction: the lower dimensional theories with maximal supersymmetry are obtainable from the higher dimensional ones after a  torus $T^n$ compactification. Finding other decompositions of $E_{d+n(d+n)}$ might be useful as a technical tool, but is of very little consequence as far as higher-dimensional theories are concerned. 
For other decompositions $G_d \times H_n \subseteq E_{d+n(d+n)}$, there is no (known) maximally supersymmetric theory (or multiplet) in $D=11-d$ dimensions with symmetry $H_n$. 
For example the existence of the subgroup $\SL(2,\mathbb{R}) \times E_{7(7)} \subseteq E_{8(8)}$ has no implications for five-dimensional physics, as there is no maximal five-dimensional theory with symmetry group $E_{7(7)}$. 

In this sense, assuming that the $\cN=(4,0)$ theory really has $E_{6(6)}$ symmetry, we see that $ E_{6(6)}$, $\SL(3,\mathbb{R})$  and $E_{8(8)}$ form a unique triple for maximally supersymmetric theories.\footnote{As mentioned, less-supersymmetric counterparts of this triple exist  with $\SL(3,\mathbb{R})$ always playing a central role. For concreteness we shall be concentrating on the maximally supersymmetric case.} 
Given that the $\SL(3,\bbR)/SO(3)$ coset is the moduli space of flat metrics on $T^3$ of fixed volume, this suggests a way of thinking about the $(4,0)$ multiplet analogous to F-theory  \cite{Vafa:1996xn}. 
A solution of three-dimensional supergravity with five non-constant scalars parametrising the coset, can be thought of as a solution of a six-dimensional theory with the left-over $E_{6(6)}$ symmetry, i.e. the $(4,0)$ theory on a $T^3$-fibered manifold satisfying certain conditions. 
Moreover, using results from earlier work on ``U-fold" torus fibrations~\cite{Liu:1997mb}, it can be shown that the geometrical information can be repackaged and presented in a form of a self-dual Weyl (SDW) tensor field, 
and differential conditions on the six-dimensional space upon linearisation 
can be reduced to the equations of motion for the SDW field. 
The details of this constructions which we call $h$-theory can be found in section~\ref{sec:h-theory}. 
A novel feature of this construction is 
that both the geometry and the SDW field on it are constructed out of the physical scalar degrees of freedom in three-dimensions. 
Our analysis also has no propagating fields along the directions of the torus, similarly to the situation in F-theory where there are no momenta in the auxiliary $T^2$ directions. This intriguing picture would thus suggest that the $(4,0)$ theory is not really six-dimensional, as the physical states are not charged under the additional momenta. 

It has been observed in \cite{Hull2} that due four-dimensional symmetry group $E_{7(7)}(\bbZ)$ not having an $E_{6(6)}(\bbZ) \times \SL(2, \bbZ)$ subgroup the $\SL(2,\bbZ)$ duality expected from six-dimensional description would act non-trivially on the graviton leading possibly to a modification of supergravity. Our picture suggests a more conservative possibility, inspired by the relations between F-theory, 11-dimensional supergravity and type IIB. We should not think of recovering the four-dimensional supergravities from $T^2$ reduction of the exotic $(4,0)$ theory any more than we expect a direct reduction of F-theory on a circle to yield the 11-dimensional supergravity, or of M-theory being simply reduced to IIB. Instead, when M-theory is put on a two-torus one can take the so called F-theory limit  that decompactifies to ten-dimensions while while retaining the $\SL(2,\bbZ)$, i.e. yields the type IIB theory. The limit holds also from M-theory on  an elliptically fibered manifold, in which case the decompactification yields type IIB on the base of the elliptic fibration. So the idea is to consider the three-dimensional maximal supergravity, i.e. the $(4,0)$ theory on a fixed volume $T^3$ in decompactication limits. Denoting the radii of circles in $T^3$ by $r_1, r_2, r_3$ and setting the $\mbox{Vol}(T^3) = 1$, up to numerical factors one has $r_1 = 1/r_2 r_3$. One can take $r_2, r_3 \ra \infty$ and hence $r_1 \ra 0$, i.e. decompactify two dimensions. 
The path
 $$E_{8(8)} \supseteq \SL(3,\mathbb{R}) \times E_{6(6)}   \supseteq  \SL(2,\mathbb{R}) \times \mathbb{R}^+ \times E_{6(6)}   \xhookrightarrow{\phantom{XX}} GL(2,\mathbb{R}) \times E_{6(6)}  \,\,\, \xrightarrow{r_2, r_3 \ra \infty} \,\,\, E_{6(6)}  \,\,\, \mbox{ in D=5}$$
 results in a five-dimensional theory with $E_{6(6)}$ symmetry, i.e. the ordinary five-dimensional supergravity. 
 Another option is $r_2, r_3 \ra 0$ and hence $r_1 \ra \infty$, i.e. decompactify a single dimension. The path now is 
 $$E_{8(8)} \supseteq \SL(3,\mathbb{R}) \times E_{6(6)}   \supseteq  \SL(2,\mathbb{R}) \times \mathbb{R}^+ \times E_{6(6)}  \xhookrightarrow{\phantom{XX}} \SL(2,\mathbb{R}) \times E_{7(7)}  \,\,\, \xrightarrow{r_1 \ra \infty} \,\,\, E_{7(7)}  \,\,\, \mbox{ in D=4}.$$
This explains the appearance of  both five-dimensional $E_{6(6)}$ and four-dimensional $E_{7(7)}$ in the decompactification limits of three-dimensional maximal supergravity. As everything else relating to the embedding of $\SL(3,\bbZ)$ in three-dimensional duality group, these chains continue to hold for theories with 16 and eight supercharges.  Calling the symmetry group $G$, we first note that 
$G^{\mbox{\tiny exotic}}_{D=6} = G_{D=5}$ and that $\SL(3,\mathbb{R}) \times G^{\mbox{\tiny exotic}}_{D=6} \subseteq G_{D=3}$ as well as $\SL(2,\mathbb{R}) \times G_{D=4} \subseteq G_{D=3}$. The deompactifications to ordinary supergravities in  four and five dimensions now work as in the maximally supersymmetric case.

Another observation which suggests that we do not think of the theory as truly six-dimensional comes from consideration of higher rank dualities. Considering the conjectured Kac-Moody symmetries $E_{8+n(8+n)}$ for $n=1,2,3$, we might expect to find that the $\SL(3,\bbR)$ commutant of $E_{6(6)}$ is extended to $\SL(3+n,\bbR)$. However, this is not the case. In particular, the $\SL(3,\bbR)\times\SL(3,\bbR)\times E_{6(6)}$ that we consider in our dimensional split (into three external dimensions, three internal dimensions and an internal $E_{6(6)}$ symmetry) does not extend to an $\SL(6,\bbR)\times E_{6(6)}$ subgroup inside $E_{11}$\footnote{We thank Guillaume Bossard for explaining these features of $E_{11}$ to us.}. However, there is a $\Spin(5,1)$ subalgebra of $KE_{11}$ corresponding to the decomposition of the $32$ component spinor representation into $4$ spinors of the same chirality in six dimensions, so that $E_{11}$ does appear to accommodate the multiplet at the level of the superalgebra. The fact that the relevant $\SL(6,\bbR)$ subgroup fails to exist indicates (unsurprisingly) that there is no six-dimensional gravity for this multiplet and potentially that the theory is not truly six-dimensional
\footnote{
One slight difference between our picture and that of F theory is that while there is no $\SL(12,\bbR)$ inside $E_{11}$, there is also no twelve-dimensional spin group or momentum charge. 
}.

\subsection{Chern-Simons couplings and anomalies}

To provide further support to this picture, we include other arguments suggesting that the naive reduction of the $(4,0)$ theory on $S^1$ or $T^2$ might not produce the dynamics of supergravity in five or four dimensions. We will also find similar statements for the $(3,1)$ theory.

Firstly, we consider the generation of the topological Chern-Simons interactions present in five-dimensional maximal supergravity~\cite{Cremmer1} 
\begin{equation} \label{cubicinvariant-i}
S_{\text{CS}} = \int k_{\Lambda\Sigma\Delta} \ A^{\Lambda} \wedge F^{\Sigma} \wedge F^{\Delta}
\end{equation}
where $k_{\Lambda\Sigma\Delta}$ is constant and the ${\Lambda, \Sigma, \Delta}$ are $E_{6(6)}$ indices running from $1$ to $27$. This interaction does not involve the metric and does not admit linearisation. By supersymmetry, failure to generate it would indicate that the equations derived from the rank three and four tensor fields will not agree with those of gravity beyond linearised level. Similar calculations have been carried out, notably in the context of theories with eight supercharges, where is was shown how triangle diagrams with massive KK modes coming from the chiral six-dimensional fields in the loop generate five-dimensional Chern-Simons terms  \cite{Bonetti1, Bonetti2, Bonetti3, Ohmori:2014kda}. 
An important point here is that while KK modes of six-dimensional fields are involved, the calculation itself is carried out in five dimensions. As we show in section~\ref{sec:CS}, under reasonable assumptions, only the reduction of the six-dimensional supergravity generates \eqref{cubicinvariant-i} consistent with the $E_{6(6)}$ cubic invariant.

Since the KK modes considered here come from chiral six-dimensional fields, the above calculation is closely related to six-dimensional anomalies and index theorems. Since the exotic multiplets feature chiral fields, questions about anomalies arise naturally. One may object that these are formulated in the flat space, and only upon reduction does (linearised) five-dimensional gravity and diffeomorphism symmetry appear. The five degrees of freedom carried by the SDW field  are to be thought of as excitations of a five-dimensional metric, so that one does not expect six-dimensional diffeomorphism symmetry, but rather exotic symmetries that give rise to five-dimensional diffeomorphisms. 

In general, diffeomorphism invariance is a critical property for quantum supergravity theories. It corresponds to the conservation of the energy momentum tensor at the quantum level and can be checked via one-loop computations with the external states being gravitons. At the same time, it can also be interpreted as the anomalous transformation of the path integral measure of chiral fields under diffeomorphism transformations of the space-time. Diffeomorphism anomalies are equivalent to anomalies for local Lorentz symmetry up to local, non-polynomial counterterms (see e.g.~\cite{Bilal}). Thus, regardless of considerations of diffeomorphism symmetry, it makes sense to ask whether the non-gravitational $(4,0)$ theory is invariant under local Lorentz transformations on arbitrary background six-dimensional manifolds. This question can be answered by computing the gravitational anomalies in the conventional sense.

We find that the exotic fields of the $(4,0)$ theory lie inside the domains of certain Dirac operators, in much the same way that self-dual $p$-forms are found inside the signature complex (see e.g.~\cite{Eguchi:1980jx}). This fact is intimately related to the exotic multiplets arising as products of matter multiplets, and is very similar to the treatment of self dual $p$-forms as part of a bispinor field. As we shall see, for the exotic fields we simply have to take higher powers of the spinor representations. The explicit calculations can be found in the section~\ref{sec:anomalies}, with further details in appendix~\ref{app:Conv}. The conclusion is that both $(4,0)$ and $(3,1)$ multiplets have non-vanishing anomalies. In a way, the decomposition of the maximally supersymmetric multiplets mentioned above gives a heuristic explanation to this. The ordinary $(2,0)$ multiplets - gravity (GM) , gravitino (GoM) and tensor (TM) - while all chiral, have fields of different chirality appearing in them, so that 
a particular combination of them even becomes a non-chiral theory\footnote{In fact all three multiplets have proportional anomaly polynomials: $I_{\mathrm TM} = \frac14 I_{\mathrm GoM} = - \frac{1}{21} I_{\mathrm GM}$. }. On the contrary, the exotic multiplets have maximally aligned chiralities so that a cancellation naively appears much less likely, and indeed does not happen.

\medskip
\noindent
The structure of the paper is as follows. 
In section~\ref{sec:multiplets} we review the structure of the exotic six-dimensional multiplets. 
In section~\ref{sec:11d} we discuss how to relate the $\cN=(4,0)$ superalgebra to that of eleven-dimensional supergravity and how to interpret its charges in terms of $E_{8(8)}$ objects, within the framework of exceptional geometry. 
Section~\ref{sec:anomalies} contains the calculation of the anomaly polynomials for the local Lorentz symmetry of exotic multiplets, which are found to be non-factorisable. 
We go on to show that there is no conventional mechanism to generate the Chern-Simons couplings of five-dimensional maximal supergravity from the circle compactification of the $\cN = (4,0)$ fields in section~\ref{sec:CS}. 
In section~\ref{sec:h-theory} we present our construction of ``h-theories" on $T^3$-fibered geometries, whose solutions are seen to match the linearised equations of motion of the exotic graviton. 
Finally, we make some concluding remarks in section~\ref{sec:discussion}. 
Appendix~\ref{app:Ch6d} contains the construction of chiral supermultiplets in six-dimensions, while appendix~\ref{app:Conv} contains some conventions and technical details such as the anomalies calculations of section~\ref{sec:exotic}.


\section{Exotic supermultiplets in six dimensions}
\label{sec:multiplets}

In this section we provide some background discussion of the six-dimensional supermultiplets, whose highest spin field is a spin-2 boson which is not a graviton. 
The supermultiplets of extended Poincar\'e supersymmetry which correspond to possible local field theories were classified in~\cite{Strathdee}. Curiously, the list provided includes the multiplet which forms the basis for the $\cN=(4,0)$ theory of~\cite{Hull1}, as well as a similar multiplet with $\cN=(3,1)$ supersymmetry. However, similar multiplets with less supersymmetry were omitted. As these will form part of our discussion later, we review the detailed construction of such multiplets with $\cN=(1,0)$, $\cN=(2,0)$ and $\cN=(4,0)$ supersymmetry in appendix~\ref{app:Ch6d}.

The Lorentz group $SO(1,5)$ admits pseudo-Majorana-Weyl spinor representations, with such chiral spinors represented as pairs of four-component complex vectors $\zeta^A$ for $A=1,2$ satisfying the pseudo-reality condition $\zeta^B = \epsilon^{AB} (\zeta^B)^*$. 
For the case of maximal supersymmetry, which will be our main focus here, one has 32 real supercharges $Q$ which are made up of four such pseudo-Majorana-Weyl spinors. 
Clearly, up to interchange of chirality, the possible combinations of chiralities are $\cN = (4,0)$, $(3,1)$ or $(2,2)$. 
The corresponding R-symmetry groups of these superalgebras are $G^R_{(p,q)} = \Symp(2p)\times\Symp(2q)$\footnote{In this article, we denote by $\Symp(2n)$ the compact symplectic group of rank $n$.} for $\cN=(p,q)$ supersymmetry. 
The physical states form representations of the little group $G_{\text{little}} = \SU(2)\times\SU(2)\times G^R_{(p,q)}$, which is the subgroup of $\Spin(5,1)\times G^R_{(p,q)}$ preserving a null-momentum vector.
Representations of $G_{\text{little}}$ will be denoted as e.g. $\rep{(3, 3; 1, 1)}$, where we use a semicolon to separate the representations of the spacetime part and the R-symmetry part of the little group. 
The representations of these superalgebras with only states of helicity at most $2$ were classified in~\cite{Strathdee}, and are presented in Table \ref{tab:6d32}.

\begin{table}[h]
\centering
\begin{tabular}{|l|l|}
\hline
\scriptsize{$D =6, (p,q)=(4,0)$}                                           &$\rep{2^8=(5,1;1)+(3,1;27)+(1,1;42)}$    \\
\scriptsize{$SU(2) \times SU(2) \times \Symp(8) $     }                      &\quad\,$\rep{+(4,1;8)+(2,1;48)}$    \\
\scriptsize{$Q$ belongs to $\rep{(2,1;8)}$ }       &    \\ \hline
\scriptsize{$D =6, (p,q)=(3,1)$                  }                        &$\rep{2^8=(4,2;1,1)+(2,2;14,1)+(3,1;6,2)}$      \\
\scriptsize{$SU(2) \times SU(2) \times \Symp(6) \times \Symp(2)$ }              &\quad\,$\rep{+(1,1;14',2)+(4,1;1,2)}$  \\
\scriptsize{$Q$ belongs to $\rep{(2,1;6,1)+(1,2;1,2)}$  }    &\quad\,$\rep{+(3,2;6,1)}$  \\
                                            &\quad\,$\rep{+(2,1;14,2)+(1,2;14',1)}$   \\ \hline
\scriptsize{$D =6, (p,q)=(2,2)$               }                           &$\rep{2^8=(3,3;1,1)+(1,3;5,1)+(2,3;4,1)}$   \\
\scriptsize{$SU(2) \times SU(2) \times \Symp(4) \times \Symp(4)$        }       & \quad\,$\rep{+(3,1;1,5)+(1,1;5,5)+(2,1;4,5)}$   \\
\scriptsize{$Q$ belongs to} $\rep{(2,1;4,1)+(1,2;1,4)}$        &\quad\,$\rep{+(3,2;1,4)+(1,2;5,4)+(2,2;4,4)}$     \\
                                           &  \quad\quad Graviton in the $\rep{(3,3;1,1)}$ \\ \hline
\end{tabular}
\caption{Six-dimensional multiplets with 32 supercharges}
\label{tab:6d32}
\end{table}

We can see that in dimension six, the chiral superalgebra $\mathcal{N}=(4,0)$ has only one massless multiplet
\begin{equation}\label{6dmultiplet}
\rep{2^8 = (5, 1; 1) + (3,1; 27) + (1,1; 42) + (4,1; 8) + (2,1;48)}.
\end{equation}
The representations $\rep{(3,1; 27)}$, $\rep{(1,1; 42)}$ and $\rep{(2,1;48)}$ are immediately identified with anti-self-dual 2-forms $B^-_{\mu\nu}$, scalars $\phi$ and chiral fermions $\lambda$.

The field in the $\rep{(5,1;1)}$ representation of the little group $SU(2) \times SU(2) \times \Symp(8)$ has been labeled the exotic graviton \cite{Hull1} and is represented as a four-index object $C_{\mu\nu\rho\sigma}$ with the same index symmetries as the Riemann tensor
\begin{equation} \label{exoticgravitonsymmetry}
C_{\mu\nu\rho\sigma} = C_{\rho\sigma \mu\nu} = C_{[\mu\nu]\rho\sigma} = C_{\mu\nu[\rho\sigma]}
\end{equation}
\begin{equation}
C_{[\mu\nu\rho]\sigma} = 0
\end{equation}
The field strength (in flat spacetime) is defined at the linearised level as 
\begin{equation} \label{exoticgravitonfieldstrength}
G_{\mu\nu\rho\sigma\tau\kappa} = \partial_{[\mu}C_{\nu\rho][\sigma\tau,\kappa]}
\end{equation}
so that 
\begin{equation}
G_{\mu\nu\rho\sigma\tau\kappa}= G_{[\mu\nu\rho][\sigma\tau\kappa]}=G_{\sigma\tau\kappa\mu\nu\rho}
\end{equation}
and self-duality is imposed on both the first three and the last three indices $G = \star G = G \star$ where we use $\star$  to denote the Hodge-star operation
\begin{equation}
\label{eq:SD}
G_{\mu\nu\rho\sigma\tau\kappa}= (\star G)_{\mu\nu\rho\sigma\tau\kappa} = \frac{1}{3!} \epsilon_{\mu\nu\rho\alpha\beta\gamma}G^{\alpha\beta\gamma}_{\;\;\;\;\;\;\;\sigma\tau\kappa}.
\end{equation}
The $\rep{(4,1; 8)}$ part of the multiplet corresponds to a chiral fermionic $2$-form-spinor field $\psi_{\mu\nu}$, which we refer to as the exotic gravitino. It is anti-symmetric and its field strength is self-dual 
\begin{equation}
\begin{aligned} \label{exotic}
\psi_{\mu\nu} &= -\psi_{\nu\mu} \\
\chi_{\mu\nu\rho} &\equiv 3 \partial_{[\mu}\psi_{\nu\rho]} \;\; \text{and} \;\; \chi = \star \;\chi, 
\end{aligned}
\end{equation}

As shown in \cite{Hull1, Hull2, Hull3},  due to the double self-duality relations \eqref{eq:SD}, the dimensional reduction of $C_{\mu\nu\rho\sigma}$ to five dimensions gives a single linearised graviton. 
This can happen because the various components of $C_{\mu\nu\rho\sigma}$ which appear in the reduction become the dual graviton and the double-dual graviton. 
This mechanism is essentially a ``squared" version of the mechanism by which a self-dual two-form in six dimensions restricts to a single vector field in five. 
Similarly, the exotic gravitino reduces to a single gravitino in five dimensions, and in total the massless degrees of freedom of the $(4,0)$ multiplet reduce to exactly the fields of five-dimensional $\mathcal{N}=8$ supergravity. 
In addition, the Kaluza-Klein tower of massive modes arising from the massless $(4,0)$ states on circle match perfectly the $\frac{1}{2}$-BPS-states of the five-dimensional maximal supergravity. 
The scalars of the $(4,0)$ multiplet transform in the correct $\Symp(8)$ representation to form a non-linear sigma model based on the coset
\begin{equation}
E_{6(6)}/\Symp(8)
\end{equation}
which is the same as that parametrised by the scalars of five-dimensional maximal supergravity. However, as discussed in the introduction, it is not clear that the $E_{6(6)}$ symmetry uplifts to the six-dimensional theory.

The little group representation corresponding to the exotic graviton has the symmetries of a self-dual Weyl tensor in four-dimensional Euclidean space. For this reason, this field and the supermultiplets for which it is the top component are often described as ``self-dual Weyl" (see e.g.~\cite{magicpyramid}), and we will use this terminology interchangeably with the label ``exotic". 

We also see that in addition to the $(4,0)$ and $(2,2)$ maximal SUSY multilplets, there is the $(3,1)$ multiplet \cite{Hull1, Hull2, Strathdee}. The highest spin field corresponds to the $(\rep{4},\rep{2};\rep{1},\rep{1})$ representation of the little group $SU(2) \times SU(2) \times \Symp(6) \times \Symp(2)$ and is a three-index object $D_{\mu\nu\rho}$ which satisfies
\begin{equation}
\label{eq:exotic-D}
D_{\mu\nu\rho}=D_{[\mu\nu]\rho}, \quad D_{[\mu\nu\rho]}=0.
\end{equation}
Its field strength is defined as 
\begin{equation}
S_{\mu\nu\rho\sigma\kappa}=\partial_{[\mu} D_{\nu \rho][\sigma, \kappa]}
\end{equation}
and constrained to satisfy the one side self-duality constraint
\begin{equation}
S_{\mu\nu\rho\sigma\kappa}=\frac{1}{6} \epsilon_{\mu\nu\rho \alpha \beta \gamma} S^{\alpha \beta \gamma}_{\;\;\quad\sigma\kappa}.
\end{equation}
It can be shown that upon a circle reduction  the $(3,1)$ multiplet also yields the linearised five-dimensional $\mathcal{N}=8$ supergravity multiplet. The scalars of this multiplet naively appear to have a coset structure \cite{Hull1}
\begin{equation}
\frac{F_{4}}{\Symp(6) \times \Symp(2)}.
\end{equation}
but the vector and two-form fields appear only to transform in a representation of $F_{4(4)}$ when combined, making it unclear that this is a symmetry of the theory.

All three of these maximal six-dimensional supermultiplets can be thought of as products of smaller supermultiplets. 
The idea that maximal supergravity can be viewed as the square of maximal super Yang-Mills theory has proved to be extremely powerful for the computation of perturbative scattering amplitudes~\cite{CK1,CK2,CK3,Bern:2019prr,Cachazo:2018hqa}.
However, this view is also useful for simply understanding the multiplet structures purely at the level of the representation theory. 
In fact, one can also obtain the supergavity multiplets with various amounts of supersymmetry by considering products of tensor multiplets with supercharges of opposite chirality~\cite{Tensorsquared, magicpyramid}
\begin{equation}
\begin{split}
\left[(2,0)_{\text {tensor}}\right] \otimes\left[(0,2)_{\text {tensor}}\right]&=\left[(2,2)_{\text {sugra}}\right]\\
\left[(2,0)_{\text {tensor}}\right] \otimes\left[(0,1)_{\text {tensor}}\right]&=\left[(2,1)_{\text {sugra}}\right]\\
\left[(1,0)_{\text {tensor}}\right] \otimes\left[(0,1)_{\text {tensor}}\right]&=\left[(1,1)_{\text {sugra}}\right]\\
 \end{split}
\end{equation}
By contrast, the exotic multiplets arise when the tensor multiplets in the product have supercharges of aligned chirality:
\begin{equation}
\label{eq:chiralsquaring}
\begin{split}
\left[(2,0)_{\text {tensor}}\right] \otimes\left[(2,0)_{\text {tensor}}\right]&=\left[(4,0)_{\text {SD-Weyl}}\right]\\
\left[(2,0)_{\text {tensor}}\right] \otimes\left[(1,0)_{\text {tensor}}\right]&=\left[(3,0)_{\text {SD-Weyl}}\right]\\
\left[(1,0)_{\text {tensor}}\right] \otimes\left[(1,0)_{\text {tensor}}\right]&=\left[(2,0)_{\text {SD-Weyl}}\right] + \left[(2,0)_{\text{tensor}}\right],\\
 \end{split}
\end{equation}
 Note that there exists also a $\left[(1,0)_{\text {SD-Weyl}}\right]$  which can be constructed using the standard methods\cite{Strathdee}. The $(2,0)_{\text {SD-Weyl}}$ case is similar to the squaring of the $(1,0)$ vector multiplet, for which the product gives $\left[(2,0)_{\text {sugra}}\right] + \left[(2,0)_{\text{tensor}}\right]$.  
 
For the non-maximally supersymmetric case, notably  $(2,0)$ and $(1,0)$ the SD-Weyl multiplets exist in parallel to the standard supergravity multiplets \cite{deWit:2002vz}, and have the same numbers of degrees of freedom as the latter, but have fields living in the different representations of the symmetry groups, as summarised in the Table \ref{tab:comp}. Their field contents upon the circle reduction match, and correspond to the five-dimensional supergravity multiplets with 16 and 8 supercharges respectively.

\begin{table}[h]
\centering
\begin{tabular}{|l|l|l|}
 \hline
 & Exotic (or SD-Weyl)  &  Gravity \\ \hline
\scriptsize{$D =6, (p,q)=(2,0)$                  }                        &$\rep{(3,1;1)\times 2^4}$    &$\rep{(1,3;1)\times 2^4}$  \\
\scriptsize{$SU(2) \times SU(2) \times \Symp(4)$ }              &$\rep{=(5,1;1)+(3,1;5)+(1,1;1)}$ &$\rep{=(3,3;1)+(1,3;5)+(2,3;4)}$ \\
\scriptsize{$Q_{\frac{1}{2}}$ in $\rep{(2,1;4)}$  }    &$\rep{+(4,1;4)+(2,1;4)+(3,1;1)}$ &  \\
                                            \hline
\scriptsize{$D =6, (p,q)=(1,0)$               }                           &$\rep{(4,1;1)\times 2^2}$  & $\rep{(2,3;1)\times 2^2}$ \\
\scriptsize{$SU(2) \times SU(2) \times \Symp(2)$        }       &$\rep{=(5,1;1)+(3,1;1)+(4,1;2)}$ & $\rep{=(3,3;1)+(1,3;1)+(2,3;2)}$ \\
\scriptsize{$Q_{\frac{1}{2}}$ in} $\rep{(2,1;2)}$        & & \\
                                           &  &  \\ \hline
\end{tabular}
\caption{Six-dimensional SD-Weyl vs. gravity multiplets}
\label{tab:comp}
\end{table}

A detailed construction and a complete list of $(1,0)$, $(2,0)$ and $(4,0)$ multiplets with low spins can be found in Appendix \ref{app:Ch6d}.

Similar considerations apply to the last  maximally supersymmetric multiplet, which receives much less attention in this paper. The $(3,1)$ multiplet can be seen as a product of tensor and vector multiplets \cite{magicpyramid}
\begin{equation}
\begin{split}
\left[(2,0)_{\text {tensor}}\right] \times\left[(1,1)_{\text {vector}}\right] =\left[(3,1) \right]_{\text{exotic}} \, .\\
 \end{split}
\end{equation}


\section{The algebraic approach}
\label{sec:11d}

The theory of eleven-dimensional supergravity can be formulated with eleven-dimensional Lorentz symmetry non-manifest, but broken to a subgroup $\SO(10-d,1) \times \SO(d)$, as one would have in dimensional reductions of the theory. Remarkably, when this is done, one finds that this group can be enhanced~\cite{deWitNicolai} to a local symmetry $\SO(10-d,1)\times \dHd$, where $\dHd$ is the (double cover of) the maximal compact subgroup of the exceptional group which would appear in the corresponding torus compactification~\cite{CremmerJulia}. 
As one increases $d$, this exceptional group becomes infinite dimensional, as does the corresponding $\dHd$, and grand proposals as to how these infinite dimensional symmetries are realised in M theory have been put forward~\cite{West, DHN}. 
Recently, work has been done constructing the exceptional field 

For $d\leq7$, these exceptional symmetries give rise to exceptional generalised geometries~\cite{Hull-EGG,PW} which can be used to describe the internal sector of the theory~\cite{CSW2,CSW3}. The full theory can then be written with these symmetries manifest and the internal sector given by the generalised geometry formulation~\cite{HS,E77EFT}. 
Further, one finds that the formulation of exceptional geometry can describe also type IIA and IIB supergravity via the exact same equations. The only change is the choice of subgroup which corresponds to the action of spacetime diffeomorphisms on tensors (i.e. the choice of ``gravity line" in the language of~\cite{Kleinschmidt:2003mf}). There are two inequivalent embeddings of $\GL(d-1,\bbR)$ into $\Edd\times\bbR^+$, giving different decompositions of the exceptional theory into ordinary tensor fields~\cite{Schnakenburg:2001he,CSW2,HS}. One of these embeddings gives type IIA and the other type IIB. In the language of~\cite{HS}, this is phrased as the choice of ``section" of a higher dimensional space. 
Such sections are subspaces $V$ of (the dual of) the generalised tangent space such that $V \otimes V$ is null in particular $E_{d(d)}$ covariant projections of the tensor product space. In generalised geometry discussions, the subspace $V$ is simply the cotangent bundle of the underlying manifold.

In this section, we explore the possibility that a third choice of spacetime subgroup could give the $\cN=(4,0)$ theory of~\cite{Hull1}. 
In the half maximal setting, it was established that both the ten-dimensional type I theory and the six-dimensional $\cN=(2,0)$ theory could be seen in this way~\cite{Ciceri:2016hup}. 
However, the $\cN=(4,0)$ theory is not a standard type of gravitational theory, so we expect that the picture will be different. 
We will see here that some hints of its known features, at least at the linearised level, can be seen from this angle of investigation, 
but these will amount more to curiosities than conclusive evidence. 
An important realisation, though, is that there is no spacetime section inside the exceptional multiplet of charges, in the way that there is for standard supergravity, but only the embedding of the momentum charge, which does not solve the section condition and carries no natural action of a special linear group. 
We will also examine the corresponding pictures for exotic multiplets with $\cN=(2,0)$ and $\cN = (1,0)$ supersymmetry, finding the same pattern of behaviour. 

We begin by studying the embedding of the spin groups into $\Cliff(10,1;\bbR)$ and the relation of this to the higher dimensional enhanced symmetries $\dHd$. 
We then comment on the interpretation of these embeddings in terms of charges and how this could correspond to different spacetime groups inside the duality group $E_{8(8)}$.

\subsection{An almost universal construction of the maximal supersymmetry algebras}
\label{sec:superalgebras}

The maximal supersymmetry algebras can all be seen as subalgebras of a Lie superalgebra $\maxalg$, which we briefly describe. The generators of $\maxalg$ consist of 32 fermionic generators $Q^\alpha$, transforming as the $\rep{32}$ representation of $GL(32,\bbR)$. The anti-commutators of these give 528 bosonic generators $X^{\alpha\beta} = X^{(\alpha\beta)} = \{ Q^\alpha, Q^\beta \}$, which have vanishing brackets with the $Q$'s. Finally, we add the generators $\mathcal{M}_\alpha{}^\beta$ of $\mathfrak{gl}(32,\bbR)$ which act on the $Q$'s and $X$'s via the adjoint action.

We can recover a maximal supersymmetry algebra from $\maxalg$ by truncating the $\mathfrak{gl}(32,\bbR)$ generators to a subalgebra of the form $\mathfrak{spin}(D-1,1) \oplus \mathfrak{k}$, where in most cases\footnote{\label{ft:exception}This pattern does not always hold, e.g. for $D=4$ the maximal compact commutant is $\mathfrak{u}(8)$ while the R symmetry is $\mathfrak{su}(8)$. This can be understood in terms of the level decomposition of $KE_{11}$, where the extra $\mathfrak{u(1)}$ can be seen to have a higher level~\cite{Bossard:2019ksx}.} $\mathfrak{k}$ is the maximal compact commutant of $\mathfrak{spin}(D-1,1)$ inside $\mathfrak{gl}(32,\bbR)$ ($\mathfrak{k}$ is the R-symmetry algebra). Decomposing $Q^\alpha$ and $X^{\alpha\beta}$ under $\mathfrak{spin}(D-1,1) \oplus \mathfrak{k}$, we recover the supersymmetry algebra. 
It is easy to see why this prescription works: 
the generators $Q^\alpha$ and $X^{\alpha\beta}$ of the algebra $\maxalg$ are simply the supertranslational part, without specifying how they transform under the Lorentz symmetry and R symmetry. This is then fixed by choosing the subalgebra $\mathfrak{spin}(D-1,1) \oplus \mathfrak{k}\subset \mathfrak{gl}(32,\bbR)$.

We now want to view the algebra $\mathfrak{gl}(32,\bbR)$ as the irreducible matrix representation of the Clifford algebra $\Cliff(10,1;\bbR)$ with $\Gamma^{(11)} = \Gamma^0 \Gamma^1 \dots \Gamma^9 \Gamma^{10} = +\id$. Choosing the natural $\mathfrak{spin}(10,1)$ subalgebra (which has no compact commutant in $\mathfrak{gl}(32,\bbR)$), the $\rep{32}$ representation is irreducible, while the $\rep{528}$ decomposes into $\rep{11}+\rep{55}+\rep{462}$, so that $X$ becomes the momentum $P_\mu$, a 2-form $Z_{\mu\nu}$ and a 5-form $Z_{\mu_1 \dots \mu_5}$. We thus recover the standard eleven-dimensional supersymmetry algebra. 

The standard (non-chiral) maximal supersymmetry algebras in lower dimensions are then obtained by taking $\mathfrak{spin}(D-1,1)$ subalgebras of this $\mathfrak{spin}(10,1)$ and then examining their compact commutants in $\mathfrak{gl}(32,\bbR)$ to find the R-symmetry (though again there are exceptions to this rule -- see footnote~\ref{ft:exception}).
We can decompose the eleven-dimensional Lorentz indices into indices $\mu,\nu = 0, 1, \dots D-1$ for the ``external spacetime" $\mathfrak{spin}(D-1,1)$ Lorentz group and $m,n = 1,\dots,d$ the orthogonal group indices for the ``internal space".

We see that the parts of $X^{\alpha\beta}$ which form the momentum charge in $D$-dimensions are completely contained in the eleven-dimensional momentum charge $P_{\mu}$, and that the $D$-dimensional Lorentz group is contained in the eleven-dimensional Lorentz group by construction. In the corresponding supergravity theories, this can be interpreted as saying that the lower-dimensional spacetime is a subspace of the higher dimensional spacetime. 

However, in some dimensions $D$ there are alternative embeddings of $\mathfrak{spin}(D-1,1)$ into $\mathfrak{gl}(32,\bbR)$, such that the resulting supercharges $Q$ have different chiralities to those in the simple embeddings above. For example, a different embedding of $\mathfrak{spin}(9,1)$ to that above gives the $\cN=(2,0)$ supersymmetry algebra of type IIB supergravity in ten dimensions. A relatively clean way to see this is to construct the embedding explicitly in terms of the $\Cliff(10,1;\bbR)$ gamma-matrices, so this is what we do next.



\subsection{Spin embeddings into higher dimensional Clifford algebras}

We start by giving a general picture of some different ways that one can embed the Lie algebra of $\Spin(s+1,t)$ into $\Cliff(s+N,t)$. The construction is very explicit, using gamma matrices and a multitude of different indices. Readers who do not wish to indulge these details could skip straight to the examples.


\subsubsection{Different embeddings of $\Spin(s+1,t)$ into $\Cliff(s+N,t)$}

Let $i,j$ be indices for the vector representation of $\SO(s,t)$ taking values in $\{ -t, \dots, -1\}$ for the timelike directions and $\{ 1, \dots , s\}$ for the spacelike directions. Let $\Gamma^M$ be the gamma matrices generating $\Cliff(s+N,t)$, with the index $M$ similarly taking values in $\{ -t, 
\dots ,-1,1,\dots,s, s+1,\dots, s+N \}$. Introducing a further set of indices $I,J$ taking values in $\{ -t, 
\dots ,-1,1,\dots,s, s+1 \}$, consider the generators 
\begin{equation}
	\left\{ \hat\gamma^{IJ} \right\} = \left\{ \begin{array}{lcl} 
		\Gamma^{ij}, & \quad & I=i, J=j \\
		\Gamma^{i \; s+1 \; s+2 \; \dots \; s+n} & & I=i, J=s+1
		\end{array}\right.
\end{equation}
in which $s+1, \dots, s+n$ label $n$ spacelike directions in the space of signature $(s+N,t)$ which are invariant under $\SO(s,t)$. One can check that these generate $\Spin(s+1,t)$ or $\Spin(s,t+1)$, where the signature of the extra direction is determined by the value of $n$ as
\begin{equation}
\begin{array}{c|cccccccccc}
	n & 0 &1 & 2&3&4&5&6&7&8& \dots\\
	\hline
	\pm &-&+&+&-&-&+&+&-&-&\dots
\end{array}
\end{equation}
In what follows, we will take $n \in \{1,2,5,6,\dots\}$ so that the extra direction is spacelike ($+$ in the table). 

If we have that $s+t+1$ is even, we can calculate the chirality matrix\footnote{In our notation, if a Clifford algebra is generated by gamma matrices $\gamma^{i}$, with the index $d$ running over $d$ values, then $\gamma^{(d)} = \prod_i \gamma^i$ is the product of the $d$ distinct gamma matrices.} $\hat\gamma^{(s+t+1)}$ for the embedded $\Cliff(s+1,t)^{\text{even}}$. This tells us how the $(s+N,t)$ spinor decomposes into $(s+1,t)$ spinors. In particular, we note that if $n=N$ then this is
\begin{equation}
	\hat\gamma^{(s+t+1)} = \Gamma^{-t \; -t+1}
		\dots \Gamma^{s-2 \; s-1} \Gamma^{s \; s+1 \; \dots \;s+N }
	= \Gamma^{-t} \dots \Gamma^{-1} \Gamma^1 \dots \Gamma^{s+N}
	= \Gamma^{(s+t+N)}
\end{equation}
which is the product of the gamma matrices in signature $(s+N,t)$ (i.e. $\pm \id$ or $\pm \ii \id$ if $s+t+N$ is odd, or the chirality matrix if $s+t+N$ is even). Thus, if in $\Cliff(s+N,t)$ we have $\Gamma^{(s+t+N)} = +\id$ then all spinors will decompose to have the same (positive) chirality. This will appear in our examples in the next section.

\subsubsection{Examples}
\label{sec:Spin-Egs}

{\bf Example 1 : Type II into eleven dimensions} We start by looking at the nine-dimensional spin group $\Spin(8,1)$, generated by $\Gamma^{ij}$, for $i,j = 0,1,\dots, 8$, inside $\Cliff(10,1)$. We then consider how we could add generators to these to enhance the group to give a $\Spin(9,1)$ inside $\Cliff(10,1)$. We see two inequivalent ways to do this, leading to decompositions of the eleven-dimensional spinor into two spinors of different chirality or into two spinors of the same chirality under the $\Spin(9,1)$ subgroups. These correspond to type IIA (non-chiral) and type IIB (chiral) respectively. 

For type IIA we simply add the spin generators corresponding to including one more direction of the eleven-dimensional space, so that our $\Spin(9,1)$ group is generated by
\begin{equation}
	\left\{ \hat\gamma^{IJ} \right\} = \left\{ \Gamma^{ij}, \Gamma^{\, i \, 9} \right\}
\end{equation}
which gives (recall that the $\Gamma$-matrices are the $\Cliff(10,1)$ gamma matrices and we take $\Gamma^{(11)} = \Gamma^0 \Gamma^1 \Gamma^2 \dots \Gamma^{10} = +\id$)
\begin{equation}
	\hat\gamma^{(10)} = \Gamma^{01} \Gamma^{23} 
		\dots \Gamma^{78} \Gamma^{89}
	= \Gamma^{(11)} \Gamma^{10}
	= \Gamma^{10} = \begin{pmatrix} \id & 0 \\ 0 & -\id \end{pmatrix}
	\hs{10pt} \text{(in an appropriate basis)}
\end{equation}
so we see that the eleven-dimensional spinor decomposes into one positive and one negative chirality ten-dimensional spinors. 

The commutant of the type IIA $\mathfrak{spin}(9,1)$ subalgebra inside $\mathfrak{sl}(32,\bbR)$ is generated by 
$\{ \Gamma^{10} \}$. This generates an $\bbR^+$ subgroup of $\SL(32,\bbR)$, and so there is no non-trivial compact commutant. This matches the R-symmetry of type IIA.

For type IIB, we instead take
\begin{equation}
	\left\{ \hat\gamma^{IJ} \right\} = \left\{ \Gamma^{ij}, \Gamma^{i \, 9 \, 10} \right\}
\end{equation}
leading to
\begin{equation}
	\hat\gamma^{(10)} = \Gamma^{01} \Gamma^{23} 
		\dots \Gamma^{8 \, 9\,10}
	= \Gamma^{(11)} = \id
\end{equation}
Thus, the 32 component spinor decomposes into only positive chirality spinors for this $\Spin(9,1)$ subgroup, as all spinors have eigenvalue $+1$ under $\hat\gamma^{(10)}$. 

The commutant of the type IIA $\mathfrak{spin}(9,1)$ subalgebra inside $\mathfrak{sl}(32,\bbR)$ is generated by 
$\{ \Gamma^{9 \,10} \}$. This generates an $\SO(2)$ subgroup, which matches the R-symmetry of type IIB.
\\~\\
{\bf Example 2 : Six-dimensional $\cN = (4,0)$ into eleven dimensions} We start with the $\Spin(4,1)$ generators $\Gamma^{ij}$, for $i,j = 0,1,\dots, 4$, inside $\Cliff(10,1)$ and look to extend this to an embedding of $\Spin(5,1)$. Taking the additional generators $\Gamma^{i5}$ would result in the $\mathfrak{spin}(5,1)$ subalgebra for standard $\cN=(2,2)$ supergravity in six-dimensions. If instead we take
\begin{equation}
\label{eq:4,0-spin51}
	\left\{ \hat\gamma^{IJ} \right\} = \left\{ \Gamma^{ij}, \Gamma^{i 56789 \, 10} \right\}
\end{equation}
then, similarly to the situation for type IIB above, we obtain
\begin{equation}
	\hat\gamma^{(6)} = \Gamma^{01} \Gamma^{23} 
		\dots \Gamma^{456789 \,10}
	= \Gamma^{(11)} = +\id
\end{equation}
so that again the 32 component spinor decomposes into only positive chirality spinors for this $\Spin(5,1)$ subgroup.

The commutant of this $\mathfrak{spin}(5,1)$ subalgebra inside $\mathfrak{sl}(32,\bbR)$ is generated by $\{ \Gamma^{m}, \Gamma^{m_1 m_2}, \dots, \Gamma^{m_1 \dots m_6} \}$ for $m,n = 5,6,\dots,10$. Of these, only the generators $\{ \Gamma^{m_1m_2}, \Gamma^{m_1m_2m_3}, \Gamma^{m_1\dots m_6} \}$ square to $-\id$ and hence are compact. The compact commutant group these generate is $\Symp(8)$, which matches the R-symmetry of the $\cN=(4,0)$ multiplet.


\subsubsection{Irreducible decomposition of charges}
\label{sec:central-charges}

In the examples of section~\ref{sec:Spin-Egs} we gave the embedding of two inequivalent $\Spin(9,1)$ groups and two inequivalent $\Spin(5,1)$ groups into $\Cliff(10,1;\bbR)$. In terms of $\Spin(10,1)$ objects the charges ($X_{\alpha\beta}$ above) can be written as an eleven-dimensional vector, two-form and five-form via
\begin{equation}
	\{Q_\alpha, Q_\beta \} = P_M (\tilde{C}\Gamma^{M})_{\alpha\beta}
		+ \tfrac12 Z_{MN} (\tilde{C}\Gamma^{MN})_{\alpha\beta}
		+ \tfrac{1}{5!} Z_{M_1 \dots M_5} (\tilde{C}\Gamma^{M_1 \dots M_5})_{\alpha\beta}
\end{equation}
where we have explicitly included the transpose intertwiner $\tilde{C}$ which satisfies $\tilde{C}\Gamma^M \tilde{C}^{-1} = -(\Gamma^{M})^T$ and $\tilde{C}^T = -\tilde{C}$. 
We can then calculate explicitly the action of our other $\Spin$ groups on the charges $(P,Z_{(2)},Z_{(5)})$, written in terms of a decomposition under the common subgroup with $\Spin(10,1)$. 
We provide a sketch of these calculations here, noting that our $\Spin$ groups are acting as subgroups of $\SL(32,\bbR)$. This means that the action of a matrix $M$ is given by
\begin{equation}
\begin{aligned}
	M \cdot (\tilde{C} \Gamma^{\dots}) 
		&= -M^T (\tilde{C} \Gamma^{\dots}) - (\tilde{C} \Gamma^{\dots}) M \\
		&= -\tilde{C} \Big( (\tilde{C}^{-1} M^T \tilde{C}) \Gamma^{\dots} + \Gamma^{\dots} M \Big) \\
\end{aligned}
\end{equation}
\\~\\
{\bf Example 1 : Type II into eleven dimensions}
\\~\\
For type IIA the generators of the relevant $\Spin(9,1)$ were found above to be $M^{\mu\nu} = \Gamma^{\mu\nu}$ and $M^{\mu 9} = \Gamma^{\mu 9}$, for $\mu,\nu = 0,1,\dots,8$ the vector indices of $\Spin(8,1)$. Clearly, these simply generate a $\Spin(9,1)$ subgroup of $\Spin(10,1)$ preserving the tenth spatial direction. As such it is clear that the $\Spin(9,1)$ irreducible combinations of charges will be
\begin{equation}
\label{eq:IIA-SUSY-charges}
\begin{aligned}
	&(P^\mu, P^9) \hs{30pt}  & & (Z_{\mu\nu}, Z_{\mu 9}) 
		\hs{30pt} && (Z_{\mu_1 \dots \mu_5}, Z_{\mu_1 \dots \mu_4 9}) \\
	& (P^{10}) & & (Z_{\mu 10}, Z_{9 \; 10}) 
		&&  (Z_{\mu_1 \dots \mu_4 10}, Z_{\mu_1 \dots \mu_3 9 \; 10})
\end{aligned}
\end{equation}
We can check this explicitly, noting that 
\begin{equation}
	M^{\mu 9} \cdot (\tilde{C} \Gamma^{\dots}) 
		= \tilde{C} [ M^{\mu 9} ,\Gamma^{\dots} ]
\end{equation}
From this, we can see that as e.g. $[M^{\mu 9}, \Gamma^{10}] = 0$ we have that $P^{10}$ is invariant under our $\Spin(9,1)$. Similarly, we see that $[M^{\mu 9}, \Gamma_{\nu 10}] = 2 \delta^\mu{}_\nu \Gamma^{9}{}_{10}$ and $[M^{\mu 9}, \Gamma_{9\; 10}] = - 2 \Gamma^\mu{}_{10}$ so that $(Z_{\mu 10}, Z_{9\; 10})$ forms a vector of $\Spin(9,1)$. 

For type IIB, the situation is more complicated as the generators of the relevant $\Spin(9,1)$ are now $M^{\mu\nu} = \Gamma^{\mu\nu}$ and $M^{\mu 9} = \Gamma^{\mu 9\;10}$. We then have 
\begin{equation}
	M^{\mu 9} \cdot (\tilde{C} \Gamma^{\dots}) 
		= -\tilde{C} \{ M^{\mu 9} ,\Gamma^{\dots} \}
\end{equation}
We must then calculate the anti-commutators to see which charges are rotated into each other by $M^{\mu 9}$. For example, $ \{ M^{\mu 9} ,\Gamma^{\nu} \} = 2 g^{\mu\nu} \Gamma^{9\; 10}$ and $\{ M^{\mu 9} ,\Gamma^{9\; 10} \} = -2 \Gamma^\mu$, so that $(P^\mu, Z_{9\; 10})$ now forms a vector of this $\Spin(9,1)$. Continuing in this way, one finds that the $\Spin(9,1)$ irreducible combinations are
\begin{equation}
\label{eq:IIB-SUSY-charges}
\begin{aligned}
	&(P^\mu, Z_{9\; 10}) \hs{30pt}  & & (Z_{\mu \nu}, Z_{\mu\nu\lambda 9\; 10}) 
		\hs{30pt} && (Z_{\mu_1 \dots \mu_5}) \\
	&  & & (Z_{\mu i}, P^{i}) 
		&&  (Z_{\mu_1 \dots \mu_4 i})
\end{aligned}
\end{equation}
where $i = 9,10$. In ten dimenions, these are a vector, a three-form, a self-dual five-form and doublets of vectors and self-dual five forms, which are precisely the charges appearing on the right hand side of the supersymmetry algebra for type IIB.
\\~\\
{\bf Example 2 : Six-dimensional $\cN = (4,0)$ into eleven dimensions}

Let us now perform the same calculations for the $\cN = (4,0)$ embedding of $\Spin(5,1)$ in~\eqref{eq:4,0-spin51}. Letting $\mu,\nu = 0,1,\dots,4$, we have the generators $M^{\mu\nu} = \Gamma^{\mu\nu}$ and $M^{\mu 5} = \Gamma^{\mu 56789\;10}$, leading to 
\begin{equation}
	M^{\mu 5} \cdot (\tilde{C} \Gamma^{\dots}) 
		= -\tilde{C} \{ M^{\mu 5} ,\Gamma^{\dots} \}
\end{equation}
Calculating the relevant anti-commutators, writing indices $m,n = 5,6,\dots, 10$, organises the charges into $1+6+6+15$ vectors of $\Spin(5,1)$
\begin{equation}
\label{eq:6d-vector-SUSY-charges}
\begin{aligned}
	&(P^\mu, Z_{\mu_1 \dots \mu_5}) \hs{30pt}  & & (Z_{\mu_1 \dots \mu_4 m}, P^m) 
	\hs{30pt}  & & (Z_{\mu m}, Z_{m_1\dots m_5}) 
		\hs{30pt} && (Z_{\mu p_1 \dots p_4}, Z_{mn}) 
\end{aligned}
\end{equation}
together with $1+15 +20$ self-dual three-forms
\begin{equation}
\begin{aligned}
	&(Z_{\mu\nu}) \hs{30pt}  & &  (Z_{\mu_1 \mu_2\mu_3 mn})  
	\hs{30pt}  & &(Z_{\mu\nu m_1 m_2 m_3}) 
\end{aligned}
\end{equation}
Of course, these charges precisely agree with the representations expected on the right hand side of the supersymmetry algebra~\eqref{eq:6d-superalg}, and one can check that they combine into representations of $\Symp(8)$ as generated by $\{ \Gamma^{mn}, \Gamma^{mnp}, \Gamma^{m_1 \dots m_6} \}$. 

\subsection{Dimensional splits, hidden symmetries and the 6d space}
\label{sec:Hd-charges}

Consider the formulation of eleven-dimensional supergravity on a product space, as considered in~\cite{deWitNicolai,CSW2,CSW3,HS}. 
Letting $\mu,\nu = 0, 1, \dots 10-d$ be spacetime indices for the external space, and $m,n = 1,\dots,d$ be those for the internal space, we have that the hidden symmetry group $\dHd$ can be realised inside $\Cliff(10,1;\bbR)$ with the generators 
\begin{equation}
\label{eq:Hd-generators}
	\mathfrak{h}_d \sim \left\{ \Gamma^{m_1m_2},  \Gamma^{m_1m_2m_3},
		\Gamma^{m_1 \dots m_6 }, \Gamma^m \Gamma^{m_1\dots m_8} \right\}
\end{equation}
for $d\leq8$, where for $d<8$ we truncate the generators which are automatically zero by antisymmetry. The first generator $\Gamma^{m_1m_2}$ is simply the generator of $\Spin(d)$, while the remaining terms correspond to the fields of the theory: the three-form $A_3$, its magnetic dual $\tA_6$ and the conjectured dual graviton~\cite{Hull:1997kt,West} $\tilde{h}_{1,8}$. 

To relate the spin embeddings of the previous section to this formalism, we need to look at the parts of the spin group which are in common in the two descriptions. 
For example, consider a dimensional split with seven external dimensions. The (continuous) U-duality group is $E_{4(4)}\simeq \SL(5,\bbR)$ and we write our theory in terms of objects transforming under $\GL(7,\bbR)\times \SL(5,\bbR)\times\bbR^+$. 
To describe eleven-dimensional supergravity in the relevant generalised geometry formalism, the generalised tangent space on the internal four-dimensional part of the space is
\begin{equation}
\label{eq:4d-charges}
	E \simeq T_4 \oplus \Lambda^2 T^*_4
\end{equation}
where $T_4$ transforms under the natural $\GL(4,\bbR)$ group of the frame bundle in four dimensions. $E$ itself transforms as a ten-dimensional representation of $\SL(5,\bbR)\times\bbR^+$. 
We view this simply as the vector space of charges of the objects living only in these four dimensions, here the four-dimensional momentum and the M2-branes wrapping directions in the four-dimensional space. 
The analogue of the spin group then becomes $\Spin(6,1)\times\Spin(5)$, which is generated by the eleven-dimensional $\Gamma$-matrices ($\mu,\nu = 0,1,\dots,6$ and $m,n = 7,8,9,10$)
\begin{equation}
	\left\{ \Gamma^{\mu\nu}, \Gamma^{m_1m_2},  \Gamma^{m_1m_2m_3} \right\}
\end{equation}
The first two sets of generators in the list generate part of the usual spacetime spin group $\Spin(6,1)\times\Spin(4) \subset \Spin(10,1)$, while the $\Gamma^{m_1m_2m_3}$ enhance the $\Spin(4)$ factor to the $\Spin(5)$ hidden symmetries which are not manifest in the standard  formulation with manifest eleven-dimensional covariance. The intersection of the $\Spin(9,1)$ groups relevant to type IIA and type IIB with this are then each isomorphic to $\Spin(6,1)\times\Spin(3)$.

With this dimensional split in place, the above discussion of extending the $\Spin(8,1)$ in nine dimensions to $\Spin(9,1)$ for type IIA or type IIB becomes a discussion of how to extend the $\Spin(6,1)\times\Spin(2)$ generated by ($\mu,\nu = 0,1,\dots,6$ and $\um,\un = 7,8$)
\begin{equation}
	\left\{ \Gamma^{\mu\nu}, \Gamma^{\um_1 \um_2} \right\}
\end{equation}
to $\Spin(6,1)\times\Spin(3)$.

In type IIA, 
the relevant $\Spin(6,1)\times\Spin(3)$ is generated by
\begin{equation}
\label{eq:IIA-spin3}
	\left\{ \Gamma^{\mu\nu}, \Gamma^{\um_1 \um_2}, \Gamma^{\um 9} \right\}
\end{equation}
and this simply corresponds to including one more of the spatial directions rotated into each other by the eleven-dimensional spin group.
To see this more explicitly, we decompose the generalised tangent space~\eqref{eq:4d-charges} under the $GL(2,\bbR)$ containing the $\Spin(2)$ factor in our $\Spin(6,1)\times\Spin(2)$, giving
\begin{equation}
\label{eq:4d-charges-decomp}
	E \simeq (T_2 \oplus \bbR_9 \oplus \bbR_{10}) \oplus (\Lambda^2 T^*_2 \oplus T^*_2 \oplus T^*_2 \oplus \bbR_{9,10})
\end{equation}
We then consider which parts of this are combined into irreducible representations of the $\Spin(3)$ factor in~\eqref{eq:IIA-spin3}, 
which is the compact subgroup of an $\SL(3,\bbR)$ with generators $(T_2 \otimes T_2^*) \oplus (T_2\otimes \bbR_9) \oplus (T_2^*\otimes \bbR_9^*)$. 
We see that this $\Spin(3)$ rotates $T_2$ into $\bbR_9$, forming $T_3 = T_2 \oplus \bbR_9$. 
This $\SL(3,\bbR)$ can be extended to a $\GL(3,\bbR)$ inside $\SL(5,\bbR)\times\bbR^+$ containing our $\Spin(3)$ and $T_3$ becomes its vector representation. We then have
\begin{equation}
\label{eq:SL5-charges-IIA}
	E \simeq (T_3 \oplus \bbR_{10}) \oplus (\Lambda^2 T^*_3 \oplus T^*_3) 
\end{equation}
with the internal momentum charges spanning the $T_3$ factor, as this is the vector representation of the corresponding general linear group. Thus, our ten-dimensional spacetime for type IIA then has directions corresponding to the seven external dimensions and the three directions in $T_3$. 
These are simply ten of the original eleven directions we started with in the first place. 
The passage from~\eqref{eq:4d-charges-decomp} to~\eqref{eq:SL5-charges-IIA} exactly mirrors the discussion of the charges in the supersymmetry algebra~\eqref{eq:IIA-SUSY-charges}, which when restricted to the singlets of $\Spin(6,1)$, reduces to the combinations
\begin{equation}
\begin{aligned}
	&(P^{\um}, P^9) \hs{30pt} && (P^{10})  \hs{30pt} & & (Z_{\um\un}, Z_{\um 9})  \hs{30pt} && (Z_{\um 10}, Z_{9 \; 10}) 
\end{aligned}
\end{equation}

For our type IIB embedding of $\Spin(9,1)$ in $\Cliff(10,1)$, the intersection with $\Spin(6,1)\times\Spin(5)$ is instead the $\Spin(6,1)\times\Spin(3)$ generated by
\begin{equation}
\label{eq:IIB-spin3}
	\left\{ \Gamma^{\mu\nu}, \Gamma^{\um_1 \um_2}, \Gamma^{\um 9 \, 10} \right\}
\end{equation}
We again look at which directions in~\eqref{eq:4d-charges-decomp} are rotated into each other by this $\Spin(3)$ group. 
In this case, the $\Spin(3)$ 
is contained in an $\SL(3,\bbR)$ with generators $(T_2 \otimes T_2^*) \oplus (T_2\otimes \bbR_9\otimes \bbR_{10}) \oplus (T_2^*\otimes \bbR_9^*\otimes \bbR_{10}^*)$
which rotates $T_2$ into $\bbR_{9,10}$ and these are combined into $T_3'$. This is again the fundamental representation of a $\GL(3,\bbR) \subset \SL(5,\bbR)\times\bbR^+$ containing our $\Spin(3)$ and the full generalised tangent space then becomes
\begin{equation}
	E \simeq T_3' \oplus T_3'^* \oplus T_3'^* \oplus \Lambda^3 T_3'^*
\end{equation}
In the type IIB case, the momentum direction we have added to $T_2$ corresponds to the charge of the $M2$-brane wrapping the $9$ and $10$ directions in the eleven-dimensional picture, as in the well-known duality between type IIB on $S^1$ and M theory on $T^2$~\cite{Schwarz:1995dk,Aspinwall:1995fw,Schwarz:1995jq}. Again, the combinations of charges which become representations of $\GL(3,\bbR)$ perfectly match those found in~\eqref{eq:IIB-SUSY-charges} restricted to the singlets of $\Spin(6,1)$: 
\begin{equation}
\begin{aligned}
	&(P^{\um}, Z_{9\; 10}) \hs{30pt}  & &  (Z_{\um i}, P^{i})   \hs{30pt}  & & (Z_{\um \un}) \\ 
\end{aligned}
\end{equation}

This discussion of type IIA and type IIB is usually presented in the exceptional geometry literature in terms of these inequivalent embeddings of the general linear groups into the exceptional groups~\cite{Schnakenburg:2001he,CSW2,CSC} (different ``gravity lines") or different solutions to a section condition~\cite{Blair:2013gqa,HS}. However, we wanted to start instead from the details of the corresponding spin groups and central charges, as in our main case of interest in this article that is the most accessible information.

Let us now consider the embedding of $\Spin(5,1)$ into $\Cliff(10,1)$ given in~\eqref{eq:4,0-spin51}. By naive comparison with~\eqref{eq:IIB-spin3} and its interpretation, one could expect that the sixth direction in this case could correspond to the charge of some six-brane in the eleven-dimensional picture. However, M-theory does not contain such an object (see~\cite{Hull:1997kt} for a full discussion of this point). We will see that in fact, the new generator can be embedded into the last generator listed in~\eqref{eq:Hd-generators}, corresponding to the dual graviton. This exists only for dimensional splits with three external dimensions or fewer. As the only case with a finite-dimensional duality group is that of three external dimensions, for convenience we choose to examine the situation in that framework.

Thus we consider a $(3+8)$-dimensional split of eleven-dimensional supergravity. The corresponding generalised geometry description would feature objects transforming under $\GL(3,\bbR)\times E_{8(8)}\times\bbR^+$ and the analogue of the spin group inside this would be $\Spin(2,1)\times\SO(16)$. In fact, for our purposes it will suffice to truncate $E_{8(8)}\times\bbR^+$ to the $\SL(9,\bbR)\times\bbR^+$ sector which contains only the graviton and dual-graviton fields~\cite{CSC}. In this subsector, the charges on the eight-dimensional part of the space transform in the rank two antisymmetric bivector representation of $\SL(9,\bbR)$, which has the $\GL(8,\bbR)$ decomposition
\begin{equation}
\label{eq:8d-charges}
	E \simeq T_8 \oplus (T^*_8 \otimes \Lambda^7 T^*_8) \oplus (\Lambda^8 T^*_8 \otimes \Lambda^8 T^*_8 \otimes T^*_8)
\end{equation}
while the decomposition of the adjoint of $\SL(9,\bbR)$ is
\begin{equation}
\label{eq:SL9-adjoint}
	\adj \SL(9,\bbR) \simeq  (T_8 \otimes T^*_8) \oplus (\Lambda^8 T_8 \otimes T_8) \oplus (\Lambda^8 T^*_8\otimes T^*_8)
\end{equation}
The corresponding spin group is $\Spin(2,1)\times\Spin(9)$ generated by ($\mu,\nu = 0,1,2$ and $\hat{m},\hat{n} = 3,4,\dots,9,10$)
\begin{equation}
\label{eq:Spin9}
	\left\{ \Gamma^{\mu\nu}, \Gamma^{\hat{m}\hat{n}}, \Gamma^{\hat{m}}\Gamma^{(8)} \right\}
	\hs{20pt} \text{where} \hs{20pt} \Gamma^{(8)} 
		= \Gamma^3 \Gamma^4 \dots \Gamma^9 \Gamma^{10}
\end{equation}
The intersection of the $\Spin(4,1)$ group from section~\ref{sec:Spin-Egs} with the $\Spin(2,1)\times\Spin(9)$ considered here is then $\Spin(2,1)\times\Spin(2)$, which is generated by %
\begin{equation}
	\left\{ \Gamma^{\mu\nu}, \Gamma^{ab} \right\}
\end{equation}
Here we define the index ranges $\mu,\nu = 0,1,2$ and $a,b = 3,4$, while $m,n = 5,6,7,8,9,10$ and $\um,\un = 6,7,8,9,10$ so that $\hat{m} = (a,m) = (a,5,\um)$. 
We seek to enhance this to the $\Spin(2,1)\times\Spin(3)$ groups which are the intersections of the $\Spin(5,1)$ groups described in section~\ref{sec:Spin-Egs} with $\Spin(2,1)\times\Spin(9)$. The $\Spin(2,1)\times\Spin(3)$ of standard $\cN=(2,2)$ supergravity in six dimensions is generated by
\begin{equation}
	\left\{ \Gamma^{\mu\nu}, \Gamma^{ab}, \Gamma^{a5} \right\}
\end{equation}
which corresponds simply to including one more 
of the standard eleven-dimensional momenta to give a total of six spacetime momenta out of the eleven. 

However, the $\Spin(5,1)$ group which corresponds to the $\cN=(4,0)$ decomposition gives rise to a $\Spin(2,1)\times\Spin(3)$ group generated by
\begin{equation}
\label{eq:4,0-into-spin9}
	\left\{ \Gamma^{\mu\nu}, \Gamma^{ab}, \Gamma^{a}\Gamma^{(8)} \right\}
\end{equation}
which are clearly contained in the generators of $\Spin(2,1)\times\Spin(9)$ in~\eqref{eq:Spin9}. 

To see how to interpret this in terms of charges, we note that this $\Spin(9)$ is contained inside the $\SL(9,\bbR)$ group generated by~\eqref{eq:SL9-adjoint}. Decomposing 
\begin{equation}
\label{eq:decomp-1}
T_8 = A_3 \oplus B_5 = C_2\oplus \bbR_5 \oplus B_5
\end{equation}
(according to $\hat{m} = (a,m) = (a,5,\um)$) we see that the $\Spin(9)$ generators featuring in~\eqref{eq:4,0-into-spin9} are inside the $\SL(3,\bbR)$ subgroup generated by
\begin{equation}
\label{eq:SL3-4,0-adjoint}
	(C_2 \otimes C^*_2) 
		\oplus (\Lambda^2 C_2 \otimes \Lambda^5 B_5 \otimes C_2) 
		\oplus (\Lambda^2 C^*_2 \otimes \Lambda^5 B^*_5 \otimes C^*_2) 
			\subset \adj \SL(9,\bbR)
\end{equation}
The five-dimensional dual graviton field (for the five-dimensional spacetime consisting of the external directions together with the momenta in $C_2$) corresponds to the term $C^* \otimes \Lambda^2 C^*$, and we see that this is the term appearing in~\eqref{eq:SL3-4,0-adjoint}. 
We then look at the decomposition of the charges~\eqref{eq:8d-charges}
\begin{equation}
\label{eq:8d-charges-decomp}
\begin{aligned}
	E &\simeq C \oplus \bbR \oplus B \\
		&\oplus ( C^* \otimes C^* \otimes \Lambda^5 B^*) 
		\oplus ( C^* \otimes \Lambda^2 C^* \otimes \Lambda^5 B^*)
		\oplus ( C^* \otimes \Lambda^5 B^*) 
		\oplus ( \Lambda^2  C^* \otimes \Lambda^5 B^*) \\
		&\oplus ( B^* \otimes C^* \otimes \Lambda^5 B^*)
		\oplus ( B^*  \otimes \Lambda^2  C^* \otimes \Lambda^5 B^*) \\
		&\oplus ( C^* \otimes \Lambda^2  C^* \otimes \Lambda^4 B^*)
		\oplus ( \Lambda^2  C^* \otimes \Lambda^4 B^*)
		\oplus ( \Lambda^2  C^* \otimes B^* \otimes \Lambda^4 B^*) \\
		&\oplus \Big[ ( \Lambda^2 C^* \otimes \Lambda^5 B^*)^2 \Big]
		\oplus \Big[ ( \Lambda^2 C^* \otimes \Lambda^5 B^*)^2 \otimes C^* \Big]
		\oplus \Big[ ( \Lambda^2 C^* \otimes \Lambda^5 B^*)^2 \otimes B^* \Big]
\end{aligned}
\end{equation}
and see which parts are combined into representations of this $\SL(3,\bbR)$. 
Here we find a very different result to the $\cN=(2,2)$ case. The terms which combine with $C$ to form an $\SL(3,\bbR)$ representation make up not a triplet but an octuplet of $\SL(3,\bbR)$:
\begin{equation}
\label{eq:SL3-octuplet}
\begin{aligned}
	C \oplus \Big( C^* \otimes C^* \otimes \Lambda^5 B^* \Big) 
		\oplus \Big[ ( \Lambda^2 C^* \otimes \Lambda^5 B^*)^2 \otimes C^* \Big]
\end{aligned}
\end{equation}
This subspace does not satisfy the section condition of $E_{8(8)}$ exceptional field theory\footnote{The $E_{8(8)}$ section condition determines whether a subspace $V \subset{E}$ has $V \otimes V$ null in the projection $\rep{248}\times\rep{248} \ra \rep{1} + \rep{248} + \rep{3875}$. This tensor product contains terms contracting $T_8$ into the $\Lambda^7 T_8^*$ factor of $T_8^* \otimes \Lambda^7 T_8^*$ and into both factors of $T_8^* \otimes \Lambda^7 T_8^*$. It is the non-vanishing of these contractions which demonstrate that several subspaces we consider in this article do not satisfy this condition.}, and thus it seems difficult to interpret it as the coordinate directions of a higher-dimensional spacetime. 
Clearly, it also does not match the naive expectation of~\eqref{eq:6d-vector-SUSY-charges}, which would suggest that the two five-dimensional momenta $P^a$ in $C$ would simply be joined by one additional charge $Z_{\mu_1 \mu_2\mu_3 ab}$ to form a triplet. 
We will examine this further in section~\ref{sec:SO3-triplet}. 
The decompositions~\eqref{eq:SL3-octuplet} and~\eqref{eq:SL3-4,0-adjoint} are essentially the same as~\eqref{eq:8d-charges} and~\eqref{eq:SL9-adjoint} and are the charges and adjoint relevant for five-dimensional pure gravity reduced to three dimensions, with the $\SL(3,\bbR)$ simply interpreted as the Ehlers symmetry. 

We note also that the $\SL(3,\bbR)$ subgroup~\eqref{eq:SL3-4,0-adjoint} is conjugate to the standard one by an $\SL(9,\bbR) \subset E_{8(8)}$ transformation. 
To see this explicitly, it is convenient to think about the action of our two $\SL(3,\bbR)$ subgroups instead on the vector representation of $\SL(9,\bbR)\times\bbR^+$
\begin{equation}
\label{eq:SL9-fund}
	V \simeq T_8 \oplus \Lambda^8 T^*_8 
		\simeq C_2 \oplus \bbR_5 \oplus B_5 \oplus (\Lambda^2 C^*\otimes \Lambda^5 B^*)
\end{equation}
The $\cN=(2,2)$ $\SL(3,\bbR)$ subgroup has $C_2 \oplus \bbR_5$ as the triplet part of the decomposition of $V$, while the $\cN=(4,0)$ $\SL(3,\bbR)$ has $C_2 \oplus (\Lambda^2 C^*\otimes \Lambda^5 B^*)$. The difference is simply the interchange of the $\bbR_5$ and $(\Lambda^2 C^*\otimes \Lambda^5 B^*)$ directions in $V$, i.e. interchange of the $\Lambda^8 T^*_8$ direction in~\eqref{eq:SL9-fund} with one of the directions in $T_8$, which can be implemented via a rotation operation inside $\SO(9)$. Thus, these two $\SL(3,\bbR)$ subgroups are conjugate via this rotation inside $\SL(9,\bbR)$. It follows that the decompositions of the charges $E$ are also related by this swapping of directions. 
As such, any triplet of this $\SL(3,\bbR)$ that we could have found would be equivalent to the standard triplet of momenta for standard $\cN=(2,2)$ supergravity by a U-duality. 

At this point, let us also make some brief remarks about the commutant groups of our $\Spin(2,1)\times\Spin(3)$ groups inside $\Spin(2,1)\times\SO(16)$, as this reveals some subtle points for consideration. 
The chains of embeddings of the spin groups we have considered so far can be summarised in the following diagram:
\begin{equation}
\label{eq:groups-diagram}
\begin{tikzpicture}[scale=1.75,baseline=(current bounding box.center)] 
\node (A1) at (0,3.5) {$\SL(32,\bbR)$}; 
\node (B1) at (-3,2.5) {$\Spin(5,1)_{(2,2)}\times\Symp(4)^2$}; 
\node (B2) at (0,2) {$\Spin(2,1)\times\Spin(3)\times\SO(16)$}; 
\node (B3) at (3,2.5) {$\Spin(5,1)_{(4,0)}\times\Symp(8)$}; 
\node (C1) at (-2,1) {$\Spin(2,1)\times\Spin(3)\times\Symp(4)^2$}; 
\node (C2) at (2,1) {$\Spin(2,1)\times\Spin(3)\times\Symp(8)$}; 
\path[->,font=\scriptsize] 
(C1) edge node[left]{} (B1)
(C1) edge node[left]{} (B2)
(C2) edge node[left]{} (B2)
(C2) edge node[left]{} (B3)
(B1) edge node[above]{} (A1)
(B2) edge node[above]{} (A1)
(B3) edge node[above]{} (A1)
; 
\end{tikzpicture}
\end{equation}
The group at the bottom right of this diagram has the generators\footnote{Recall that we defined the index ranges $\mu,\nu = 0,1,2$ and $a,b = 3,4$, while $m,n = 5,6,7,8,9,10$ and $\um,\un = 6,7,8,9,10$ so that $\hat{m} = (a,m) = (a,5,\um)$.}
\begin{equation}
\label{eq:spin21-spin3-sp8}
	\left\{ \Gamma^{\mu\nu}, \Gamma^{ab}, \Gamma^{a}\Gamma^{(8)},
	\Gamma^{m_1 m_2}, \Gamma^{m_1 m_2 m_3}, \Gamma^{m_1 \dots m_6} \right\}
\end{equation}
while the group at the bottom left has the generators
\begin{equation}
\label{eq:spin21-spin3-sp4-sp4}
	\left\{ \Gamma^{\mu\nu}, \Gamma^{ab}, \Gamma^{a5} ,
	\Gamma^{\um_1\um_2}, \Gamma^{\um_1\um_2\um_3} \right\}
\end{equation}
The first three terms of each generate their respective $\Spin(2,1)\times\Spin(3)$ factors, and are related by exchanging $\Gamma^5$ and $\Gamma^{(8)}$ as one would expect from the discussion of the $\SL(9,\bbR)$ rotation operation above. However, one can perform this exchange on the remaining generators in~\eqref{eq:spin21-spin3-sp8} to obtain generators for a $\Spin(2,1)\times\Spin(3)\times\Symp(8)$ group containing~\eqref{eq:spin21-spin3-sp4-sp4}:
\begin{equation}
\label{eq:spin21-spin3-sp8-B}
	\left\{ \Gamma^{\mu\nu}, \Gamma^{ab}, \Gamma^{a5} ,
	\Gamma^{\um_1\um_2}, \Gamma^{\um_1\um_2\um_3},
	\Gamma^{\um} \Gamma^{(8)}, \Gamma^{\um_1\um_2}\Gamma^{(8)}, 
	\Gamma^{\um_1 \dots \um_5}\Gamma^{(8)} \right\}
\end{equation}
Very naively, one might then wonder why the group $\Spin(5,1)_{(2,2)}\times\Symp(4)^2$ in~\eqref{eq:groups-diagram} is not $\Spin(5,1)_{(2,2)}\times\Symp(8)$. The reason is because the generators added to those in~\eqref{eq:spin21-spin3-sp4-sp4} do not commute with the generators $\Gamma^{i5}$ which are present in $\Spin(5,1)_{(2,2)}$, but which are not part of its $\Spin(2,1)\times\Spin(3)$ subgroup. 

This shows that one should be careful about making conclusions when imposing dimensional splits in the way that we have done in this section. Indeed, there is an apparent paradox in our work here. The embeddings of $\Spin(5,1)$ into $\SL(32,\bbR)$ really are inequivalent as they give different decompositions of the $\rep{32}$ representation into irreducible parts. However, on imposing the dimensional split that we have done, the corresponding $\Spin(2,1)\times\Spin(3)$ subgroups have been found to be conjugate by an $\SO(9)$ transformation. Thus, this inequivalence is not apparent from the point of view of our dimensional split. Similarly, the corresponding $\SL(3,\bbR)$ subgroups inside $\SL(9,\bbR) \subset E_{8(8)}$ also appear to be equivalent, unlike in the case of the type IIA vs type IIB embeddings. 
From our analysis it thus remains unclear exactly how the inequivalent decompositions of the spinor can be seen within the framework of exceptional groups. To learn more, one would need to include the full external $\Spin(5,1)$ group as well as the dual graviton charges, which would be contained only in a full $E_{11}$ analysis. The details go beyond the scope of our current investigation, though the resolution appears to be that there simply does not exist an $\mathfrak{sl}(6,\bbR)$ subalgebra containing our $\mathfrak{spin}(5,1)_{(4,0)}$ 
whose possible equivalence one can ask about~\cite{Guillaume}.

Let us now turn to a comparison of what we have found with the construction of~\cite{Hull1}. In that picture, one examines the five-dimensional maximal supersymmetry algebra
\begin{equation}
	\{ Q_{\alpha A}, Q_{\beta B} \} = C_{AB} P_\mu {\gamma^\mu}_{[\alpha\beta]} 
		+ K C_{AB} C_{\alpha\beta} + \dot{Z}_{[AB]}C_{\alpha\beta}
			+ \dot{Z}_{\mu[AB]}{\gamma^\mu}_{[\alpha\beta]}  
				+ Z_{[\mu\nu](AB)} \gamma^{\mu\nu}_{(\alpha\beta)} 
\end{equation}
The central charge $K$ is singled out as it is a singlet of the bosonic subalgebra $\mathfrak{spin}(4,1)\times\mathfrak{sp}(8)$, and it is remarked that it is not the charge of any of the five-dimensional vector fields, but becomes the magnetic charge of the gravi-photon on reduction to four dimensions. To identify the higher-dimensional physical object carrying the charge $K$, it is useful to consider that, in terms of the eleven-dimensional charges, it is the five-form charge $Z_{(5)}$ carried by the M5-brane but with all indices in the five-dimensional external space. (This was shown to be paired with the five-dimensional momentum to form a vector of $\Spin(5,1)_{(4,0)}$ in~\eqref{eq:6d-vector-SUSY-charges}.) Possibly the simplest picture of this arises from the type IIA decomposition. We think of the fifth direction of the five-dimensional external space as the M theory circle and note that the charge $K$ can then be seen as a D6-brane with legs along the six internal directions. 

In terms of the decomposition~\eqref{eq:8d-charges-decomp}, the D6-brane is part of the M-theory dual graviton, but to see this, we need to decompose further. Thus we go back to~\eqref{eq:decomp-1}, and this time give explicit labels to three one dimensional subspaces spanning $A_3$
\begin{equation}
	A_3 = \lc_3 \oplus \lc_4 \oplus \lc_5
\end{equation}
where our previous $C_2 = \lc_3 \oplus \lc_4$. We then imagine $\lc_4$ to correspond to the M theory circle direction. In terms of these labels, the internal D6-charge corresponds to the dual graviton charge $\lc_4^* \otimes (\lc_4^* \otimes \lc_5^* \otimes \Lambda^5 B^*) \subset T^* \otimes \Lambda^7 T^*$. The momentum charge around the M theory circle becomes the D0-brane charge in the IIA picture and corresponds to $\lc_4 \subset T_8$. Thus, naively it appears\footnote{See section~\ref{sec:SO3-triplet} for a more complete discussion.} that the charges
\begin{equation}
\label{eq:Hull-section}
	\lc_3 \oplus \lc_4 \oplus \big[\lc_4^* \otimes (\lc_4^* \otimes \lc_5^* \otimes \Lambda^5 B^*) \big] \subset E
\end{equation}
are thought of as the three momenta which, in conjunction with the three momenta in the external space, make up the momenta in the six-dimensional spacetime of~\cite{Hull1}. 

While the smaller subspaces $\lc_3 \oplus \lc_4$ or $\lc_3 \oplus \big[\lc_4^* \otimes (\lc_4^* \otimes \lc_5^* \otimes \Lambda^5 B^*)\big]$ solve the section constraint of $E_{8(8)}$ exceptional field theory, the three charges~\eqref{eq:Hull-section} together do not. This is because the charge $\lc_4$ has a non-zero contraction with the charge $\lc_4^* \otimes (\lc_4^* \otimes \lc_5^* \otimes \Lambda^5 B^*)$ in the relevant tensor product. Thus, these charges fail to satisfy the usual requirements to be a spacetime section.

Further, in~\cite{Hull2}, the conjectured six-dimensional theory is compactified on $T^2$ to give a maximally supersymmetric four-dimensional theory with an $\SL(2,\bbR)$ internal symmetry. It was noted there that this $\SL(2,\bbR)$ symmetry must be outside of the usual $E_{7(7)}$ symmetry of four-dimensional maximal supergravity\footnote{The lack of this $\SL(2,\bbR)$ is related to the absence~\cite{Godazgar:2013pfa,LSW2} of uplifts of the deformed $\SO(8)$ gauged supergravities of~\cite{SO8-gaugings}. It is also related to the missing $U(1)$ factor of footnote~\ref{ft:exception}.}. However, if we view the two momenta on $T^2$ as the D0 and D6 charges $\lc_4 \oplus \big[\lc_4^* \otimes (\lc_4^* \otimes \lc_5^* \otimes \Lambda^5 B^*)\big]$, then we see that in fact there is also no $SL(2,\bbR)$ subgroup of $E_{8(8)}$ which rotates these charges into each other, as this would have to contain a generator $\lc_4^* \otimes \lc_4^* \otimes (\lc_4^* \otimes \lc_5^* \otimes \Lambda^5 B^*)$. Thus, 
the $\SL(2,\bbR)$ symmetry of~\cite{Hull2} also appears to lie outside of the $E_{8(8)}$ duality group.

A strongly related fact is that there is also no $\SL(3,\bbR)$ subgroup of the $E_{8(8)}$ duality group for which the charges~\eqref{eq:Hull-section} form a triplet representation. As we found above, these can only be combined into an octuplet of $\SL(3,\bbR)$. The D0 and D6 charges then sit inside this octuplet in such a way that there is no $\SL(2,\bbR)$ subgroup under which they form a doublet. 

One then wonders if there is a different triplet of charges for our $\SL(3,\bbR)$ group~\eqref{eq:SL3-4,0-adjoint}, which could form the six-dimensional space of the $\cN=(4,0)$ theory. One quickly see that there is precisely such a set: writing 
$C = \lc_3 \oplus \lc_4$ as before, we have the triplet
\begin{equation}
\label{eq:SL3-triplet}
	\lc_5 \oplus \big( C^* \otimes \Lambda^2 C^* \otimes \Lambda^5 B^* \big)
\end{equation}
comprising one of the spatial momenta in M theory together with the six-dimensional dual gravitons with no leg along that direction. This set of charges thus solves the section condition of $E_{8(8)}$ exceptional geometry.
However, as noted above, 
the same $\SO(9)$ transformation which related the $\SL(3,\bbR)$ subgroup~\eqref{eq:SL3-4,0-adjoint} to the standard one relates this section to the standard one spanned by $\lc_3 \oplus \lc_4 \oplus \lc_5$. As such, the charges~\eqref{eq:SL3-triplet} are simply U-dual to the three momentum charges along $\lc_3$, $\lc_4$ and $\lc_5$. 
This would indicate that something has gone wrong, as the corresponding theories are supposed to be very different, as are the relevant spinor decompositions. 
Further still, by considering the orbits of the charges in the supersymmetry algebra under $\Spin(5,1)_{(4,0)}$ and how these are mapped into the $\rep{248}$ representation of $E_{8(8)}$ we can see that~\eqref{eq:SL3-triplet} does not match the momenta of the six-dimensional space. We will do this explicitly in the next section.


\subsection{Charges in $E_{8(8)}$ and the triplet of $\SO(3)$}
\label{sec:SO3-triplet}

In this section we will see that our identification of charges in~\eqref{eq:Hull-section} is not quite right. Unlike the lower rank exceptional groups, in $E_{8(8)}$ the internal charges appearing in the anti-commutator of supersymmetries do not map onto the $\rep{248}$ representation. Rather, they span only the subspace forming the $\rep{120}$ representation of the maximal compact subgroup $\SO(16)$. As such, the momentum charge $P^{\hat{m}}$ of eleven-dimensional supergravity in the eight internal directions, embeds into not just the obvious vector $T_8$ in~\eqref{eq:8d-charges}, but it also has a component along $T_8^* \otimes (\Lambda^8 T_8^*)^2$. 
The interpretation of this is that the supersymmetry algebra closes not just onto local translations, but a combination of these with higher gauge transformations of the dual gravitons. We also note that the subspace of the charges into which the momentum directly embeds does not solve the section condition. 

For standard supergravities, one could identify the spacetime section from the momentum charge coming from the supersymmetry algebra in the following way.
The embedded momentum charge in fact lives in a subspace of the sum of two isomorphic vector representations of the orthogonal group inside $E$. For the momentum $P_{\hat{m}}$ above, these two become the $T_8$ and $T_8^* \otimes (\Lambda^8 T_8^*)^2$ representations of the $\GL(8,\bbR)$ subgroup fo $E_{8(8)}$ containing $\SO(8)$. One can project onto these two subspaces in a $\GL(8,\bbR)$ covariant way. More generally, there are $\SO(8)$ covariant projectors onto any linear combination of them. The property that picks out the subspace $T_8$ (or $T_8^* \otimes (\Lambda^8 T_8^*)^2$ which is the same up to an automorphism of $\SL(9,\bbR)$) is that it solves the section condition (while any linear combination does not). Thus, even though the momentum charge does not directly live in the directions $T_8$ of the spacetime section, it is fairly simple to identify the spacetime section and project onto it.

Let us contrast this with the situation for the momentum charge of the $\cN=(4,0)$ theory. 
There, the result~\eqref{eq:6d-vector-SUSY-charges} tells us that two of the five-dimensional momenta are combined with the charge labelled $K$ above into a triplet, which makes up the three internal momenta of the six-dimensional spacetime. This triplet is invariant under the $\Symp(8)$ R-symmetry, which uniquely identifies it inside the $\rep{248}$ of $E_{8(8)}$ as the generators of $\SO(3)_{(4,0)}$ (see~\eqref{eq:ad-sl3-e66} later). In terms of the charges in~\eqref{eq:8d-charges-decomp} this triplet consists of $\Lambda^2 C^* \otimes \Lambda^5 B^*$ together with a two-dimensional subspace of $C \oplus (C^*\otimes\Lambda^2 C^* \otimes \Lambda^5 B^*)$. We would then like to project this onto a triplet of an $\SL(3,\bbR)$ group containing $\SO(3)_{(4,0)}$, as we did for the standard supergravity case. Naively it would even seem reasonable that the projected subspace could be similar to the charges~\eqref{eq:Hull-section}. 
However, here there is no such projection. 
The $\SL(3,\bbR)$ group containing $\SO(3)_{(4,0)}$ makes the triplet of $\SO(3)_{(4,0)}$ into an octuplet. It is not a subspace of the sum of two triplets.

What we have learned here is that there is no spacetime section for the $\cN=(4,0)$ theory in the standard sense. Rather, the momentum charge is the triplet of $\SO(3)_{(4,0)}$ which is invariant under $\Symp(8)$, and like the embedded momentum charge in other cases, this does not solve the section condition. Moreover, the identification of this subspace appears to require the decomposition under $\SO(3)_{(4,0)}\times\Symp(8)$, which requires knowledge of the physical fields. Thus, very differently to the case of standard supergravity, it appears that the momentum charge, or even a relevant subspace of the correct dimension, can only be identified once a field configuration is specified. This picture also resonates with the earlier mentioned observation that the $\Spin(5,1)_{(4,0)}\times\Symp(8)$ group is present inside $KE_{11}$, but there appears to be no $\SL(6,\bbR)\times E_{6(6)}$ subgroup which contains it, suggesting that a description of the $\cN=(4,0)$ theory in the $E_{11}$ formalism must make explicit use of the Lorentz symmetry. 


\subsection{Interpretation of $\SL(3,\bbR)\times E_{6(6)}$ inside $E_{8(8)}$}
\label{sec:E8}

In the previous section, we argued that the role of $\SO(3)_{(4,0)} \subset \SL(3,\bbR)$ is very different for the $\cN=(4,0)$ theory compared with the role of the Lorentz and general linear groups in standard supergravity. 
In particular, there is no three-dimensional spacetime section satisfying the section condition, but only the analogue of the embedding of the momentum charge in the $\rep{248}$ of $E_{8(8)}$. 
Noting that any $\SL(3,\bbR)$ subgroup of $E_{8(8)}$ with commutant $E_{6(6)}$ will be conjugate as $\SL(3,\bbR)\times E_{6(6)} \subset E_{8(8)}$ is a maximal subgroup, we now examine the decompositions of the generalised tangent space and the adjoint of $E_{8(8)}$ under $\SL(3,\bbR)\times E_{6(6)}$. Remarkably, despite all that has been said in the previous sections, some aspects of the $\cN=(4,0)$ theory do fit into this picture as we now discuss. 

We start from the $\GL(8,\bbR)$ decomposition of the $E_{8(8)}\times\bbR^+$ multiplet of charges related to eleven-dimensional supergravity on an eight-dimensional internal space~\cite{CSC}
\begin{equation}
\begin{aligned}
	E \simeq \rep{248}_{+1} 
		&\simeq T \oplus \Lambda^2 T^* \oplus \Lambda^5 T^* 
			\oplus (T^* \otimes \Lambda^7 T^*) \\
			& \qquad \oplus (\Lambda^8 T^* \otimes \Lambda^3 T^*) 
			\oplus (\Lambda^8 T^* \otimes \Lambda^6 T^*) 
			\oplus ((\Lambda^8 T^*)^2 \otimes T^*) \\
\end{aligned}
\end{equation}
This corresponds to the decomposition of the adjoint representation of $E_{8(8)}$
\begin{equation}
\begin{aligned}
	\rep{248}_{0} &\simeq (T \otimes T^*) \oplus \Lambda^{3} T \oplus \Lambda^{3} T^*
		\oplus \Lambda^{6} T \oplus \Lambda^{6} T^*
		\oplus (\Lambda^8 T \otimes T) \oplus (\Lambda^8 T^* \otimes T^*) \\
\end{aligned}
\end{equation}
together with the embedding of $\GL(8,\bbR)$ into $E_{8(8)}\times\bbR^+$ such that $\rep{1}_{+1} = (\Lambda^8 T^*)$. These expressions do not provide a generalised geometry in the usual way due to problems with diffeomorphism covariance associated to the dual graviton field (see~\cite{CSC} for a discussion) but one can argue that using additional section conditions to constrain certain compensator fields in the tensor hierarchy it is possible to write an exceptional field theory construction based on them~\cite{HS-E88}. 

We now wish to study further splits of the dimensions. In particular, we choose three of the eight dimensions to join the three external dimensions, leaving 5 remaining internal dimensions (in the eleven-dimensional picture). This mirrors our study of the spin groups in section~\ref{sec:Hd-charges}.

As such, let us decompose under $\GL(3,\bbR)\times\GL(5,\bbR) \subset \GL(8,\bbR)$ so that
\begin{equation}
	T_8 = A_3 \oplus B_5 .
\end{equation}
as before. 
We reiterate that the straightforward $\SL(3,\bbR)$ subgroup of the $\GL(3,\bbR)$ factor is appropriate for our purposes here, as the choice which seems most naturally related to the six-dimensional $\cN=(4,0)$ theory is equivalent to this one (as shown explicitly in section~\ref{sec:Hd-charges}).
Indeed, whichever $\SL(3,\bbR)$ subgroup we chose, we would wish to write our eventual decompositions in terms of its triplet representation and tensor products thereof. As $\SL(3,\bbR)\times E_{6(6)}$ is a maximal subgroup, the result of doing this will be the same whichever $\SL(3,\bbR)$ we chose initially. 

The $\GL(5,\bbR)$ factor can be seen to be a subgroup of a $\Spin(5,5)\times\bbR^+$ group inside $E_{8(8)}\times\bbR^+$ which commutes with our $\GL(3,\bbR)$. Identifying the $\Spin(5,5)\times\bbR^+$ representations as is familiar from five-dimensional exceptional generalised geometry via
\begin{equation}
\begin{aligned}
	(B\otimes B^*) \oplus \Lambda^3 B \oplus \Lambda^3 B^* &\simeq \mathfrak{spin}(5,5) \\	
	\Lambda^5 B^* &\simeq \rep{1}_{+4} \\
	B \oplus \Lambda^2 B^* \oplus \Lambda^5 B^* & \simeq \rep{16}_{+1} \\
	B^* \oplus \Lambda^4 B^* & \simeq \rep{10}_{+2} 
\end{aligned}
\end{equation}
we find the $\GL(3,\bbR)\times\Spin(5,5)\times\bbR^+$ decompositions
\begin{equation}
\label{eq:adj-gl3-spin55}
\begin{aligned}
	\rep{248}_{0} 
		\simeq \;
		 \mathfrak{gl}(3,\bbR)
		&\oplus \mathfrak{spin}(5,5) \\	
		&\oplus (A \otimes \rep{16^-_{-1}})
		\oplus (A^* \otimes \rep{16^+_{+1}}) \\
		& \oplus (\Lambda^2 A \otimes \rep{10_{-2}}) 
		\oplus (\Lambda^2 A^* \otimes  \rep{10_{+2}}) \\
		& \oplus (\Lambda^3 A \otimes \rep{16^+_{-3}}) 
		\oplus (\Lambda^3 A^* \otimes \rep{16^-_{+3}}) \\
		&\oplus (\Lambda^3 A \otimes A \otimes \rep{1_{-4}})
		\oplus (\Lambda^3 A^* \otimes A^* \otimes \rep{1_{+4}})
\end{aligned}
\end{equation}
and
\begin{equation}
\label{eq:E-gl3-o55-R}
\begin{aligned}
	E \simeq \rep{248}_{+1} 
		\simeq \; A 
		&\oplus \rep{16^+_{+1}} 
		\oplus (A^* \otimes \rep{10_{+2}}) \\
		&\oplus (\Lambda^2 A^* \otimes \rep{16^-_{+3}}) 
		\oplus (\Lambda^3 A^* \otimes \rep{45_{+4}}) 
		\oplus (\Lambda^2 A^* \otimes A^* \otimes \rep{1_{+4}}) \\
		&\oplus (\Lambda^3 A^* \otimes A^* \otimes \rep{16^+_{+5}})  \\
		&\oplus (\Lambda^3 A^* \otimes \Lambda^2 A^* \otimes \rep{10_{+6}}) \\
		&\oplus ((\Lambda^3 A^*)^2 \otimes \rep{16^-_{+7}}) \\
		&\oplus ((\Lambda^3 A^*)^2 \otimes A^* \otimes \rep{1_{+8}}) \\
\end{aligned}
\end{equation}
From this, we see explicitly that the commutant of $\GL(3,\bbR)$ inside $E_{8(8)}\times\bbR^+$ cannot be enhanced further than $\Spin(5,5)\times\bbR^+$, as~\eqref{eq:adj-gl3-spin55} contains no trivial $\GL(3,\bbR)$ singlets beyond the $\mathfrak{spin}(5,5)$ summand. This agrees with the standard picture in supergravity, where we expect six-dimensional $\cN=(2,2)$ supergravity to have global symmetry $\Spin(5,5)$. 

However, we expect the six-dimensional $\cN=(4,0)$ theory to have global symmetry $E_{6(6)}$, and thus it would be desirable if we could see a way to make $E_{6(6)}$ the commutant of our spacetime subgroup inside $E_{8(8)}$. To match this to the above, we decompose the above under $\SL(3,\bbR)\times\Spin(5,5) \times \bbR^+\subset \GL(3,\bbR)\times\Spin(5,5)\times\bbR^+$. Under $\SL(3,\bbR)$ we have additional identifications $\Lambda^3 A \simeq \Lambda^3 A^* \simeq \rep{1}$ and $\Lambda^2 A \simeq A^*$ 
and thus we have the decompositions
\begin{equation}
\begin{aligned}
	\rep{248}_{0} 
		&\simeq\mathfrak{sl}(3,\bbR)
		\oplus \Big( \bbR \oplus  \mathfrak{spin}(5,5) \oplus \rep{16^+_{-3}} \oplus \rep{16^-_{+3}} \Big) \\	
		&\oplus \Lambda^2 A^* 
			\otimes \Big(\rep{1_{-4}} \rep{ \oplus \rep{10_{+2}}  \oplus16^-_{-1}} \Big)
		\oplus \Lambda^2 A 
			\otimes \Big( \rep{1_{+4}} \oplus \rep{10_{-2}} \oplus \rep{16^+_{+1}} \Big)
\end{aligned}
\end{equation}
\begin{equation}
\begin{aligned}
	E \simeq \rep{248}_{+1} 
		&\simeq 
		\rep{1_{+4}} \otimes \Big[ \Big(  \bbR \oplus \mathfrak{spin}(5,5) \oplus \rep{16^+_{-3}} \oplus \rep{16^-_{+3}} \Big)
		\oplus A^* \otimes \Big( \rep{1_{+4}} \oplus \rep{10_{-2}} \oplus \rep{16^+_{+1}} \Big) \\
		& \qquad \oplus \Lambda^2 A^* 
			\otimes \Big(\rep{1_{-4}}  \oplus \rep{10_{+2}}  \oplus \rep{16^-_{-1}}\Big)
		\oplus  (\Lambda^2 A^* \otimes A^*)_0 \Big]
\end{aligned}
\end{equation}
where $(\Lambda^2 A^* \otimes A^*)_0$ denotes the irreducible part of $(\Lambda^2 A^* \otimes A^*)$ whose totally anti-symmetric part is zero. The summands $\bbR \oplus  \mathfrak{spin}(5,5) \oplus \rep{16^+_{-3}} \oplus \rep{16^-_{+3}}$ form an $\mathfrak{e}_{6(6)}$ subalgebra of $\mathfrak{e}_{8(8)}$ and we recognise the decompositions 
\begin{equation}
\begin{aligned}
	 \mathfrak{e}_{6(6)} &\ra \bbR \oplus  \mathfrak{spin}(5,5) \oplus \rep{16^+_{-3}} \oplus \rep{16^-_{+3}}\\
	\rep{27} &\ra  \rep{1_{-4}} \oplus \rep{10_{+2}} \oplus \rep{16^-_{-1}} \\
	\rep{27'} & \ra \rep{1_{+4}}  \oplus \rep{10_{-2}}  \oplus \rep{16^+_{+1}}
\end{aligned}
\end{equation}
Ignoring the overall $\bbR^+$ weight 
(as there is no non-trivial homomorphism $\SL(3,\bbR)\times E_{6(6)} \ra \bbR^+$) 
and choosing to use the isomorphisms $\Lambda^3 A \simeq\Lambda^3 A^* \simeq \bbR$ and $\Lambda^2 A \simeq A^*$ to write the result in a suggestive way, we find the standard decompositions
\begin{equation}
\label{eq:ad-sl3-e66}
\begin{aligned}
	\rep{248}_{0} 
		&\ra \mathfrak{sl}(3,\bbR)
		\oplus \mathfrak{e}_{6(6)} \oplus (\Lambda^2 A^* \otimes \rep{27} )
		\oplus (\Lambda^2 A  \otimes \rep{27'})
\end{aligned}
\end{equation}
\begin{equation}
\label{eq:E-sl3-e66}
\begin{aligned}
	E \simeq \rep{248}_{+1} 
		&\ra (A^* \otimes \rep{27'}) 
		\oplus (\Lambda^2 A^* \otimes A^*)_0
		\oplus (\Lambda^2 A^* \otimes \Lambda^3 A^* \otimes \rep{27})
		\oplus (\Lambda^3 A^* \otimes \rep{78})
\end{aligned}
\end{equation}
We could have written these down at the outset. The reason for presenting this chain of decompositions and recombinations at this level of detail is to keep track of all of how the different charges combine into the $E_{6(6)}$ representations, and to show very explicitly that all that is needed to realise $E_{6(6)}$ is to break $\GL(3,\bbR)$ to $\SL(3,\bbR)$. 

Naively applying the usual assignment of forms in the adjoint to physical fields and scalars to a sigma model, one would suspect that the six-dimensional parent theory would have two-forms in the $\rep{27}$ of $E_{6(6)}$ and scalars in the coset $E_{6(6)} / \Symp(8)$, exactly as one would hope for the $\cN=(4,0)$ theory.

However, this is also problematic, as one would also like to interpret the forms in the generalised vector as their charges.
The one-forms in $E$ are in the wrong $E_{6(6)}$ representation to be the charges of the two-forms in the adjoint. This is because in the adjoint the $\Lambda^2 A$ and $\Lambda^2 A^*$ terms also live in different representations. In the usual Kac-Moody prescription we would want to interpret the corresponding charges in $E$ as being dual in some higher sense. However, a possible resolution is that under the maximal compact subgroup $\Symp(8)$, these become equal. This suggests that really the symmetry of any theory underlying these observations is $\Symp(8)$ rather than $E_{6(6)}$ (c.f. the situation for $F_{4(4)}$ in the $\cN=(3,1)$ multiplet as discussed in the introduction). An alternative resolution would be to decompose under $\SO(3) \subset \SL(3,\bbR)$, which allows the identification of vectors and two-forms, so that the third term in~\eqref{eq:E-sl3-e66} could be viewed as the charges of the two-forms. 

Further signs in this direction come from comparison of~\eqref{eq:E-sl3-e66} with the charges in the superalgebra~\eqref{eq:6d-superalg}. We expect to find vector charges in the $\rep{1}\oplus\rep{27}$ of $\Symp(8)$ together with (anti-self-dual) three forms in the $\rep{36}$. These objects are present inside~\eqref{eq:E-sl3-e66}, but to see them we must decompose under $\SO(3)\times\Symp(8)$, as we noted in the previous section. In order to see $E_{6(6)}$ we have to combine the magnetic charges of the scalars with the three-form central charges, while the singlet vector momentum charge becomes part of a non-vector representation of $\SL(3,\bbR)$. This again shows that moving from Lorentz to special linear group is be problematic in this context, and that to identify a subspace for the momentum of the correct dimension we must decompose under $\SO(3)$.

However, there are also encouraging signs in this, in that the non-vector representation of $\SL(3,\bbR)$ which absorbs the singlet vector central charge has the correct index structure to be a charge for the exotic graviton $C_{\mu\nu\rho\sigma}$ from section~\ref{sec:multiplets}, as a charge $\Lambda_{m[np]}$ can give a gauge transformation $\delta C_{mnpq} \sim \der_{[m} \Lambda_{n][pq]} + \der_{[p} \Lambda_{q][mn]} - 2\der_{[m} \Lambda_{npq]}$, where the last term vanishes identically in a three-dimensional restriction. 

Indeed, one can see that this does in fact appear in the following way. 
If we consider $\bbR^3$ with standard Euclidean metric (and now take $m,n = 1,2,3$) and define
\begin{equation}
	\der^m{}_n = \epsilon^m{}_n{}^p \der_p
	\hs{30pt}
	\Lambda^m{}_n = \tfrac12 \epsilon^{mpq} \Lambda_{n[pq]}
\end{equation}
we can then compute the part of the projection of $\der \Lambda$ into the $\mathfrak{sl}(3,\bbR)$ part of the adjoint in~\eqref{eq:ad-sl3-e66}:
\begin{equation}
	[\der,\Lambda]^m{}_n = \der_q \Lambda_n{}^{qm} - 3 \delta^m{}_{[n} \der^p \Lambda^q{}_{pq]}
\end{equation}
If we then define a dualised variable
\begin{equation}
	\tLambda_{m,pq} = \tfrac12 \epsilon_m{}^{rs}\epsilon_{pq}{}^t \Lambda_{t[rs]}
\end{equation}
and restrict to considering $\tLambda$ in the $\rep{5}$ representation of $\SO(3)$ then we find
\begin{equation}
\label{eq:Dorfman-exotic-gauge}
	[\der,\Lambda]^m{}_n = -\epsilon^{mpq}\epsilon_n{}^{rs} \Big( \der_{[p} \tLambda_{q]rs} 
		+ \der_{[r} \tLambda_{s]pq} \Big)
\end{equation}
Considering a variation of the exotic graviton $C_{[mn][pq]}$ to transform in the adjoint of $\SL(3,\bbR)$ via defining
\begin{equation}
	\delta C^m{}_n = \epsilon^{mpq}\epsilon_n{}^{rs} \delta C_{[mn][pq]}
\end{equation}
we find
\begin{equation}
\label{eq:SO3-deltaC}
	\delta C_{[mn][pq]} = -  \Big(  \der_{[p} \tLambda_{q]rs} + \der_{[r} \tLambda_{s]pq} \Big)
\end{equation}
The projection of $\der\Lambda$ we have calculated would naively become part of the action of the generalised Lie derivative or exceptional Dorfman derivative as introduced in~\cite{CSW2}. 
Recall that this object has the general form\footnote{In fact, for $E_{8(8)}$ it has been argued that one must add additional terms to this formula, including a second constrained gauge parameter, in order to correctly account for the tensor hierarchy and address issues with closure of the gauge algebra and covariance~\cite{HS-E88}. Here we consider only a local patch of flat space and ignore these issues, as we are merely looking for signs of agreement in the core part of the object.}
\begin{equation}
\label{eq:Dorfman}
	L_V = \der_V - (\der \proj{\adj} V) \cdot
\end{equation}
where $V \in E$ is a generalised vector. The first term is a straightforward derivative, while the second term term gives the action of the appropriate derivatives of the gauge parameter. 
What we have discovered here is that, with the definitions made above, we seem to be able to recover the gauge transformation of the exotic graviton as part of this object. 
In particular, the derivative~\eqref{eq:Dorfman-exotic-gauge} which would be the only place where $\tLambda$ would appear in~\eqref{eq:Dorfman}, appears to give the correct gauge transformation~\eqref{eq:SO3-deltaC}. 
This gives us some confidence in our interpretation of the momentum charge and that our assertion of the necessity of working under the Lorentz group $\SO(3)$ is justified. 

Overall, it seems that there is some hope of identifying the terms in~\eqref{eq:ad-sl3-e66} and~\eqref{eq:E-sl3-e66} in the usual way. In~\eqref{eq:ad-sl3-e66}, the $\mathfrak{sl}(3,\bbR)$, $\mathfrak{e}_{6(6)}$ and $\Lambda^2 A^*$ terms correspond to the exotic graviton, scalar sigma model and two-forms respectively, while in~\eqref{eq:E-sl3-e66} the terms match the charges of the two-forms, the exotic graviton, higher duals of the two-forms, the three-form charges in~\eqref{eq:6d-superalg} and the magnetic duals of the scalars in that order. However, as discussed, it is really only under $\SO(3) \subset \SL(3,\bbR)$ that we can identify the triplet $A$ with spacetime, which makes these apparent matches at least slightly surprising. 

All of these comments should be taken as suggestive but in no way conclusive. However, they are in harmony with other proposals made in this article concerning the importance of a fixed volume $T^3$ fibred manifold, leaving only an action of $\SL(3,\bbR) \subset \GL(3,\bbR)$ and the absence of a six-dimensional ``section". The observation that one needs to work under $\SO(3)$ to identify the six-dimensional momentum charge is also curious, as it suggests that knowledge of the exotic graviton field configuration is needed to identify the six-dimensional space. They also fit a pattern of behaviour shared by multiplets with less supersymmetry, as we explore next.


\subsection{Exotic gravity with less supersymmetry}

In this section, we examine the versions of the decompositions~\eqref{eq:ad-sl3-e66} and~\eqref{eq:E-sl3-e66} relevant to the cases of theories with less than maximal supersymmetry. In all cases we see that a special role is played by the five-dimensional Ehlers symmetry $\mathfrak{sl}(3,\bbR)$, which becomes the terms relevant to the exotic graviton in our decompositions. In a sense, the decompositions for these theories are built by adding additional terms to this $\mathfrak{sl}(3,\bbR)$ base in a similar sense to the way that conventional generalised geometries are built as extensions of ordinary geometry with frame bundle group $\GL(d,\bbR)$. 


\subsubsection{$\cN = (2,0)$ supersymmetry and $\SO(8,8+n)$}
\label{sec:alg-N=2}

If, instead of looking at eleven-dimensional supergravity, we look at type I supergravity (which has half-maximal supersymmetry in ten-dimensions) the analagous group to $E_{8(8)}$ appearing in reductions to three dimensions (with Abelian gauge symmetry) is $\SO(8,8+n)$, where $n$ is the number of vector multiplets in ten dimensions.

We can then ask if the same procedure outlined above for the charges and adjoint representation of $E_{8(8)}$ will go through to match the field content of half-maximal exotic gravity. In this section we will show that it does.

Rather than examining first the decompositions under a standard spacetime $\GL(7,\bbR)$ group (corresponding to the spatial directions on the seven-torus in a type I comactification), let us 
assume that exotic gravity will correspond to an $\SL(3,\bbR)$ subgroup as in the previous section and
simply decompose under the product of $\SL(3,\bbR)$ with a suitable commutant inside $\SO(8,8+n)\times\bbR^+$. As such, consider the maximal subgroup $\SO(3,3)\times\SO(5,5+n)\times\bbR^+$, noting that $\Spin(3,3) \simeq \SL(4,\bbR)$. We then decompose the adjoint under the $\SL(3,\bbR)\times\SO(5,5+n)\times\bbR^+$ subgroup and give the two presentations of the result corresponding to~\eqref{eq:ad-sl3-e66} and~\eqref{eq:ad-sl3-e66}
\begin{equation}
\label{eq:spin88-adj}
\begin{aligned}
	\mathfrak{spin}(8,8+n) 
		&\simeq\mathfrak{sl}(3,\bbR)
		\oplus  \bbR \oplus  \mathfrak{spin}(5,5+n) \\	
		&\qquad \oplus \Big[ \Lambda^2 A^* 
			\otimes \big(\rep{1_{+2}} \oplus \ctab{1}_{\;\rep{-1}} \big) \Big]
		\oplus \Big[ \Lambda^2 A 
			\otimes \big( \rep{1_{-2}} \oplus \ctab{1}_{\;\rep{+1}} \big) \Big]
\end{aligned}
\end{equation}
\begin{equation}
\label{eq:spin88-E}
\begin{aligned}
		E &\simeq 
		\Big[ A^* \otimes \big( \rep{1_{-2}} \oplus \ctab{1}_{\;\rep{+1}} \big) \Big] 
		\oplus  (\Lambda^2 A^* \otimes A^*)_0 \\
		&\qquad \oplus \Big[ \big(  \bbR \oplus \ctab{1,1} \big) \otimes \Lambda^3 A^* \Big]
		\oplus  \Big[ \Lambda^2 A^* \otimes \Lambda^3 A^*
			\otimes \big(\rep{1_{+2}} \oplus \ctab{1}_{\;\rep{-1}} \big) \Big]
\end{aligned}
\end{equation}
This would correspond to having two-forms transforming in the $\big(\rep{1_{+2}} \oplus \ctab{1}_{\;\rep{-1}} \big)$ representation of $\SO(5,5+n)$ together with scalars in the coset $\SO(5,5+n)\times\bbR^+ /\SO(5)\times\SO(5+n)$. Together with the exotic graviton, this would precisely match the bosonic field content of one $\cN=(2,0)$ exotic graviton multiplet together with $(5+n)$ $\cN=(2,0)$ tensor multiplets. However, again we see that the representation of the $A^*$ charges in~\eqref{eq:spin88-E} does not quite match that of the fields $\Lambda^2 A^*$ in~\eqref{eq:spin88-adj} as the $\bbR^+$ weights do not match. Thus again we see a sign that the full $\SO(5,5+n)\times\bbR^+$ may not be a symmetry of any corresponding theory, or that we may not be able to move from $\SO(3)$ to $\SL(3,\bbR)$ in the usual way. 


\subsubsection{$\cN=(1,0)$ supersymmetry}
\label{sec:alg-N=1}

We can also consider what happens for various theories with eight supercharges which (on reduction to three dimensions) have scalars living in symmetric spaces as for the maximal and half-maximal theories considered above.
A list of such theories and their corresponding coset manifolds can be found in~\cite{VanProeyen}. 

For example, let us first consider pure five-dimensional supergravity. On reduction to three dimensions, we obtain scalars living in the coset space $G_{2(2)} / \SU(2)\times\SU(2)$, thus the analogue of the group $E_{8(8)}$ from the maximal case here is $G_{2(2)}$. This has an $\SL(3,\bbR)$ subgroup, under which the decomposition of the adjoint representation is
\begin{equation}
\label{eq:g22}
\begin{aligned}
	\mathfrak{g}_{2(2)}
		\simeq \; &\mathfrak{sl}(3,\bbR)
		\oplus \Lambda^2 A^* 
		\oplus \Lambda^2 A 
\end{aligned}
\end{equation}
which would match a theory in six-dimensions with an exotic graviton and a single self-dual two-form. Thus, as expected, this matches the field content of the $\cN=(1,0)$ exotic graviton multiplet. 

Next, consider pure $\cN=(1,0)$ supergravity in six-dimensions, which upon reduction to three-dimensions has scalar manifold $\SO(4,3) / \SO(4)\times\SO(3)$. The group $\SO(4,3)$ again has an $\SL(3,\bbR)$ decomposition of the relevant type:
\begin{equation}
\label{eq:so43}
\begin{aligned}
	\mathfrak{so}(4,3) 
		\simeq \; &\mathfrak{sl}(3,\bbR)
		\oplus  \bbR  \\	
		&\oplus \Big[\Lambda^2 A^* \otimes (\rep{1_{+2}} \oplus \rep{1_{-1}})  \Big]
		\oplus \Big[  \Lambda^2 A \otimes (\rep{1_{-2}} \oplus \rep{1_{+1}}) \Big]
\end{aligned}
\end{equation}
This matches a theory with an exotic graviton, two self-dual two-forms and one scalar, which is the bosonic field content of an exotic graviton multiplet together with one tensor multiplet.

This pattern continues for the other theories outlined in~\cite{VanProeyen}. A more involved example is six-dimensional minimal supergravity coupled to two vector multiplets and two tensor multiplets. On reduction to three dimensions, one obtains the scalar manifold $F_{4(4)} / \Symp(6)\times\Symp(2)$. One then looks at the decomposition
\begin{equation}
\begin{aligned}
	\mathfrak{f}_{4(4)}
		\simeq \; &\mathfrak{sl}(3,\bbR)
		\oplus  \mathfrak{sl}(3,\bbR) \\	
		&\oplus \Big[ \Lambda^2 A^* 
			\otimes \rep{6} \Big]
		\oplus \Big[ \Lambda^2 A 
			\otimes \rep{6'} \Big]
\end{aligned}
\end{equation}
Thus we hypothesise an exotic graviton, self-dual two-forms in the $\rep{6}$ representation of $\SL(3,\bbR)$ and five scalars in the coset manifold $\SL(3,\bbR) / \SO(3)$. This field content matches an exotic graviton multiplet together with five tensor multiplets, and we expect a global symmetry group $\SL(3,\bbR)$, modulo the same problems with charges and fields living in different representations. 

Table~\ref{tab:magic} summarises the corresponding results for this collection of theories. In all cases, the $\SL(3,\bbR)$ subgroup gives a decomposition which exactly matches a combination of an exotic graviton multiplet and some number of tensor multiplets, identifying the conjectured global symmetry group as its commutant. This global symmetry and its coset are precisely those of the corresponding six-dimensional conventional supergravity theory on $S^1$. If one assumes that the reduction of these theories on $S^1$ should give the same five-dimensional theory as reducing the standard $\cN=(1,0)$ supergravity then this is inevitable, since the five-dimensional scalars must come only from the six-dimensional scalars of the exotic theory. Below we explain why other features of this table inevitably must work out. 

We also note that in all cases but the first row, the charges of the two-forms do not match the representation for the two-forms, as we found in the cases considered in sections~\ref{sec:E8} and~\ref{sec:alg-N=2}. Thus, we again see that the numerator group of the scalar coset may not be a true symmetry of the corresponding theory, or that really one must work under $\SO(3)$ to make these match. 

Finally, we explain why the decomposition of the the duality group in 3d inevitably has the form 
\begin{equation}
\label{eq:general-decomp}
	\mathfrak{g} = \mathfrak{sl}(3,\bbR) \oplus \mathfrak{k} \oplus (\rep{3}\otimes \rep{r})
		\oplus (\rep{3'}\otimes \rep{r'})
\end{equation}
if the three-dimensional theory can be written as a torus reduction of a five-dimensional supergravity theory. The existence of the $\mathfrak{sl}(3,\bbR)$ is the usual Ehlers symmetry appearing in the reduction of 5d gravity to three dimensions. Under $\GL(2,\bbR)$ this has the form
\begin{equation}
	\mathfrak{sl}(3,\bbR) = (C\otimes C^*) \oplus (\Lambda^2 C^*\otimes C^*) 
		\oplus (\Lambda^2C\otimes C)
\end{equation}
If the three-dimensional theory comes from the reduction of a five-dimensional supergravity theory, then the only other degrees of freedom are standard scalars and $p$-form fields. Thus the adjoint can only contain $\GL(2,\bbR)$ representations of the form $\Lambda^p C^*\oplus \Lambda^p C^*$ together with scalars and $\mathfrak{sl}(3,\bbR)$ as above. The only options for $p$ are $p=0,1,2,3$. Any $\SL(3,\bbR)$ representation in~\eqref{eq:general-decomp} other than $\rep{1}$, $\rep{3}$ or $\rep{3'}$ would give other types of $\GL(2,\bbR)$ representations and thus is not allowed. Thus the decomposition~\eqref{eq:general-decomp} is universal. 
Further, once it is known that the degrees of freedom of pure five-dimensional supergravity lift to $\cN=(1,0)$ exotic gravity and both vector and tensor multiplets lift to $\cN=(1,0)$ tensor multiplets, it is clear that this decomposition will match the decomposition of an $\cN=(1,0)$ exotic gravity. 
Thus the matching of the degrees of freedom between the $\SL(3,\bbR)$ decompositions and the exotic gravity theories is inevitable once one assumes that they reduce to those of standard gravitational theories in three dimensions. 

\begin{table}[!ht]
\begin{tabular}{c|ccc|ccc|}
3d coset & \multicolumn{3}{|c|}{6d Supergravity} & \multicolumn{3}{|c|}{ 6d Exotic gravity} \\
& $n_V$ & $n_T$ & 6d coset & $n_T$ & $B_{\mu\nu}^-$ rep & 6d coset \\
\hline
$\cfrac{G_{2(2)}}{\SU(2)\times\SU(2)}$ & \multicolumn{2}{|c}{5d sugra} & - & $0$ & $\rep{1}$ & - \\
& & & & & & \\
$\cfrac{\SO(4,3)}{\SO(4)\times\SO(3)}$ & $0$ & $0$ & - & 1 & $\rep{1_{+2} \oplus \rep{1_{-1}}} $ & $\bbR^+$ \\
& & & & & & \\
$\cfrac{\SO(4,4+n)}{\SO(4)\times\SO(4+n)}$ & $n$ & $1$ & $\bbR^+$ & $n\!+\!2$ 
	& $\rep{1}_{+2} \oplus \rep{n\!+\!2}_{\,-1}$ 
	& $\cfrac{\SO(1,n+1)}{\SO(n+1)}\times \bbR^+ $ \\
& & & & & & \\
$\cfrac{F_{4(4)}}{\Symp(6)\times\Symp(2)}$ & $2$ & $2$ 
	& $\cfrac{\SO(2,1)}{\SO(2)}$
	& $5$ & $\rep{6}$
	& $\cfrac{\SL(3,\bbR)}{\SO(3)}$ \\
& & & & & & \\
$\cfrac{E_{6(2)}}{\SU(6)\times\SU(2)}$ & $4$ & $3$ 
	& $\cfrac{\SO(3,1)}{\SO(3)}$ 
	& $8$ & $\;\,\; [\rep{3}\otimes\rep{\bar{3}}]_{\bbR}$
	& $\cfrac{\SL(3,\bbC)}{\SU(3)}$ \\
& & & & & & \\
$\cfrac{E_{7(-5)}}{\SO(12)\times\SU(2)}$ & $8$ & $5$ 
	& $\cfrac{\SO(5,1)}{\SO(5)}$ 
	& $14$ & $\rep{15}$
	&  $\cfrac{\SU^*(6)}{\Symp(6)}$ \\
& & & & & & \\
$\cfrac{E_{8(-24)}}{E_7\times\SU(2)}$ & $16$ & $9$ 
	& $\cfrac{\SO(9,1)}{\SO(9)}$ 
	& $26$ & $\rep{27}$
	& $\cfrac{E_{6(-26)}}{F_4}$  \\
\end{tabular}
\caption{$\cN=(1,0)$ supergravity and exotic supergravity theories and their duality groups}
\label{tab:magic}
\end{table}


\section{Anomalies of exotic multiplets}
\label{sec:anomalies}

Since the exotic multiplets contain chiral fields, they may suffer from anomalies. This section is aimed at extending the results known for chiral spin $\frac{1}{2}$, spin $\frac{3}{2}$  and self-dual fields to the SD Weyl field and the exotic gravitino.

We start by considering a Dirac operator coupled to a vector bundle $V$, and briefly review the relation between the index theory and anomalies. Some choices of $V$ are  well-understood and relate to the standard anomalies for fields that appear in supergravity multiplets \cite{AGW, AGG1, AGG2, ASZ}.  These cases, i.e. the chiral spin $1/2$ and $3/2$ fermions, and selfdual tensor fields, will be reviewed in subsection \ref{sec:standard}, mostly following the conventions of a recent review  \cite{Bilal}.  As we shall show, the curvatures of all relevant exotic fields, i.e. the SD Weyl field of $(4,0)$ multiplet, its counterpart  in the $(3,1)$ multiplet as well as the exotic gravitino can be found in the domain of the Dirac operators for appropriate choices of $V$. The index calculation for the fields in the $(4,0)$  multiplet will be presented in subsection \ref{sec:exotic}. The anomaly polynomials for other six-dimensional exotic multiplets with different number of supercharges will be given in subsection \ref{sec:exo_a}.


\subsection{Anomalies in standard supergravity fields}
\label{sec:standard} 

The anomalies in $4k+2$ dimensional theories are encoded by characteristic classes in $4k+4$ dimensions, which can be computed using the index theorems for the Dirac operators. 

Suppose our space-time manifold with Euclidean signature has a spin structure and let $S$ be the spinor bundle. Then the Dirac operator on the smooth section of the spinor bundle $\mathcal{C}^{\infty}(S)$
is defined as the composition 
\begin{equation} \label{Cliffordmultiplication}
D = cl \circ \nabla^{S} : \mathcal{C}^{\infty}(S) \xrightarrow{\nabla^{S}} \mathcal{C}^{\infty}(T^*M \otimes S) \xrightarrow{cl} \mathcal{C}^{\infty}(S),
\end{equation}
where $\nabla^{S}$ is the spin connection and $cl$ is the Clifford multiplication. In local coordinates, this is the Dirac trace of the covariant derivative in some representations of the gamma matrices\footnote{In section~\ref{sec:anomalies} we use indices $\mu,\nu$ for the $4k+2$ dimensional spacetime.} 
\begin{equation}
D = cl(e^{\mu}) \nabla^{S}_{e_{\mu}} = \gamma^{\mu} \nabla^{S}_{\mu} \,.
\end{equation}
In space-time dimension $4k+2$, the spinor bundle decomposes into subbundles of definite chiralities with respect to the Euclidean chirality operator $\Gamma = i^{2k+1} e^1 e^2 \cdots e^{4k+2}$, i.e. $S=S^+ \oplus S^-$.  Consequently,  $D$ takes an off-diagonal form:
\begin{equation}
D = \begin{pmatrix}
0 & D^-  \\
D^+ & 0 \\
\end{pmatrix} 
\end{equation}
and the relevant positively projected Dirac operator flips the chirality of the spinor field
\begin{equation}
D^+: \mathcal{C}^{\infty}(S^+) \longrightarrow \mathcal{C}^{\infty}(S^-) .
\end{equation}
Since the full Dirac operator $D$ is self-adjoint, it has always vanishing index. It is $D^+$, whose adjoint is $D^- : \mathcal{C}^{\infty}(S^-) \longrightarrow \mathcal{C}^{\infty}(S^+)$, that has a non-trivial index. For the rest of the paper we shall omit the superscript $+$ and use $D$ to denote the appropriate Dirac operator.

The Dirac operator can be twisted by some vector bundle  $V$ (i.e. act on spinors coupled to some vector gauge field)
\begin{equation} \label{Diracform}
D : \mathcal{C}^{\infty} (S^+ \otimes V) \longrightarrow \mathcal{C}^{\infty}(S^- \otimes V)
\end{equation}
Applying the index theorem \cite{ATS},  its index density is given by \footnote{See appendix~\ref{app:Conv} for the definitions and conventions.}
\begin{equation}
\label{eq:IDirac}
\text{Ind}(D) = \Hat{A}(M) \text{ch}(V).
\end{equation}
Furthermore, one can also generalize the definition of the Dirac operator to the Clifford module $E$, a vector bundle whose fiber admits a Clifford action. In the definition \eqref{Cliffordmultiplication} we just replace the Clifford multiplication $cl$ by the Clifford action and replace the spin connection $\nabla^{S}$ by the connection $\nabla^{E}$ on $E$. 

To talk about the index theory of the generalized Dirac operator we would like to put it in the twisted form \eqref{Diracform}. If our even-dimensional base manifold is spin and oriented, then every $E$ Clifford module has a product structure $E= S \otimes V$, where $V$ is a vector bundle determined by $E$, $S$ and the Clifford action on $E$ \cite{BNV}. By making use of the chiral decomposition of $S$ we define $E^{\pm}:=  S^{\pm} \otimes V$ and thus 
\begin{equation}
D_{E} : \mathcal{C}^{\infty} (E^+) \longrightarrow \mathcal{C}^{\infty}(E^-). 
\end{equation}

A pertinent example of Clifford module is  given by the bundle of differential forms $\Lambda^\bullet T^*M$, which is a tensor product of spinor bundles $\Lambda^\bullet T^*M= S \otimes S$. The sections of $S$ are spinors transforming in the spinor representation of $SO(4k+2)$. One could further restrict to the chiral $S^+$ and anti-chiral $S^-$ subrepresentations and obtain
\begin{equation}
 S^{\pm} \otimes S^{\pm}= \Lambda^1 T^*M \oplus \Lambda^3 T^*M \oplus \ldots \oplus \Lambda^{2k+1}_{\pm} T^*M
\end{equation}
and 
\begin{equation}
S^{+} \otimes S^{-}= \Lambda^0 T^*M \oplus \Lambda^2 T^*M \oplus \ldots \oplus \Lambda^{2k} T^*M
\end{equation}
where $\Lambda^{2k+1}_{\pm} T^*M$ are the self-dual (anti-self-dual) forms. In Euclidean signature a $n$-form $F_{\mu_1 \ldots \mu_n}$ is self-dual it it obeys
$F_{\mu_1 \ldots \mu_n} = \frac{i}{n!} \epsilon_{\mu_1 \ldots \mu_{2n}} F^{\mu_{n+1} \ldots \mu_{2n}}$.

The Hirzebruch signature operator is given by 
\begin{equation}
\tau : \mathcal{C}^{\infty} (S^+\otimes (S^+ \oplus S^-)) \longrightarrow \mathcal{C}^{\infty}(S^- \otimes (S^+ \oplus S^-))
\end{equation}
with $V = S^{+} \otimes S^{-}$ (cf \eqref{eq:IDirac}), and its index is given by the Hirzebruch $L$-polynomial.  From other side, the complexifications of self-dual even forms and anti-self-dual odd forms are given by $S^{+} \otimes S^{-}$ and $S^{-} \otimes S^{-}$ respectively, and we are interested in the index of 
\begin{equation}
D_A : \mathcal{C}^{\infty} (S^+\otimes  S^-) \longrightarrow \mathcal{C}^{\infty}(S^- \otimes S^- )
\end{equation}
with  $V = S^{-}$. It can be shown that the result for the index is equal to half of the Hirzebruch $L$-polynomial with an additional $-$ sign due to Bose rather than Fermi statistics, and is given by 
\begin{equation}
I^{A}_{2n+2} =\left(-\frac{1}{2}\right)\left(\frac{1}{4}\right)[\Hat{A}(M) \ch(\tilde{R})]_{2n+2} =\left(-\frac{1}{2}\right)\left(\frac{1}{4}\right)[L(M)]_{2n+2}
\end{equation} 
where $\tilde{R} = \frac{1}{2} R_{\mu\nu} \gamma^{\mu\nu}$ with $R$ being the Riemann tensor of $M$ and $\gamma^{\mu\nu}$ the generator in the spinor representation. 
The pre-factor factorizes $\frac{1}{4} = \frac{1}{2} \times \frac{1}{2}$, where the first $\frac{1}{2}$ due is to the chirality projector of the second spinor and the second $\frac{1}{2}$ comes from the constraint that we consider $F$ as a real field when analytically continuing to Lorentzian signature. 

For the gravitino field, the relevant  Rarita-Schwinger complex is given by
\begin{align} \label{RSC}
 \mathcal{C}^{\infty}(S^+\otimes T^*M) 
\longrightarrow \mathcal{C}^{\infty}(S^-\otimes T^*M)
\end{align}
The gravitino anomalies are actually given by the map:
\begin{align}  \label{RSCP}
D: \mathcal{C}^{\infty}(S^+\otimes (T^*M-1)) 
\longrightarrow \mathcal{C}^{\infty}(S^-\otimes (T^*M-1)) \, .
\end{align}
The origin of this formal shift is explained in \cite{AGW} and we shall come back to it in the next subsection. 
The tensor product  $S^+\otimes T^*M$ contains an anti-chiral spinor $S^-$ that needs to be projected out. In addition  a vector potential  in $D$ dimensions has $D-2$ physical degrees of freedom. These together lead to the $-\Hat{A}(M)$ in the expression for the index:
\begin{equation} \label{normalgravitino}
I^{\mathrm{spin}\frac{3}{2}}_{2n+2} =\Hat{A}(M) \ch(R) - 2 \Hat{A}(M) + \Hat{A}(M) =[\Hat{A}(M) (\ch(R) - 1)]_{2n+2} ,
\end{equation}
where $R$ is the curvature two-form in the vector representation of $SO(4k+2)$.


\subsection{Anomalies for product multiplets}
\label{sec:product}

Many supergravity theories can be seen as products of Yang-Mills multiplets with less supersymmetry. In cases when the resulting supergravity is chiral, the anomalous part of the spectrum can be analysed like in the previous subsection. All the fields are in the domain of a Dirac operator with choices of $V$ being given by the tangent bundle or (products of) spin bundles. As a result all standard supermultiplets have anomalies of very constrained form.

Type IIB supergravity is a prime example of such a product theory, and can be obtained as a double copy  two $(1,0)$ Yang-Mills multiplets. The anomalous part (a couple of left gravitini, two right dilatini and a tensor field with a self-dual five-form field strength)  is given by 
$$
   \lambda^L \circ A_{\mu} +   A_{\mu} \circ  \lambda^L +  \lambda^L \circ \lambda^L
$$
Hereafter we shall use $\circ$ to denote the products of fields. As already mentioned $ \lambda^L \circ A_{\mu} $ projects into the left gravitino and a right spin $1/2$ field. Note that both are in the IIB spectrum, and one only needs to worry about the subtraction of 2 vectorial degrees of freedom. The whole IIB anomalous complex can the be thought of as 
\begin{align}  \label{IIB}
\mathcal{C}^{\infty}(S^+\otimes ( 2 \times (T^*M-2) \oplus S^+))
\longrightarrow \mathcal{C}^{\infty}(S^-\otimes ( 2 \times (T^*M-2) \oplus S^+)) \, .
\end{align}
 with the resulting anomaly given by the 12-form
\begin{equation} \label{IIBano}
I^{\mathrm{IIB}} =  - \Bigl[\Hat{A}(M) \left( \ch(R) - 2 -  \frac18 \ch(\tilde{R}) \right) \Bigr]_{12}
\end{equation}
that vanishes \cite{AGW}.

The reduction of IIB on a $K3$ surface yields a six-dimensional $(2,0)$ theory that contains a supergravity multiplet and 21 tensor multiplets and is also anomaly free. One can also see that the non-chiral and obviously non-anomalous maximal $(2,2)$ supergravity can be decomposed into $(2,0)$ multiplets and contains a $(2,0)$ gravity multiplet, together with four gravitino multiplets and five tensor multiplets. Hence the three standard $(2,0)$ multiplets have anomaly polynomials that are proportional
\begin{equation} \label{eq:X8}
- \frac{1}{21} I_{\text{gravity}} = \frac14 I_{\text{gravitino}} = I_{\text{tensor}} := X_8 = \frac{1}{48}\left(\frac{p_1^2}{4} - p_2 \right). 
\end{equation}
Because of the M5-brane anomalies and inflow, the $X_8$ polynomial appears in the M-theory action via gravitational Chern-Simons couplings. The contraction structure in $X_8$ is given by the $t_8$ tensor that appears naturally in the string amplitudes.

Working directly with six-dimensional multiplets,  we note that 
the product of two $\mathcal{N}=(1,0)$ vector multiplets is a sum of  $\mathcal{N}=(2,0)$ gravity and tensor multiplets:

\begin{table}[h]
\centering
\label{table1}
\caption{6d $\mathcal{N}=(1,0)$ Yang-Mills squared}
\begin{tabular}{|l|l|l|}
\hline
$\mathcal{N}=(1,0)$ & $A_{\mu}$ vector $\rep{(2,2;1)}$ & $\lambda^{L}$ chiral fermion $\rep{(1,2;2)}$
              \\ \hline
               \multirow{4}{*}{$A_{\mu}$ vector $\rep{(2,2;1)}$}   & \multicolumn{1}{l|}{$g_{\mu\nu}$ $\rep{(3,3;1)}$} & \multicolumn{1}{l|}{$\psi^{L}_{\mu}$ $\rep{(2,3;2)}$} 
                                                                      \\ & \multicolumn{1}{l|}{${B_{\mu\nu}}^{-}$ $\rep{(3,1;1)}$} & \multicolumn{1}{l|}{$\lambda^{R}$ $\rep{(2,1;2)}$} \\

                                  & \multicolumn{1}{l|}{${B_{\mu\nu}}^{+}$ $\rep{(1,3;1)}$} & \multicolumn{1}{l|}{}\\

                                  & \multicolumn{1}{l|}{$\phi$ $\rep{(1,1;1)}$} & \multicolumn{1}{l|}{}\\
         \hline
\multirow{2}{*}{$\lambda^{L}$ chiral fermion $\rep{(1,2;2)}$} & \multicolumn{1}{l|}{$\psi^{L}_{\mu}$ $\rep{(2,3;2)}$} & \multicolumn{1}{l|}{$\phi$ $\rep{(1,1;4)}$} \\

                                  & \multicolumn{1}{l|}{$\lambda^{R}$ $\rep{(2,1;2)}$} & \multicolumn{1}{l|}{${B_{\mu\nu}}^{+}$ $\rep{(1,3;4)}$} \\
                    \hline

\end{tabular}
\end{table}

The anomalous part of the product is given by 
 \begin{equation}
(A_{\mu}+2 \times \lambda^L) \circ (A_{\mu}+2 \times \lambda^L)  \Rightarrow 2   \lambda^L \circ A_{\mu} +  2 A_{\mu} \circ  \lambda^L + 4 \lambda^L \circ \lambda^L,
\end{equation}
and like in the ten-dimensional case,  $ \lambda^L \circ A_{\mu} $ and  $  A_{\mu}  \circ \lambda^L   $  contains the  left-moving gravitini and the right-moving tensorini, which are in the spectrum with a net 
contribution to the anomaly given by  $-\frac{1}{2} \Hat{A}(TM) [\ch(R)-2]$. The product of the two chiral spinors $\lambda^L$ results in a self-dual 2-form. The total anomaly of is 
\begin{equation}
\begin{aligned}
&- 2 \cdot 2 \cdot  \frac{1}{2} \Hat{A}(TM) [\ch(R) -2] - 4 I^{A}  \\
=& - \frac{1}{5760} \left(536 {p_1}^2 - 1952 p_2 \right) - 4 \cdot \frac{1}{5760} \left(16 {p_1}^2 - 112 p_2 \right) = - 20 X_8 \\
\end{aligned}
\end{equation}

Another six-dimensional example dimensions is the studied in \cite{YangMillssquared}, where the tensor product of super Yang-Mills multiplets with $\mathcal{N}=(1,0)$ and $\mathcal{N}=(1,1)$ supersymmetries is shown to yield the gravity multiplet in $\mathcal{N}=(2,1)$. The details of the tensor product are summarised in the following table:

\begin{table}[!ht]
\centering
\caption{ 6d $\mathcal{N}=(1,0)$ Yang-Mills tensor with $\mathcal{N}=(1,1)$ Yang-Mills}
\label{table2}
\begin{tabular}{|l|llll|}
\hline
                                         & \multicolumn{4}{l|}{$\mathcal{N}=(1,1)$}                                                                                                                                                                                                   \\
                                         & $A_{\mu}$ $\rep{(2,2;1,1)}$                               & $\lambda^{R}$ $\rep{(2,1;1,2)}$                                          & $\lambda^{L}$ $\rep{(1,2;2,1)}$                                          & $\phi$ $\rep{(1,1;2,2)}$                         \\ \hline
$\mathcal{N}=(1,0)$                      & \multicolumn{1}{l|}{$g_{\mu\nu}$ $\rep{(3,3;1,1)}$}       & \multicolumn{1}{l|}{\multirow{2}{*}{$\psi^{R}_{\mu}$ $\rep{(3,2;1,2)}$}} & \multicolumn{1}{l|}{\multirow{2}{*}{$\psi^{L}_{\mu}$ $\rep{(2,3;2,1)}$}} & \multirow{4}{*}{$A_{\mu}$ $\rep{(2,2;2,2)}$}     \\
\multirow{3}{*}{$A_{\mu}$ $\rep{(2,2;1)}$}     & \multicolumn{1}{l|}{${B_{\mu\nu}}^{-}$ $\rep{(3,1;1,1)}$} & \multicolumn{1}{l|}{}                                              & \multicolumn{1}{l|}{}                                              &                                            \\
                                         & \multicolumn{1}{l|}{${B_{\mu\nu}}^{+}$ $\rep{(1,3;1,1)}$} & \multicolumn{1}{l|}{\multirow{2}{*}{$\lambda^{L}$ $\rep{(1,2;1,2)}$}}    & \multicolumn{1}{l|}{\multirow{2}{*}{$\lambda^{R}$ $\rep{(2,1;2,1)}$}}    &                                            \\
                                         & \multicolumn{1}{l|}{$\phi$ $\rep{(1,1;1,1)}$}             & \multicolumn{1}{l|}{}                                              & \multicolumn{1}{l|}{}                                              &                                            \\ \cline{2-5} 
\multirow{2}{*}{$\lambda^{L}$ $\rep{(1,2;1)}$} & \multicolumn{1}{l|}{$\psi^{L}_{\mu}$ $\rep{(2,3;2,1)}$}   & \multicolumn{1}{l|}{\multirow{2}{*}{$A_{\mu}$ $\rep{(2,2;2,2)}$}}        & \multicolumn{1}{l|}{$\phi$ $\rep{(1,1;4,1)}$}                            & \multirow{2}{*}{$\lambda^{L}$ $\rep{(1,2;4,2)}$} \\
                                         & \multicolumn{1}{l|}{$\lambda^{R}$ $\rep{(2,1;2,1)}$}      & \multicolumn{1}{l|}{}                                              & \multicolumn{1}{l|}{${B_{\mu\nu}}^{+}$ $\rep{(1,3;4,1)}$}                &                                            \\ \hline
\end{tabular}
\end{table}
\noindent
The resulting  $\mathcal{N}=(2,1)$ supergravity multiplet \cite{Strathdee}  contains one graviton $g_{\mu\nu}$, 4 left-handed gravitini $\psi^{L}_{\mu}$, 2 right-handed gravitini $\psi^{R}_{\mu}$, 8 vectors $A_{\mu}$, one anti-self-dual 2-form ${B_{\mu\nu}}^{-}$, 5 self-dual 2-form ${B_{\mu\nu}}^{+}$, 4 right-handed fermions $\lambda^{R}$, 10 left-handed fermions $\lambda^{L}$ and 5 scalars $\phi$. We may once more consider the anomaly as the sum of the anomalies from individual terms in the product
\begin{equation}
\begin{aligned}\label{21product1}
&(A_{\mu}+2 \times \lambda^L) \circ (A_{\mu}+2 \times \lambda^R +2 \times \lambda^L + 4 \times \phi)  \\
\Rightarrow \quad & 2 A_{\mu} \circ  \lambda^R +  2 A_{\mu} \circ  \lambda^L  + 2   \lambda^L \circ A_{\mu} +4 \lambda^L \circ \lambda^R + 4 \lambda^L \circ \lambda^L + 8  \lambda^L \circ \phi \\
= \quad & 2   \lambda^L \circ A_{\mu} +4 \lambda^L \circ \lambda^L + 8  \lambda^L \circ \phi  \, ,
\end{aligned}
\end{equation}
where only the anomalous terms are kept. The total anomaly is given by
\begin{equation}
\begin{aligned}\label{21product2}
I^{\text{product}}& = -  2 \cdot  \frac{1}{2} \Hat{A}(TM) [\ch(R) -2] - 4 I^{A} - 8 \cdot \frac{1}{2}I^{\mathrm{spin}\frac{1}{2}} \\
&=- \frac{1}{5760} \left(268 {p_1}^2 - 976 p_2 \right) - 4 \cdot \frac{1}{5760}\left(16 {p_1}^2 - 112 p_2 \right)- 4 \cdot \frac{1}{5760}\left(7 {p_1}^2 - 4 p_2 \right)\\
&= - \frac{360}{5760} (p^{2}_1-4p_2)
\end{aligned}
\end{equation}
and  agrees with the direct calculation
\begin{equation}
\begin{aligned}\label{21gravity}
I^{\text{gravity}}_{\mathcal{N}=(2,1)} &= \frac{1}{2} \cdot (-4+2) I^{\mathrm{spin}\frac{3}{2}} + (1-5) I^{A} + \frac{1}{2}(4 - 10)I^{\mathrm{spin}\frac{1}{2}} \\
&= -I^{\mathrm{spin}\frac{3}{2}}-4 I^{A} - 3I^{\mathrm{spin}\frac{1}{2}} =  -12 X_8 \, .
\end{aligned}
\end{equation}

\subsection{ Index densities of exotic Dirac operators}
\label{sec:exotic}

The indices of the exotic fields (and multiplets) can be computed using \eqref{eq:IDirac}. The only essential difference from the calculations reviewed above is that $V$ is now given by a product of bundles. 
In the $\mathcal{N}=(4,0)$ multiplet \eqref{6dmultiplet} there are two exotic anomalous objects, namely the exotic gravitino $\psi_{\mu\nu}$ in  
 $\rep{(4,1; 8)}$ and the exotic graviton $C_{\mu\nu\rho\sigma}$ in $\rep{(5,1;1)}$. We treat each in turn.
 
 \subsubsection{Exotic gravitino}
 
 We start with the fermion $\psi_{\mu\nu}$.  The field strength $\chi$ is anti-self-dual with respect to $SO(6)$:\footnote{Note that in the Euclidean space-time it is anti-self-dual and it is self-dual in the Minkowskian cases. Similarly, left-handed spinors have negative chirality in the Euclidean space-time, while they are right-handed with positive chirality in Minkowskian.}
\begin{equation}
\chi = - \star_{\text{E}} \;\chi \Longleftrightarrow \chi_{\mu\nu\rho} = - \frac{i}{3!} \epsilon_{\mu\nu\rho\alpha\beta\gamma} \chi_{\alpha\beta\gamma}
\end{equation}
where the Hodge-star $\star_{\text{E}}$ is taken in the Euclidean convention. Because of the (anti)-self-duality of the field strength $\chi_{\mu\nu\rho}$, it can be viewed as transforming in the representation $[0,0,3]$ of $\mathfrak{su}^*(4)$. The advantage of  working directly with $\chi$ is that we do not need to worry about the ghost contribution and the calculation follows the treatment of the self-dual forms  \cite{AGW}. The potential $A$ and its self-dual field strength $F^+$ are viewed as independent variable in the path-integral formalism. Since there is no gauge freedom in $F^+$, there is no need to  subtract ghost contributions.

Recall that the Dynkin label of the negative chiral spinor representation of $\mathfrak{su}^*(4)$ is $[0,0,1]$\footnote{Our conventions for the Dynkin labels are outlined in appendix~\ref{app:Conv}.}. The field strength $\chi_{\mu\nu\rho}$ is viewed as an irreducible piece $[0,0,3]$ in the tri-spinor product $(S^-)^{\otimes 3}$. 
\begin{align*}
[0,0,1] \otimes [0,0,1] \otimes [0,0,1] 
&= [0,0,3] \oplus [0,1,1] \oplus [0,1,1] \oplus [1,0,0] 
\end{align*}
We can recast the result for representations of   $\mathfrak{su}^*(4)$  in terms of the sections of the corresponding bundles:
\begin{align*}
D_{\chi}: \mathcal{C}^{\infty}(S^+\otimes S^- \otimes S^-) - \mathcal{C}^{\infty}(S^+ \otimes T^*M) - \mathcal{C}^{\infty}(S^+ \otimes T^*M) +\mathcal{C}^{\infty}(S^-)\\
\longrightarrow \mathcal{C}^{\infty}(S^-\otimes S^- \otimes S^-) - \mathcal{C}^{\infty}(S^- \otimes T^*M) - \mathcal{C}^{\infty}(S^- \otimes T^*M) + \mathcal{C}^{\infty}(S^+)
\end{align*}
leading to the definition of the complex for the exotic gravitino
\begin{align} \label{arrow}
\Rightarrow D_{\chi}: \mathcal{C}^{\infty}(S^+\otimes[ S^- \otimes S^- -  T^*M^{\oplus 2} -1]) 
\longrightarrow \mathcal{C}^{\infty}(S^-\otimes[ S^- \otimes S^- -  T^*M^{\oplus 2} -1]).
\end{align}

The formal manipulation above is allowed in K-theory \cite{ATS}, and effectively we have the index theorem for the index density of $D_{\chi}$
\begin{equation} \label{Dchi}
\begin{split}
\text{Ind}(D_{\chi})&= \Hat{A}(M) [\text{Ch}((S^-)^{\otimes 2}) - \text{Ch}(T^*M^{\oplus 2})-1]\\
&=\Hat{A}(M) [\text{Ch}(S^-)^2 - 2 \text{Ch}(T^*M) - 1]
\end{split}
\end{equation}
According to the famous results \cite{ATS}, we have 
\begin{equation}
\text{Ch}(S^+ \oplus S^-) = \prod^{n}_{j=1} 2 \cosh \frac{x_j}{2}  \quad \text{and} \quad \text{Ch}(S^+ - S^-) = \prod^{n}_{j=1} 2 \sinh \frac{x_j}{2} 
\end{equation} 
for the space-time manifold in $2n$ dimensions. 
It follows that 
\begin{align}
\text{Ch}(S^+) &=\frac{1}{2} \left( \prod^{n}_{j=1} 2 \cosh \frac{x_j}{2} + \prod^{n}_{j=1} 2 \sinh \frac{x_j}{2}\right) \\
 \text{Ch}(S^-) &= \frac{1}{2} \left( \prod^{n}_{j=1} 2 \cosh \frac{x_j}{2} - \prod^{n}_{j=1} 2 \sinh \frac{x_j}{2}\right)
\end{align}
Inserting this into \eqref{Dchi} and using the relation \eqref{eigenvalueR}, we arrive at
\begin{equation}\label{indexdensityexoticgravitino}
\text{Ind}(D_{\chi}) = \frac{1}{5760}(501 p_1^2 +3828 p_2) \, .
\end{equation}
The contribution to the gravitational anomaly from $\chi$ is obtained from the above result by multiplying it by $(-1)^2 \frac{1}{2}$. The first $-1$ comes from the fact that $\chi$ is fermionic and the second $-1$ is because the map in \eqref{arrow} is actually in the opposite direction \cite{ATS}. The division by  $2$ is due to the fact  that self-dual tensor in Lorentz signature satisfies the reality condition.
\begin{equation} \label{exoticgravitino}
I_{\chi} = (-1)^2 \frac{1}{2} \text{Ind}(D_{\chi}) =  \frac{1}{5760}(\frac{501}{2} p_1^2 +1914 p_2).
\end{equation}


\subsubsection{SD Weyl field}
\label{sec:a-SDW}
We now turn to the index density of the field strength of the exotic graviton defined in~\eqref{exoticgravitonfieldstrength}, $G_{\mu\nu\rho\sigma\tau\kappa} = \partial_{[\mu}C_{\nu\rho][\sigma\tau,\kappa]} $. The field strength $G$ is in the $[0,0,4]$ of $\mathfrak{su}^*(4)$, and in order to obtain it from a tensor product, one can take a pair of the field strengths $F_3^-$ of self-dual 2-forms:
\begin{equation} \label{2x2}
[0,0,2] \otimes [0,0,2] =[0,0,4]  \oplus [0,1,2]   \oplus [0,2,0].
\end{equation}
For the $[0,1,2]$ part, 
\begin{equation}
[0,1,0] \otimes [0,0,2] = [0,1,2] \oplus [1,0,1].
\end{equation}
The representations $[0,2,0]$ and $[1,0,1]$ are immediately recognised  as the metric $g_{(\mu\nu)}$ and the two-form $B_{[\mu\nu]}$ respectively. 
The individual $[0,0,2]$ appears also as an irreducible part  in the tensor product of 2 negative chirality spinors:
\begin{equation}
[0,0,1] \otimes [0,0,1] = [0,1,0] \oplus [0,0,2].
\end{equation}
We can consider a product of four chiral spinors $[0,0,1]$ and,  applying the tensor product decomposition, obtain 
\begin{equation}
\begin{split}
[0,0,1]^{\otimes 4} & = \left( [0,1,0] \oplus [0,0,2] \right) \otimes \left( [0,1,0] \oplus [0,0,2] \right) \\
&= \left( [0,1,0] \otimes [0,1,0] \right) \oplus \left( [0,1,0] \otimes [0,0,2] \right) \oplus \left( [0,0,2] \otimes [0,1,0] \right) \\
& \;\;\;\;\oplus  [0,0,4]  \oplus [0,1,2]   \oplus [0,2,0], 
\end{split}
\end{equation}
where  \eqref{2x2} is used to get the last three terms.

The $[0,0,4]$ can now be extracted, and the result can be recast in terms of sections of corresponding bundles. The details of this calculation can be found in Appendix \ref{sec:calcul}. The resulting complex for $D_G$ operator is given by
\begin{equation}
\begin{aligned} 
D_{G}&: \mathcal{C}^{\infty}\left(S^+\otimes [S^- \otimes S^-  \otimes S^- - (S^- \otimes T^*M )^{\oplus 3} + (S^+)^{\oplus 2} ] \right)  + B + g \\
&\longrightarrow \mathcal{C}^{\infty}\left(S^-\otimes [S^- \otimes S^-  \otimes S^- - (S^- \otimes T^*M )^{\oplus 3} + (S^+)^{\oplus 2} ] \right)  + B + g.
\end{aligned}
\end{equation}
At this stage, we can state that the sections to which $B$ and $g$ belong do not contribute to the index density. Simply said, the metric and a generic two-form field are anomaly free. It follows the relevant complex is
\begin{equation}
\begin{aligned} 
 D_{G}&: \mathcal{C}^{\infty}\left(S^+\otimes [S^- \otimes S^-  \otimes S^- - (S^- \otimes T^*M )^{\oplus 3} + (S^+)^{\oplus 2} ] \right) \\
&\longrightarrow \mathcal{C}^{\infty}\left(S^-\otimes [S^- \otimes S^-  \otimes S^- - (S^- \otimes T^*M )^{\oplus 3} + (S^+)^{\oplus 2} ] \right),
\end{aligned}
\end{equation}
and is again in the form \eqref{Diracform}. The index density for $D_G$ is then 
\begin{equation} 
\text{Ind}(D_G)= \Hat{A}(M) \left( \text{Ch}((S^-)^3) - 3  \text{Ch}(S^-) \text{Ch}(T^*M) +2  \text{Ch}(S^+) \right)
\end{equation}
Every individual factor is known and one can show that
\begin{equation} 
\text{Ind}(D_R)= \frac{1}{3} (2 p_1^2 + 10 p_2) = \frac{1}{5760}(3840 p_1^2 + 19200 p_2).
\end{equation}
Since $G$ is bosonic and  the reality condition  is imposed on it in order to move to the Minkowski signature, the anomaly for the field strength is 
\begin{equation} \label{DR}
I_G = (-1) (\frac{1}{2})\text{Ind}(D_G) = \frac{1}{5760}  (-1920 p_1^2 - 9600 p_2) \,.
\end{equation}\\

\subsubsection{Exotic graviton in the $(3,1)$ multiplet}
The field strength of the three-index exotic graviton $D$ in the $(3,1)$ multiplet $S_{\mu\nu\rho\sigma\kappa}=\partial_{[\mu} D_{\nu \rho][\sigma, \kappa]}$ is also subject to self-duality condition, and hence the field is expected to have a non-vanishing index. The discussion follows closely  the previous section and  we focus on the field strength $S$ which is in the $[1,0,3]$ representation. Due to the absence of residual gauge symmetry, one can avoid the discussion of ghosts and quantisation.

The relevant Dirac operator for $S$ is given by  (details of the computation can be found in the Appendix \ref{sec:calcul}):
\begin{equation}
\begin{aligned} 
 D_{S}&: \mathcal{C}^{\infty}\left(S^+\otimes [S^- \otimes S^-  \otimes S^+ - (S^+ \otimes T^*M )^{\oplus 2} - (S^-)^{\oplus 2} ] \right) \\
&\longrightarrow \mathcal{C}^{\infty}\left(S^-\otimes [S^- \otimes S^-  \otimes S^+ - (S^+ \otimes T^*M )^{\oplus 2} - (S^-)^{\oplus 2} ]\right).
\end{aligned}
\end{equation}
It follows that
\begin{equation} 
\text{Ind}(D_S)= \Hat{A}(M) \left( (\text{Ch}(S^-))^2\text{Ch}(S^+) - 2  \text{Ch}(S^+) \text{Ch}(T^*M) -2  \text{Ch}(S^-) \right) \, ,
\end{equation}
and the anomaly polynomial can be computed as
\begin{equation} 
I_S= (-1) (\frac{1}{2})\text{Ind}(D_S) = \frac{1}{5760}  ( -3808 p_1^2 - 7904 p_2).
\end{equation}


\subsection{Anomalies of the exotic multiplets with different supersymmetries}
\label{sec:exo_a}

Now we are able to collect everything together and present the anomaly formulae for different multiplets.

The anomalous objects among the  6d $\mathcal{N}=(4,0)$ multiplet are the exotic graviton $\rep{(5, 1; 1)}$, the self-dual 2-forms  $\rep{(3,1; 27)}$, the exotic gravitini $\rep{(4,1; 8)}$ and the chiral fermions $\rep{(2,1;48)}$.  Taking into account signs due to chirality the total anomaly is given by
\begin{equation} \label{totalanomalies}
I_{(4,0)} = I_R + 27 I_A +  8 I_{\chi} + \frac{1}{2} \times 48 I_{\frac{1}{2}} = \frac{1}{5760} (684 p_1^2 + 2592 p_2) \neq 0
\end{equation}
Since this multiplet is a product \cite{Tensorsquared, magicpyramid}, we could obtain the same result by following the section \ref{sec:product}.  A concrete product construction is described in Table 12  of \cite{magicpyramid}, here we just give the construction of exotic graviton in the light-cone
\begin{equation}
\rep{(3,1) \otimes (3,1) = (5,1) \oplus (3,1) \oplus (1,1)}.
\end{equation}
It follows that if we view the exotic graviton field strength as product of field strengths of a pair of chiral 2 forms and apply the same product construction to other fields, we end up with the same equation \eqref{totalanomalies} for the total anomaly.

The anomaly contributions in the $(3,1)$ multiplet are given by
\begin{equation} 
I_{(3,1)}=  I_S + 12 I_A +  2 I_{\chi} + \frac{1}{2} \times 6 I_{\frac{3}{2}}  + \frac{1}{2} \times (28-14) I_{\frac{1}{2}}
= \frac{1}{5760} \left(-2241 p_1^2 -8388p_2 \right).
\end{equation}

The  $(2,0)$ SD Weyl multiplet  consists of {$C_{2,2}; 6 B_2^{-}; \phi; 4 \psi_{\mu\nu}^R; 4 \psi^R$}. Its anomaly is given by 
\begin{equation}
I_{(2,0) \ exotic} =  I_R + 6 I_A +  4 I_{\chi} + \frac{1}{2} \times 4 I_{\frac{1}{2}} = \frac{1}{5760}( -808 p_1^2  - 2624 p_2) 
\end{equation}

Finally, the $(1,0)$ SD Weyl multiplet comprises $C_{2,2}; B_2^{-}; \phi; 2 \psi_{\mu\nu}^R$ and has an anomaly polynomial
\begin{equation}
I_{(1,0) \ exotic} =  I_R +  I_A +  2 I_{\chi} = \frac{1}{5760}( -1403 p_1^2  - 5884 p_2) 
\end{equation}

\subsection{An anomaly-inspired  proposal for the ghost structure}
\label{sec:ghost}

So far we have performed the index-theoretical computation of the exotic field anomalies. While a universal feature is that all fields that appear in these supermultiplets are in the domain of a Dirac operator for some choice of vector bundle $V$, for the exotic fields $V$ is given by a product of spin bundles. Hence, much like for the self-dual tensor fields (and unlike the gravitino) the computation involves the field strengths rather than potentials. This comes with a certain advantage - since there is no gauge freedom of the field strength, there is no need to manually add any ghost field contributions to the anomaly as one would do for the gravitino.

To the best of our knowledge,  there is no complete quantisation scheme at hand for the free classical field $\psi_{\mu\nu}$ (with self-duality constraint on the field strength), i.e. the exotic gravitino. To make a educated guess, we carry out an alternative computation of the anomaly via the potential $\psi_{\mu\nu}$ without excluding the ghosts contributions. Then by comparing the result with \eqref{exoticgravitino}, we should be able to deduce the ghost structure. 

First, we follow the method in \cite{AGW, AGG1} and consider a more general gravitino $\psi_A$ in a tensor representation of $SO(6)$ (or of $SO(1,5)$) with the tensor index $A$. Then, similarly to \eqref{normalgravitino}, we get
\begin{equation} \label{wholetensor}
 \Hat{A}\left[\tr \left( e^{\frac{i}{2\pi}R} \right) \right]  = \Hat{A}\left[\tr \left( e^{\frac{i}{2\pi} \frac{1}{2} R_{ab} {(T^{ab})}_{AB}} \right) \right],
\end{equation}
where $a,b = 1,...,6$ is the orthogonal frame indices of $SO(6)$ and $(T^{ab})_{AB}$ is the generator. This is the anomaly of the whole tensor product. Contributions from unwanted fields that appear in the tensor product and the ghost contributions are yet to be subtracted.

To compute the anomaly of $\psi_{\mu\nu}$ (or equivalently $\psi_{ab}$ in the orthogonal frame) we set $A$ to $[ab]$ and ${(T^{ab})}_{AB} = {(T^{ab})}_{cd,ef}={(T^{ab})}_{[cd], [ef]}$.  However, two questions need to be answered before moving forward.
\begin{itemize}
\item  \textbf{the (anti-)self-duality}\\
A generic tensor field by itself has no contribution to the gravitational anomaly. It is the self-dual or anti-self-dual part that are individually anomalous, and  their anomalies cancel  when they are combined to an unconstrained tensor field.  For us it is necessary to impose the anti-self-dual condition \eqref{exotic} by hand. \\
There is a generalized Rarita-Schwinger action of the fermionic two-form proposed in \cite{Action}
\begin{equation} \label{Rarita-Schwinger}
S = \int d^6x \; \bar{\psi}_{\mu\nu} \Gamma^{\mu\nu\rho\sigma\tau}\partial_{\rho}\psi_{\sigma\tau}.
\end{equation}
The equation of motion derived from this action is 
\begin{equation} \label{eom}
\Gamma^{\alpha\beta\mu\nu\rho}\chi_{\mu\nu\rho}=0,
\end{equation}
which is shown \cite{Action} to be equivalent to the anti-self-dual condition \eqref{exotic} and the constraint 
\begin{equation}  \label{constraint}
\Gamma^{i j k l} \chi_{jkl}=0.
\end{equation}
Here $i,j,k,l$ are the spacelike indices. \\
Since the fields that appear in the Noether conserved currents must satisfy their equations of motion, we can take $(T^{ab})_{cd,ef}$ in the rank $2$-tensor representation of $SO(1,5)$
\begin{equation}
\label{generator}
(T^{ab})^{cd}_{\;\;\;ef} = 2\left( (T^{ab})^{c}_{\;\;[e} \delta^{d}_{\;\;f]} + (T^{ab})^{d}_{\;\;[f} \delta^{c}_{\;\;e]}  \right).
\end{equation}
This way the anti-self-duality constraint on the field strength of ${\psi}_{\mu\nu}$ is automatically satisfied, provided it is on-shell.

\item  \textbf{projection}

An anti-symmetric two-form of $SO(5,1)$ corresponds to the $\mathfrak{su}^*(4)$ highest weight 
$[1,0,1]$. We take the product of it with a chiral spinor
\begin{equation} \label{tensorexoticgravitino}
[1,0,1] \otimes [1,0,0] = [2,0,1] \oplus [0,1,1] \oplus [1,0,0]. 
\end{equation}
The exotic gravitino is in the $[2,0,1]$ and $[0,1,1]$ is describing an ordinary gravitino with opposite chirality.\\
We can also check this using the little group $SO(4) \equiv SU(2) \times SU(2)$.  The physical degrees of freedom of a two-form $B_{\mu\nu}$ are given by $B_{ij}$ with $i, j = 1, 2, 3, 4$ runing over the $SO(4)$ indices. Then, if we take the self-dual and anti-self-dual part together
\begin{equation}
\bf [(3,1) + (1,3)] \otimes (2,1) = (4,1) \oplus  (2,3) \oplus (2,1), 
\end{equation}
which is nothing but the decomposition \eqref{tensorexoticgravitino} translated in the little group. 
For the anti-self-dual part of $B$
\begin{equation}
\bf (3,1) \otimes (2,1) = (4,1)  \oplus (2,1) \, .
\end{equation}
Hence when computing the anomaly of the exotic gravitino, contributions of an ordinary anti-chiral gravitino and a chiral fermion should be subtracted from the expression for the index.

\end{itemize}
With these two subtleties in mind we have
\begin{equation}
D: \mathcal{C}^{\infty}(S^+\otimes \Lambda^2T^*M) 
\longrightarrow \mathcal{C}^{\infty}(S^-\otimes \Lambda^2T^*M)
\end{equation}
\begin{equation}
\text{Ind}_{\text{total}} =  \Hat{A}\left[\tr \left( e^{\frac{i}{2\pi} \frac{1}{2} R_{ab} {(T^{ab})}_{cd,ef}} \right) \right]
\end{equation}
Inserting the generator \eqref{generator} in the above equation gives
\begin{equation}
\text{Ind}_{\text{total}} = \Hat{A}\left[\tr \exp \left( \frac{i}{2\pi}  \left(R_{ce} \delta_{df} +R_{df} \delta_{ce} -R_{cf} \delta_{de}-R_{de} \delta_{cf}  \right) \right) \right],
\end{equation}
The exponent is a  $15\times 15$ matrix  which is is anti-symmetric in $c,d$ and $e,f$.  Defining
\begin{equation}
M_{cd,ef} \equiv \frac{1}{2\pi} \left(R_{ce} \delta_{df} +R_{df} \delta_{ce} -R_{cf} \delta_{de}-R_{de} \delta_{cf}  \right), 
\end{equation}
the trace can be computed in $2n$ dimensions for  $n(2n-1) \times n(2n-1)$ matrix M.
\begin{equation}
\tr e^{i M} =  \text{dim}(T) - \frac{1}{2}\tr M^2 + \frac{1}{4!} \tr M^4 + \ldots
\end{equation}
where $\text{dim}(T)= n(2n-1)$ is the dimension of the representation as well as the number of the independent components of a generic $2$ form in $2n$ dimensions. 

Since $M$ is an anti-symmetric matrix, both itself and its odd powers are traceless.  The first non-vanishing contributions to the trace are
\begin{equation}
\tr M^2 = \frac{1}{2} \sum_{a, b} M^2_{ab,ab} = \frac{1}{(2\pi)^2} (2n-2) \tr R^2, 
\end{equation}
where the factor of $\frac{1}{2}$ accounts for the fact that we are summing over independent pairs of indices $a, b$ instead of taking them as anti-symmetric double indices, and  
\begin{equation}
\tr M^4 = \frac{1}{(2\pi)^4} \left( (2n-8) \tr R^4 + 3 (\tr R^2)^2 \right).
\end{equation}
Details of the computation can be found in Appendix \ref{sec:exotic}. The final expression for the index
in six dimensions, $2n=6$, is given by
\begin{equation} \label{I_total1}
\text{Ind}_{\text{total}} = \frac{1}{5760}(1065p^{2}_1 + 1860 p_2). 
\end{equation}

This is still not the final result for the anomaly of $\psi_{\mu\nu}$. Firstly, in accordance with the contributions of an ordinary anti-chiral gravitino and a chiral fermion need to be subtracted:
\begin{equation} 
\widetilde{\text{Ind}} = \text{Ind}_{\text{total}} - \Hat{A}(M) \ch{(T^*M)} = \frac{1}{5760}(783p^{2}_1 + 2844 p_2). 
\end{equation}
The above index density can be compared with \eqref{indexdensityexoticgravitino} 
\begin{equation}
\Delta = \widetilde{\text{Ind}} - \text{Ind}(D_{\chi})  =  \frac{1}{5760}(282 p^{2}_1 - 984  p_2) =I^{\mathrm{spin}\frac{3}{2}}  +  I^{\mathrm{spin}\frac{1}{2}}
\end{equation}
One recognise that $\Delta$ equals the index density contribution of a chiral spinor and a chiral gravitino. While the precise ghost structure needed in quantising the classical exotic gravitino field $\psi_{\mu\nu}$ is not completely fixed by this argument,  the net degrees of freedom that need to be removed from it are given by a chiral gravitino and a chiral spinor both of the same chirality as the exotic gravitino.

There is a quick check of the above proposal just by counting degrees of freedoms. In six dimensions, a generic off-shell chiral spinor has four components while an off-shell  unconstrained (meaning not required to satisfy $\gamma_{\mu} \psi^{\mu}=0$) chiral gravitino has $24=4 \times 6$ components. Given that $\psi_{\mu\nu}$ transform in $[2,0,1]$ of $SO(1,5)$ (see \eqref{tensorexoticgravitino}), it has generically $36$ components. 
Removing the components of a spinor and a gravitino leaves us with 8 components. As already mentioned, if one starts from the action \eqref{Rarita-Schwinger} self-duality and being on-shell are equivalent conditions for the free field $\psi_{\mu\nu}$ \cite{Action}. Hence the number of on-shell degrees of freedom of the self-dual exotic gravitino is four, 
consistently with the $\psi_{\mu\nu}$ field in the light cone transforming as $(4,1)$ of the space time little group $SO(4)$.


\section{Five-dimensional Chern-Simons interactions}

\label{sec:CS}

The non-triviality of the index bundle discussed without any obvious anomaly cancellation mechanism in view (at least for the maximally supersymmetric $(4,0)$ and $(3,1)$ cases) might be just one of the signs of trouble with the multiplets involving the SD Weyl field $C$ or its three-index counterpart $D$.
Given the lack of general covariance, this might appear to be neither too surprising nor lethal if mechanisms for reproducing the non-linear dynamics of lower dimensional gravitational theories can be established.

When compactified on a circle the degrees of freedom of these multiplets can be arranged into the fields of the five-dimensional supergravity~\cite{Hull1}. The SD Weyl field $C$ can in five dimensions be represented in terms of a symmetric field $h_{\mu \nu}$, while $D$  reduces to $h_{\mu \nu}$ plus a vector\footnote{In this section we will use five-dimensional indices $\mu,\nu$ and six-dimensional indices $M,N$.}. The six-dimensional (linearised) equations of motion are consistent with the interpretation of $h$ as the linearised excitation around the flat metric. A direct study of the dynamics of $C$ or $D$ fields beyond linearisation, and hence the comparison with the non-linear five dimensional gravity, is very difficult and this is the key problem in establishing whether interacting $(4,0)$ and $(3,1)$ theories exist.

From other side, the maximal five-dimensional supergravity is unique, and contains interactions that do not involve the metric. The topological Chern-Simons term \cite{Cremmer1} 
\begin{equation} \label{cubicinvariant}
S_{\text{CS}} = \int k_{\Lambda\Sigma\Delta} \ A^{\Lambda} \wedge F^{\Sigma} \wedge F^{\Delta}
\end{equation}
where $k_{\Lambda\Sigma\Delta}$ is constant and the ${\Lambda, \Sigma, \Delta}$ are $E_{6(6)}$ indices running from $1$ to $27$, does not admit linearisation. Hence probing its origin could be the first step towards understanding the interaction in six-dimensional $(4,0)$ and $(3,1)$ theories, while avoiding the complications associated with the $C$ and $D$ fields.

All  vectors of the five-dimensional maximal supergravity are in the \textbf{27} representation of $E_{6(6)}$.  The interaction \eqref{cubicinvariant} is possible due to the fact that there is a $E_{6(6)}$ singlet in the cubic tensor product of the fundamentals $\textbf{27} \otimes \textbf{27} \otimes \textbf{27} = \textbf{1} \oplus \cdots$. There is a more refined structure: under $E_{6(6)} \, \longrightarrow \, \SL(6,\bbR) \times \SL(2,\bbR)$,  we have $\textbf{27} \ra  (\textbf{15}, \textbf{1}) + (\textbf{6},\textbf{2})$ and the only allowed trilinear couplings involve either three fields in $\textbf{15}$ of $\SL(6,\bbR)$ that are $\SL(2,\bbR)$ singlets or a single vector field in  $\textbf{15}$ and a doublet of $\SL(2,\bbR)$  in  $\textbf{6}$ of $\SL(6,\bbR)$. This structure is perfectly consistent with eleven-dimensional origin of the Chern-Simons interactions, and arises in the reduction of the six-dimensional $(2,2)$ supergravity on a circle. The $\textbf{15}-\textbf{15}-\textbf{15}$ interaction can be seen directly from the $T^6$ reduction of eleven-dimensional Chern-Simons terms. The doublet of $\textbf{6}$ corresponds to the metric and the three-form field having one leg along the torus.  Note that even if the Chern-Simons interactions do not involve five-dimensional gravitons, $6$ of the $27$ vector fields have eleven-dimensional gravitational origin. 

In theories with 16 and 8 supercharges, the intimate connections between the six-dimensional anomalies and five-dimensional Chern-Simons couplings has been studied, and it is expected that only the anomaly-free theories yield gauge invariant Chern-Simons interactions upon circle reduction~\cite{Corvilain:2017luj, Corvilain:2020tfb}. In the maximally supersymmetric case, the refined structure of the Chern-Simons couplings makes their compatibility with a non-vanishing gravitational index in the $(4,0)$ or  $(3,1)$ multiplets very unlikely.


It is instructive to review the five-dimensional Chern-Simons terms in theories with 8 supercharges \cite{ FMS, Bonetti1, Bonetti2, Ohmori:2014kda} and their six-dimensional $\mathcal{N}= (1,0)$ origin (the case with 16 supercharges and six-dimensional $\mathcal{N}= (2,0)$ is very similar). There are two ways of generating these upon the circle reduction. The first involves either simple dimensional reduction of existing six-dimensional Chern-Simons terms, or field redefinitions involving the graviphoton field $A^0$ coming form the six-dimensional metric 
\begin{equation} \label{KKansatz}
d s^2_6 = d s^2_5 + g_{55}(dx^5 + A^0_{\mu} dx^{\mu})^2
\end{equation}
where $\mu=0, 1, ..., 4$. In the reduction of the eleven-dimensional supergravity to five dimensions, the entire \eqref{cubicinvariant} can be generated in this fashion.
The second mechanism involves integrating out at one loop the massive spin $1/2$, $3/2$ and two-form, i.e. potentially anomalous, fields coupled to $A^0$ or six-dimensional vector fields.

A generic six-dimensional $(1,0)$ theory has $n_T$ tensor multiplets with an anti-selfdual three-form in each, and a self-dual three-form in the gravity multiplet, leading to an $O(1, n_T)$ symmetry, and gauge multiplets with a gauge group of dimension $n_V$. The six-dimensional interactions lead via reduction to the following triple interactions
\begin{equation}
\label{eq:tensorCS}
A^0 \wedge F^{\alpha} \wedge F^{\beta} \eta_{\alpha \beta} + k_{\alpha i j} A^{\alpha} \wedge F^i \wedge F^j 
\end{equation}
with $\alpha, \beta$ being  $O(1,n_T)$ index and $i$ running over the Cartan subalgebra of the six-dimensional gauge group.

The $O(1,n_T)$ symmetry does not allow generation of any terms cubic in $A^{\alpha}$ \cite{FMS}, but couplings 
\begin{equation} \label{graviphotonCS}
k_0 A^0 \wedge F^0 \wedge F^0 + k_{0 ij}  A^0 \wedge F^i \wedge F^j + k_{ijk} A^i \wedge F^j \wedge F^k
\end{equation}
are allowed, and are in fact  a part of the five-dimensional low energy effective action arising after integrating out the massive fields. For example, the first term in \eqref{graviphotonCS}, can be traced to a triangle diagram with three external legs being graviphotons with some massive fields running in the loop. 

By taking the ansatz \eqref{KKansatz}, all six-dimensional fields that are coupled minimally to graviton will provide massive fields in five dimensions that couple minimally to the graviphoton with charges given by the corresponding Kaluza-Klein level. We list the minimal five-dimensional  coupling between the $U(1)$ vector fields and massive spin $1/2$ and spin $3/2$ fermions and complex two-forms:
\begin{equation} \label{minimalcoupling}
\begin{split}
& i q \bar{\psi} \gamma^{\mu} A_{\mu} \psi\\
& i q \bar{\psi}_{\rho} \gamma^{\rho\mu\nu} A_{\mu} \psi_{\nu}\\
& \pm \frac{1}{4} i q \epsilon^{\mu\nu\rho\sigma\tau} \bar{B}_{\mu\nu} A_{\rho}B_{\sigma\tau},
\end{split}
\end{equation}
where the sign in the last line is correlated with the six-dimensional chirality of the $B$-field. We have followed the conventions of \cite{ Bonetti1, Bonetti2}.
A lengthy one-loop computation indeed leads to the appearance of the cubic interactions of the form \eqref{graviphotonCS}. 

The five-dimensional theory also has non-minimal couplings 
\begin{equation}
\begin{split}
& \frac{1}{2} i \tilde{q}_{1/2} F^{\mu\nu} \bar{\psi} \gamma^{\mu\nu} \psi\\
& \frac{1}{2} i \tilde{q}_{3/2} F^{\mu\nu} \bar{\psi}_{\rho} \gamma^{\mu\nu\rho\sigma} \psi_{\sigma} + \frac{1}{2} i \tilde{q}'_{3/2} F^{\mu\nu} \bar{\psi}_{\mu}  \psi_{\nu}\\
& \tilde{q}_{B} \bar{B}_{\mu\nu} F^{\nu\rho} {B_{\rho}}^{\mu} +\tilde{q}'_{B} \bar{B}_{\mu\nu} F^{\nu\rho} B_{\rho\sigma} F^{\sigma\mu}. \\
\end{split}
\end{equation}
However, as shown in \cite{Bonetti1} these can be used to cancel divergences in relevant diagrams and do not affect the Chern-Simons couplings.

The five-dimensional Chern-Simons interactions \eqref{eq:tensorCS} and \eqref{graviphotonCS} do not contain any scalars and are gauge invariant by virtue of six-dimensional anomaly cancellation \cite{Corvilain:2017luj, Corvilain:2020tfb}.  We shall not establish any direct relation between the non-vanishing index for the $(4,0)$ and $(3,1)$ multiplets and the impossibility of recovering the gauge invariant Chern-Simons couplings of the maximal five-dimensional supergravity. Instead we shall show that there are no diffeomorphism invariant couplings compatible with the structure of these multiplets that can be reduced on the circle or give rise to interactions like \eqref{minimalcoupling} that are needed in order to generate the five-dimensional Chern-Simons terms.


\subsection{Testing the (4,0) multiplet}

We should recall that the six-dimensional multiplets do not contain gravity, and while the five-dimensional Planck length is given by the radius of the compactification circle, the reduction procedure is by no means the conventional Kaluza-Klein. The most notable difference is the absence of the ``graviphoton", i.e. the KK vector that usually arises from the reduction of the metric.

The 27 chiral two forms $B_{MN}^{\alpha}$ in the $(4,0)$ multiplet are in the \textbf{27} of $E_{6(6)}$ \cite{Hull1}. Due to self-duality each six-dimensional $B_{MN}$ yields a five-dimensional vector $A_{\mu}$ and there are no KK vectors arising in the reduction of other fields in the $(4,0)$ multiplet.
In other words the 27 five-dimensional vectors $A_{\mu}^{\Lambda}$ in  five-dimensional $\mathcal{N}=8$ supergravity  all originate form the six-dimensional tensor fields. 

In order to explore the possibility of the coupling \eqref{cubicinvariant} governed by the $E_{6(6)}$ cubic invariant being generated via loop integration of the massive states with three external five-dimensional vector fields,  the $E_{6(6)}$ invariant three-vertices involving six-dimensional $B_{MN}$-fields have to be examined.

The first immediate observation is that these tests do not involve the SD Weyl field. Indeed, in order to get a contribution from  the exotic graviton $C_{MNPQ}$ running in the loop,  $\textbf{1} \otimes \textbf{27} \otimes \textbf{1}$ needs to contain an $E_{6(6)}$ singlet, which is clearly not possible.

Turning to the fermions we start from the chiral spin $1/2$ fields in the \textbf{48} of $E_{6(6)}$. The  minimal five-dimensional coupling is of the form
\begin{equation}
 i q c_{\Lambda i j} \bar{\psi}^i \gamma^{\mu} A_{\mu}^{\Lambda} \psi^j,
\end{equation}
where  $\Lambda$ is in $\textbf{27}$, $i, j$ are \textbf{48} indices and $c_{\Lambda i j}$ is a constant. Such a tri-vertex is allowed since there is a singlet contained in $\textbf{48} \otimes \textbf{27} \otimes \textbf{48}$.  However, in order to lift this coupling to six dimensions we must complete the term
\begin{equation}
i q c_{\Lambda i j} \bar{\psi}^i \gamma^{\mu} B_{\mu5}^{\Lambda} \psi^j
\end{equation}
to a Lorentz scalar. The easiest way is to put a derivative on $B$ and thus yielding
\begin{equation}
\label{eq:psi2B}
i q c_{\Lambda i j} \bar{\psi}^i \Gamma^{M} \partial^N B_{MN}^{\Lambda} \psi^j. 
\end{equation}
However $\partial^N B_{MN}^{\Lambda}=0$ serves like the Lorenz gauge just as in the case for Abelian vector field, and \eqref{eq:psi2B} vanishes.
Another option is to increase the rank of the gamma matrix sandwiched by the fermions
\begin{equation}
i q c_{\alpha i j} \bar{\psi}^i \Gamma^{MN}  B_{MN}^{\Lambda} \psi^j. 
\end{equation}
This could give rise to the wanted minimal coupling when the index $N=5$, but a chiral fermion bilinear in six dimensions with two fermions of the same chirality does not contain any two forms. Hence, the above expression is identically zero. Further possible  six-dimensional couplings  dimensions lead to non-minimal couplings in five dimensions, which as already explained do not  give quantum contributions to the one-loop Chern-Simons terms.

The exotic gravitino $\psi^a_{MN}$ is in the \textbf{8} of $E_{6(6)}$,  and the trilinear coupling with the vector $\textbf{8} \otimes \textbf{27} \otimes \textbf{8}$ contains a singlet. The gravitino-vector coupling as listed in \eqref{minimalcoupling}
\begin{equation} \label{exoticgravitinocoupling}
i q \tilde{k}'_{\Lambda ab}\bar{\psi}^a_{\rho} \gamma^{\rho\mu\nu} A^{\Lambda}_{\mu} \psi^b_{\nu}
\end{equation}
 is lifted to
\begin{equation}
i q \tilde{k}'_{\Lambda ab}\bar{\psi}^a_{\rho5} \Gamma^{\rho\mu\nu} B^{\Lambda}_{\mu5} \psi^b_{\nu5} 
\end{equation}
There are three six-dimensional candidates that do not have any derivatives acting on $B_{MN}$ or $\psi_{MN}$
\begin{equation}
\begin{split}
& i q \tilde{k}'_{\Lambda ab} \bar{\psi}^a_{PA}  \Gamma^{PMNABC}  \psi^b_{NB}  B^{\Lambda}_{MC}, \quad i q \tilde{k}'_{\Lambda ab}\bar{\psi}^a_{PA}  \Gamma^{PMNB}  {\psi^b_{N}}^A  B^{\Lambda}_{MB},  \\ 
 &i q \tilde{k}'_{\Lambda ab}\bar{\psi}^a_{PA}  \Gamma^{PMNB}  \psi^b_{NB}  {B^{\Lambda}_{M}}^A.
 \end{split}
\end{equation}
The first one is allowed by $SO(5,1)$ representation, but to achieve \eqref{exoticgravitinocoupling} we would have to set $A=B=C=5$ so the gamma matrix vanishes by antisymmetry. The other two vanish due to the tensor product decomposition of the exotic gravitini.

The trilinear coupling of vectors with the massive two-forms $B$ also need to be considered. Such couplings for the reduction of $(1,0)$ theory in \eqref{minimalcoupling} contain the graviphoton and originate from self-duality of the six-dimensional tensor fields \cite{Bonetti3}. This is no longer the case, and one should be looking for a  six dimensional  cubic invariant built solely from the bare potentials $B_{MN}^{\alpha}$   
\begin{equation}\label{eq:6dB}
k_{\Lambda \Delta \Gamma} \ B^{\Lambda} \wedge B^{\Delta} \wedge B^{\Gamma} \Longrightarrow k_{\Lambda \Delta\Gamma} \epsilon^{MNPQRS} \ B_{MN}^{\Lambda} B_{PQ}^{\Delta} B_{RS}^{\Gamma} \,,
\end{equation}
the reduction of which would contain a minimal term proportional to 
\begin{equation}\label{eq:minB}
+ i q k_{\Lambda \Delta \Gamma} \epsilon^{\mu\nu\rho\sigma\tau} \bar{B}^{\Lambda}_{\mu\nu} A^{\Delta}_{\rho}B^{\Gamma}_{\sigma\tau}.
\end{equation}
This product contains an $E_{6(6)}$ singlet and  hence is allowed. As discussed,  under $E_{6(6)} \, \longrightarrow \, \SL(6,\bbR) \times \SL(2,\bbR)$,  $\textbf{27} \ra  (\textbf{15}, \textbf{1}) + (\textbf{6},\textbf{2})$ 
and the trilinear couplings are either between three $\textbf{15}$ or between one $\textbf{15}$ and two different $\textbf{6}$. This means that any possible contribution to \eqref{cubicinvariant}  from \eqref{eq:minB} should have a massive two form in $\textbf{15}$ in the loop. 
From other side as shown in  \cite{BOS}, the only two-forms allowed to enter the five-dimensional action are in one of the  $\textbf{6}$ representations. Hence contributions from the massive two-forms to \eqref{cubicinvariant} seem to be ruled out by supersymmetry. At any rate it would be very hard to imagine a  gauge invariant completion of  \eqref{eq:6dB}.\footnote{Note that in \cite{Bonetti1, Bonetti2} the five-dimensional couplings generated using the vertex with a massive two-form in the loop and an external vector field involve the graviphoton. As mentioned, in the reduction of the $(4,0)$ theory this field does not even arise.}

This seems to exhaust the possibilities for generating the five-dimensional Chern-Simons couplings using the five-dimensional massive modes  coupled to the fields of the five-dimensional maximal supergravity in a way that can be lifted to six-dimensional Lorentz and gauge invariant interactions.

Finally, one may entertain the possibility  of a coupling like
\begin{equation} \label{cubicinvariant2}
S_{\text{CSE}} = \int k_{\Lambda \Delta \Gamma} \ B_{MN}^{\Lambda} \wedge H_{PQV}^{\Delta} \wedge H_{RSW}^{\Gamma} \eta^{VW} \epsilon^{MNPQRS} \, .
\end{equation}
and its direct reduction to five-dimensions.
Clearly this coupling is not gauge invariant in six dimensions. But that is not the only problem - upon reduction only the part involving $\eta^{55}$ gives a sensible and  gauge invariant five dimensional coupling. On the other hand, five-dimensional interactions cannot contain $\eta^{\mu \nu}$. Hence constraints need to be imposed on \eqref{cubicinvariant2} in order to eliminate the unwanted  parts. This would come at the expense of the Lorentz invariance, and we do not consider this possibility here. Running slightly ahead, we remark that the possibility of even such - however dubious - 
cures is not available for  the $(3,1)$ multiplet.


\subsection{Testing the (3,1) multiplet}
The (3,1) multiplet, written in the representation of $\mathfrak{su}(2)\times\mathfrak{su}(2)\times\mathfrak{sp}(6)\times\mathfrak{sp}(2)$ is
\begin{equation}
\begin{split}
&{\bf (4,2;1,1)+(2,2;14,1)+(3,1;6,2)+(1,1;14',2) }\\
&{\bf +(4,1;1,2)+(3,2;6,1)+(2,1;14,2)+(1,2;14',1) }.
\end{split}
\end{equation}
It follows \cite{Hull1} that, the $\bf (4,2;1,1)$ field $D_{MNP}$ gives a linearised metric $h_{\mu\nu}$ and a vector $A^0_{\mu}=D_{\mu55}$, which we denote with a superscript $0$ to distinguish from other vectors.

There are also five-dimensional vectors $A^i_{\mu}$ given by $\bf (2,2;14,1)$ and those $A^{\alpha}_{\mu}$ from the chiral two-form $\bf (3,1;6,2)$. With respect to the chain of groups
\begin{equation}
\Symp(6)\times \Symp(2) \subset F_{4(4)} \subset E_{6(6)},
\end{equation}
the $\textbf{27}$ of $E_{6(6)}$ has a decomposition under $\Symp(6)\times \Symp(2)$
\begin{equation}
{\bf 27 = (1,1) +(14,1) +(6,2)}.
\end{equation}
To build a six-dimensional vertex with $(3,1)$ field content,  a $\Symp(6)\times \Symp(2)$ singlet needs to be constructed. Recalling the structure of the $E_{6(6)}$ cubic invariant  \eqref{cubicinvariant}, it is not hard to see that 
\begin{equation}
\begin{split}
&(\alpha, \beta, i) \quad (\alpha, \beta, 0)\\
&(0, 0, 0) \quad (0, 0, i) \quad (0, i, j) \quad (i, j, k)
\end{split}
\end{equation}
trilinear couplings need to be generated. Note that this structure is rather different from that of the trilinear couplings arising in the reduction of $(1,0)$ theory which has e.g. $(\alpha, i, j)$ couplings obtainable by direct reduction, and does not have  $(\alpha, \beta, i)$ couplings. It can be shown, using arguments from the previous subsection, that it is not possible either to directly lift this structure to Lorentz and gauge invariant couplings of $(3,1)$ multiplet, or to generate them by integrating out massive modes in the loop.



\section{Evidence for h-theories}
\label{sec:h-theory}

In this section we shall discuss how trying to solve the equations of motion for the SD Weyl field $C_{MNPQ}$ may suggest an alternative way of thinking about some of the six-dimensional exotic multiplets. The basic construction works for theories with 32, 16 or 8 supercharges, but for the latter two a number of $(2,0)$ tensor and $(1,0)$ vector multiplets respectively need to be added. Bellow, we shall mostly discuss the maximally supersymmetric case of $(4,0)$ exotic multiplet.

The SD Weyl field has 5 physical degrees of freedom. While this is the same number of degrees of freedom as that of five-dimensional metric, we argued (however indirectly) that the dynamics of this field when reduced on a circle is unlikely to be the same as that of gravity. From other side, this number also matches  the number of the parameters of the  $\SL(3,\mathbb{R})/SO(3)$ coset, i.e. a three-torus of of fixed volume. Moreover, as discussed in section \ref{sec:E8} this coset is closely related to the SD Weyl field, both in terms of the degrees of freedom of the field itself and its gauge transformation parameters. So one may wonder if the system  of five scalars parametrising the coset coupled to thee-dimensional gravity, which carries no dynamical degrees of freedom, may be related to the solutions of the equations of motion for the SD Weyl field. This system is familiar, and it has been shown in~\cite{Liu:1997mb} that its solutions can be summarised by a Ricci-flatness condition of a semi-classical metric on a six-dimensional space $X$ obtained as a $T^3$ fibration over the three-dimensional base.\footnote{We shall work with a Euclideanised version of the three-dimensional theory.} 

As we shall review shortly, the Ricci-flatness condition is equivalent to a  real two-form $k$ on $X$, constructed from the coset element of $\SL(3,\mathbb{R})/SO(3)$, being covariantly constant. One can think of $k$ as the K\"ahler form on $X$, but in the context of supersymmetric theories, one cannot establish a duality between any six-dimensional supergravity (with 32, 16 or 8 supercharges) on $X$ and solutions of the above three-dimensional system, preserving a quarter of supersymmetry. On the contrary, the latter are consistent with the $(4,0)$ and exotic $(2,0)$ and $(1,0)$ supersymmetry respectively, and $k$ can be squared to a SD Weyl field satisfying its flatness condition. It can be shown that at the linearised level, the differential conditions on $k$ reduce to the equations of motion for the SD Weyl field.

In order to see this we just need to examine the duality groups of these theories. Let us start from the 32 supercharge case. The three-dimensional theory has $E_{8(8)}$ symmetry and it's scalar manifold is the coset space $E_{8(8)}/SO(16)$. All but five of these scalars are set to zero in the solution, leaving the $E_{6(6)}$ symmetry, which is stabilised by $\SL(3,\bbR)$ inside $E_{8(8)}$ intact. So when geometrising the $\SL(3,\bbR)$ symmetry and thinking of the solutions of the three-dimensional system of gravity and five scalars in terms of solutions of some six-dimensional theory on $X$, one expects the latter to have manifest $E_{6(6)}$ symmetry. This is not the case for the maximal six-dimensional supergravity, but it is for the $(4,0)$ SD Weyl multiplet.  As it is clear from the details of the construction in subsection \ref{ssec:coset}, on three-dimensional bases one can construct at most $T^3$ and hence geometrise  only $\SL(3,\bbR)$ this way. In many ways the construction is reminiscent of geometrisation of $\SL(2,\bbR)$ in type IIB and F-theory.  Moreover since the construction involves two groups of intersecting co-dimension two defects, the $\SL(3,\bbR)$ arises as the group generated by two $\SL(2,\bbR)$ subgroups. This is reminiscent of the two pairs of charges in~\eqref{eq:Hull-section} which have accompanying $\SL(2,\bbR)$ actions, even though the full triplet~\eqref{eq:Hull-section} has no $\SL(3,\bbR)$, as discussed in sub-section~\ref{sec:Hd-charges}. However, to make closer comparison with that discussion, it may be better to consider the points made in section~\ref{sec:SO3-triplet} and look at the two $\SO(2)$ groups generating an $\SO(3)$ under which the momenta transform as a triplet.

Similarly in the case with 16 supercharges, embedding $SL(3,\bbR)$ in the duality group $SO(8,8+n)$ leaves invariant a $SO(5,5+n)\times\mathbb{R}^+$ suggesting that this is the symmetry on the resulting six-dimensional theory. While all $(2,0)$ multiplets have the same R-symmetry,  this symmetry  is a bit bigger than that admitted by the $(2,0)$ gravity plus $n+5$ tensor multiplets, and the $\mathbb{R}^+$ factor accounts for the extra 
scalar in the $(2,0)$ SD Weyl multiplet.

Theories with 8 supercharges are a bit harder to analyse, notably because there are many options for the scalar manifolds available. Yet the most ``typical" quaternionic coset is given by $SO(4,4+n)/SO(4)\otimes SO(4+n)$. Under $\SL(3,\bbR)$ embedding one gets a coset space $SO(1,1+n)/SO(1+n)$ which can describe the moduli space of  $n_T = 1+n$ six-dimensional $(1,0)$ tensor multiplets coupled either to $(1,0)$ gravity multiplet or to $(1,0)$ SD Weyl multiplet. Note however that the latter case has fewer degrees of freedom (a  $(1,0)$ gravity multiplet  is ``worth" a $(1,0)$ SD Weyl multiplet + a tensor multiplet).

In this section, we will discuss the equations of motion of the SD Weyl field and their $T^3$ reduction, looking to capture the solutions of three-dimensional gravity with varying scalars in terms of a six-dimensional geometric construction involving the SD Weyl field. 
In particular, we will construct a $T^3$ fibered manifold together with a tensor field $\cC_{MNPQ}$ using five scalar fields with dependence only on the three-dimensional base. On imposing that the five scalars solve the supersymmetry conditions of three-dimensional supergravity, this field will solve an equation 
\begin{equation}
\cG^{Q}_{\, \, NPQRS}= 0.
\hs{30pt}
\cG_{MNPQRS} = \LC_{[M} \cC_{NP][QR;S]}
\label{eq:SDW-FS2}
\end{equation}
However, the field $\cC_{MN,PQ}$ cannot directly be interpreted as the SD Weyl field on the curved space. To identify the SD Weyl field of~\cite{Hull1} we should linearise the system by thinking of the three-dimensional scalar fields underlying the construction as small fluctuations. The corresponding geometry will then  be seen as a small fluctuation of a flat manifold $\bbR^3 \times T^3$, with the five scalars determining  the metric on the fibres. As we shall see shortly, in the expansion of $\cC$ in the powers of the scalar fields
\begin{equation}
\cC_{MNPQ} = \cC^{(0)}_{MNPQ} + C_{MNPQ} + \dots
\end{equation}
the linear fluctuations of the SD Weyl field of~\cite{Hull1} will be identified with the the first order term, denoted $C_{MNPQ}$. To the first order in fluctuations, \eqref{eq:SDW-FS2} reduces to the standard equation for the SD Weyl field of~\cite{Hull1}
\begin{equation}
G^{Q}{}_{ NPQRS}= 0.
\hs{30pt}
G_{MNPQRS} = \der_{[M} C_{NP][QR,S]}
\label{eq:SDW-FS3}
\end{equation}
provided that  $C_{MNPQ}$ is taken to have no dependence on the $T^3$ directions of the geometry. 

Note that geometric fluctuations around the flat geometry $\bbR^3 \times T^3$ would only affect the non-linear parts in the expansion of \eqref{eq:SDW-FS2} without spoiling the agreement of it with the linearised equation \eqref{eq:SDW-FS3}.
Thus one could view equation~\eqref{eq:SDW-FS2} as a non-linear extension of the SD Weyl field equation of motion. Here we only check that the two agree at the linearised level.
Another feature of this construction which mirrors comments made in section~\ref{sec:E8} is that the $T^3$ fibered geometry uses the physical degrees of freedom in its definition, and thus the six-dimensional space requires the physical fields for its definition.

This reformulation of three-dimensional theories in terms of the (diffeomophism non-invariant) six-dimensional one on (non-compact) manifolds with certain geometric properties, suggests the interpretation of the latter as lower-dimensional cousins of F-theory.


\subsection{SD Weyl field on $\bbR^3 \times T^3$}
\label{sec:SDW-T3}

Since to the linear order in scalar fields, the equation of motion for the SD Weyl field $C_{MNPQ}$ is not sensitive to the metric fluctuations, in this section we will examine the reduction of the equations on a flat $\bbR^3 \times T^3$. 

Following~\cite{Hull1, Hull2}, we define the SD Weyl field strength in flat space as
\begin{equation}
G_{MNPQRS} = \der_{[M}C_{NP][QR,S]}
\label{eq:SDW-FS}
\end{equation}
subject to self-duality constraint (we are working in Euclidean signature, hence the factor of $i$):
\begin{equation}
G_{MNPQRS} = \frac{i}{3!} \epsilon_{MNPTUV} G^{TUV}{}_{QRS} = \frac{i}{3!} \epsilon_{QRSTUV} G_{MNP}{}^{TUV}
\end{equation}
with $M,N,P... = 1, ..., 6$. 
The equation of motion for $C$ in 6 dimensions is then given by
\begin{equation}
G^{Q}{}_{ NPQRS}= 0.
\label{eq:Ceom}
\end{equation} 

We shall now assume that the SD Weyl field $C$ depends only on three of the coordinates. We  can separate the the coordinates into the $\bbR^3$ part $x^{\alpha}$ with $\alpha = 1,2,3$ and $T^3$ part $\xi^i$ with $i= 1, 2, 3$, and allow the dependance only on the $x^\alpha$-coordinates of $\bbR^3$, i.e. take
\begin{equation}
\partial_i C_{ABCD}= 0,
\end{equation}
This, together with the self-dual condition of the field strength of $C$ eliminates the some of the components
\begin{equation}
\begin{split}
G_{ijkABC} &= \partial_{[i} C_{j k] [AB;C]} = 0 \\
G_{\alpha \beta \gamma ABC} &= \epsilon_{\alpha \beta \gamma  i j k} G^{i j k}{}_{ABC} = 0
\end{split}
\end{equation}
The non-vanishing components field strength of $C$ in the product ansatz are of the type
\begin{equation}
G_{\alpha i j ABC} \quad \text{or} \quad G_{\alpha \beta i ABC}, 
\end{equation}
where because of the exchanging symmetry $C_{ABCD}=C_{CDAB}$ we only need to focus on the first 3 indices of the field strength $G$.
Since these two types of components are related again by the self-duality of $G$
\begin{equation}
G_{\alpha i j ABC} = \frac{i}{3!}\epsilon_{\alpha i j \beta \gamma k} G_{\beta \gamma k ABC} \quad \text{or} \quad 
G_{\beta \gamma k ABC} = \frac{i}{3!}\epsilon_{\beta \gamma k \alpha i j} G_{\alpha i j ABC}. 
\end{equation}
it is sufficient to consider the components $G_{\alpha\beta i \gamma\delta j}$ arising from the potentials $C_{\mu i \nu j}$. In each pair of indices on $C$ there is one $\bbR^3$ coordinate index and one $T^3$ coordinate index. The equation of motion for $C$ in 6 dimensions \eqref{eq:Ceom}  reduces to 
\begin{equation}
\delta^{\gamma\delta} \partial_{[\gamma} C_{\alpha i] [\beta j,  \delta]} = 0
\label{eq:Ceom*}
\end{equation}
This is a set of (linear) three-dimensional equations for five degrees of freedom contained in the SD Weyl field. As already mentioned the $\SL(3,\mathbb{R})/SO(3)$ coset has the same number of degrees of freedom. 
In what follows, we will construct a $T^3$ fibred geometry together with a tensor field $\cC_{MNPQ}$ satisfying similar equations to those for $C_{MNPQ}$ above (but on the curved geometry), such that the linearisation of the total system reduces to~\eqref{eq:Ceom*}. The fact that the geometry is given by $\bbR^3\times T^3$ only at zeroth order in fluctuations does not affect the linearised equation~\eqref{eq:Ceom*}.


\subsection{The $\SL(3,\mathbb{R})/SO(3)$ sigma-model and the SD Weyl field}
\label{ssec:coset}
The symmetric space $\SL(3,\bbR) / \SO(3)$ has dimension five. 
Thus, we need 5 real scalars to parametrize the non-linear sigma-model with target $SL(3,\mathbb{R})/SO(3)$. Its vielbein $V_{ai}$  in Borel gauge can be written as follows
\begin{equation} 
V = e^{\Phi_1/\sqrt{3}}
\begin{pmatrix}
1 & a & b \\
0 & e^{-(\sqrt{3}\Phi_1-\Phi_2)/2} & ce^{-(\sqrt{3}\Phi_1-\Phi_2)/2} \\
0 & 0 & e^{-(\sqrt{3}\Phi_1+\Phi_2)/2} 
\end{pmatrix} , 
\label{eq:coset}
\end{equation}
where $a$ is a $SO(3)$ index, while $i$ is a $\SL(3,\mathbb{R})$ index. The two dilatonic  scalars $\Phi_1$ and $\Phi_2$ correspond to the two Cartan generators of $\mathfrak{sl}(3,\mathbb{R})$ and the three other scalars $a$, $b$ and $c$ are nilpotent generators which complete the coset.

The Mauer-Cartan form $dVV^{-1}$ can be split into the part symmetric in $SO(3)$ indices, $P^{ab}=P^{(ab)}$,  and the anti-symmetric part $Q^{ab}=Q^{[ab]}$
\begin{equation}
\label{eq:VdV-1}
(\partial_{\alpha}V V^{-1})^{ab} =  P_{\alpha}^{ab}+Q_{\alpha}^{ab}.
\end{equation}
Here the partial derivative $\partial_{\alpha}$ is taken with respect to the 3 dimensional space on which we put the $\SL(3,\mathbb{R})/SO(3)$ sigma-model. (We think of this as the base space in what follows.) 
The involution $\sigma$ under which the $\mathfrak{so}(3)$ subalgebra is invariant corresponds to taking minus the matrix transpose, and thus~\eqref{eq:VdV-1} splits $\partial_{\alpha}V V^{-1}$ into its $\sigma$-eigenvector parts. 
The symmetric part $P_{\alpha}$ transforms covariantly under the action of base-coordinates dependent $SO(3)$ elements while $Q_{\alpha}$ transforms like a connection.

The action of this sigma-model coupled to three-dimensional gravity is given by
\begin{equation}
S = \int d^3 x \ \frac{1}{2 \kappa^2}\sqrt{-g} (R-g^{\alpha\beta} \Tr  P_{\alpha}P_{\beta}),
\end{equation}
with the field equations
\begin{equation}
\begin{split}
\mathcal{D}_{\alpha} P^{\alpha} &=g^{\alpha\beta} (\nabla_{\alpha}  P_{\beta} + [Q_{\alpha}, P_{\beta}] )\\
R_{\alpha\beta} &= \Tr P_{\alpha}P_{\beta}.
\end{split}
\end{equation}
The in terms of the scalar fields \eqref{eq:coset}, the Lagrangian can be brought into a simple form
\begin{equation}
\begin{split}
\mathcal{L} = \frac{1}{2 \kappa^2}\sqrt{-g} (R &- \frac{1}{2}(\partial \Phi_1)^2- \frac{1}{2}(\partial \Phi_2)^2 -\\
&-\frac{1}{2} e^{{\sqrt{3}} \Phi_1 -  \Phi_2} (\partial a)^2 -\frac{1}{2} e^{2\Phi_2}(\partial c)^2 -\frac{1}{2} e^{{\sqrt{3}} \Phi_1+ \Phi_2}(\partial b - c \partial a)^2).
\end{split}
\label{eq:Sl/SO}
\end{equation}
The Lagrangian \eqref{eq:Sl/SO} can be embedded into three-dimensional supersymmetric theories with the varying amounts of the supersymmetry. The relevant part of the supersymmetry transformations for the spin $3/2$ and $1/2$ fermions\footnote{Note that the spin $1/2$ fields transform under the maximal compact subgroup of the symmetry group. Here we restricted to the relevant $SO(3)$ subgroup.} is given by
\begin{equation}
\begin{split}
\delta \psi_{\alpha} & = \mathcal{D}_{\alpha} \epsilon = (\nabla_{\alpha}  + \frac14 Q^{ab}_{\alpha} T^{ab})  \epsilon \\
\delta \chi^a &= - \frac12 P^{ab}_{\alpha} T^a \epsilon
\end{split}
\end{equation}
where $T^{ab}$ and $T^a$ are the $SO(3)$ generators in the adjoint and spin representation respectively.


Solutions with varying moduli consistent with $\SL(3, \mathbb{Z})$ were constructed in  \cite{Liu:1997mb}. The solution takes the form of overlapping codimension two objects. One can start by solving for each such object, which will be picking a specific $SO(2)$ inside the $SO(3)$ automorphism group. A $\frac12$-BPS projector for a brane with transverse $x^{\bar a_1} - x^{\bar a_2}$  plane can be written as\footnote{The barred indices ${\bar a_i} = 1,2,3$ refer to tangent space}  
\begin{equation}
P = \frac12 \left(1+ \gamma^{ \bar a_1\bar a_2} \Lambda_{a_1 a} \Lambda_{a_2 b} T^{ab} \right)
\label{eq:project}
\end{equation}
where $\Lambda_{ab}(x)$ is an $SO(3)$ rotation matrix. Solving the BPS conditions for a single codimension-two object yields a solution very much like the standard seven-branes in ten dimensions. On a three-dimensional base there is room for two groups of intersecting objects with a net quarter of supersymmetry preserved.\footnote{It is not hard to verify that there are only two independent projectors of the type \eqref{eq:project} on a three-dimensional space.} Two groups of such overlapping objects will now fill out the entire $SO(3)$, and the solution geometrically realises a $T^3$ fibration over the three-dimensional base space.

Using the $\SL(3,\bbR)$-invariant form of metric on $T^3$ one can summarise the solution using a six-dimensional metric of the form 
\begin{equation}
ds_6^2 = ds_{\text{base}}^2 + (V^T V)_{ij}(x) d\xi^i d\xi^j \, 
\end{equation}
with the metric on the three-dimensional base space taken as
\begin{equation}
ds_{\text{base}}^2 = e^{2\phi_1(x)} dx_1^2 + e^{2\phi_2(x)} dx_2^2+e^{2\phi_3(x)} dx_3^2.
\end{equation}

As shown in  \cite{Liu:1997mb},  solution of the killing spinor equations is equivalent to the following two-form on the six-dimensional space
\begin{equation}
k_{MN} = 
\begin{pmatrix}
0 & -e^{-\phi_{\alpha}} \delta^{\alpha}_{a} V_{ai}\\
e^{\phi_{\beta}} V^{T}_{ib}\delta^{\beta}_{b}& 0 \\
\end{pmatrix} 
\end{equation}
being covariantly constant
\begin{equation}
\nabla_{M}k_{NP}= 0.
\label{eq:cck}
\end{equation}

One could think of $k$ as the fundamental form on the resulting six-dimensional manifold. However, 
 as mentioned the solution to \eqref{eq:Sl/SO} cannot be lifted to a solution of six-dimensional supergravity on any six-manifold since its symmetry group is $SO(5,5)$ and not the $E_{6(6)}$ group that is stabilised by $\SL(3,\bbR)$ inside $E_{8(8)}$ for the case of maximal supersymmetry. The group stabilised by $\SL(3,\bbR)$ inside the three-dimensional duality group is compatible with the six-dimensional $(4,0)$ theory or less supersymmetric exotic theories.

In order to describe the six-dimensional lift in terms of the exotic graviton, one can build a four-index object with the properties of Riemann tensor: 
\begin{equation}
{\mathcal C}_{MNPQ} = k_{MN}k_{PQ} - k_{[MN}k_{PQ]},
\end{equation}
which has the non-trivial components ${\mathcal C}_{\alpha i \beta j}$. 
The algebraic symmetries of ${\mathcal C}$ are manifestly the same as in \eqref{exoticgravitonsymmetry}. By virtue of \eqref{eq:cck} ${\mathcal C}$  satisfies 
\begin{equation}
\label{eq:Ceom**}
\nabla_{\gamma} {\mathcal C}_{\alpha i \beta j; \gamma} = 0 .
\end{equation}
Notice that the three-dimensional covariant derivatives are used here. Consistently \eqref{eq:Ceom**} and with the self-duality properties the field strength of ${\mathcal C}$, ${\mathcal C}_{ijkl}$ components can be taken to zero.

In order to compare with   \eqref{eq:Ceom*}, we need to consider the linearisation of \eqref{eq:Ceom**}.  Using
\begin{equation}
k_{MN} = k_{MN}^{(0)} + k_{MN}^{(1)} + ... = 
\begin{pmatrix}
0 & - \delta^{\alpha}_{i}\\
\delta^{\alpha}_{i}& 0 \\
\end{pmatrix}  + k_{MN}^{(1)} + ...
\end{equation}
one can expand ${\mathcal C}$ in a similar fashion, with ${\mathcal C}^{(0)}_{MNPQ} =  k_{MN}^{(0)}k_{PQ}^{(0)} - k_{[MN}^{(0)}k_{PQ]}^{(0)}$ having only constant components.  It is  the linear term in the expansion of $\mathcal C$ that is taken to be equal to the  SD Weyl field  $C_{MNPQ} $ 
\begin{equation}
{\mathcal C}_{MNPQ} =  {\mathcal C}_{MNPQ}^{(0)} +  C_{MNPQ} + ...
\end{equation}
It can be checked then, that the linearised equations of motion for the five  three-dimensional scalar fields \eqref{eq:Sl/SO}  imply $\partial_{[\alpha} C_{\beta i] [\gamma j, \alpha]} = 0$.

In other words the three-dimensional gravity coupled to scalars in $\SL(3,\bbR)/SO(3)$ coset is solved at linearised level by $(4,0)$ SD Weyl supersymmetry on a $T^3$ fibered Ricci-flat manifold $M$. Note that on $M$, the covariantly constant tensor ${\mathcal C}$ is globally defined. This is not the case for the SD Weyl field $C$ which is obtained by picking the part linear in scalar fields in the expansion of ${\mathcal C}$.

From other side the conspiracy between $\SL(3, \mathbb{Z})$ and the duality groups in three and six dimensions makes this construction  unique. One could construct a $T^2$ fibered four-manifold in a similar fashion, but the group $E_{7(7)}$ group that is stabilised by $\SL(2,\bbR)$ inside $E_{8(8)}$ is too big for any five-dimensional theory. This has a well-known realisation in terms of codimension-two objects with a deficit angle. The $T^3$ fibered construction corresponds to two sets of intersecting codimension-two objects, each realising an $SO(2)$ within $SO(3)$. As mentioned, on a three-dimensional base there are only two independent such groups each  preserving half supersymmetry (any other half-supersymmetric projector can be built out of the above two). In agreement with this, no other $\SL(n,\bbR)$ group (for $n >3$) inside $E_{8(8)}$ stabilises any known duality group for an $(n+3)$-dimensional theory (since the stabiliser is $E_{9-n(9-n)}$ as can easily be seen from the extended Dynkin diagram). 

Here we have concentrated on the maximally supersymmetric theory in three dimensions and its lift to the six-dimensional $(4,0)$. From other side, there is very little dependance on the details of the multiplet or amount of supersymmetry, and as discussed above similar relation exists between three-dimensional theories with 16 and 8 supercharges, and $(2,0)$ and $(1,0)$ SD-Weil multiplets completed by matter multiplets.


\section{Discussion}
\label{sec:discussion}

We conclude by briefly mentioning some of the many aspects of the exotic supersymmetric multiplets that we have not addressed.

The algebraic structure of the exotic six-dimensional multiplets and the embedding into the exceptional geometry framework appears to be an interesting story, which we have only scratched the surface of here.  
It is clear that the six-dimensional momenta and spin group can be described in the algebraic framework, such that they agree with the supersymmetry algebra, but there is no spacetime section in the usual sense. This should not be a great surprise as these are not standard gravitational multiplets. However, the wider interpretation of the matching of momentum charges and section condition is subtle issue for the higher-rank exceptional groups which perhaps deserves further study in its own right. One could wonder whether the presence of the additional $\rep{248}$ constrained fields needed to accommodate the gauge algebra and tensor hierarchy in~\cite{HS-E88} could play a role in this. Naively, one would expect some modification to the usual generalised Lie derivative picture would be needed in order for the gauge algebra to close in the absence of a spacetime solving the section condition.

One could also wonder whether there is a similar story for the $D_{[\mu \nu] \lambda}$ exotic graviton of~\eqref{eq:exotic-D}. In the case of the $\cN=(3,1)$ theory, the decomposition of the adjoint of $E_{8(8)}$ under $\SL(3,\bbR)\times F_{4(4)} \subset G_{2(2)} \times F_{4(4)}$ is
\begin{equation}
\label{eq:E8-F4}
		\rep{248} 
		\ra \Big( \mathfrak{sl}(3,\bbR) \oplus \repp{3}{1} \oplus \repp{3'}{1} \Big)
		\oplus \mathfrak{f}_{4(4)} \oplus \repp{3}{26} \oplus \repp{3'}{26} \oplus \repp{1}{26}
\end{equation}
where the three terms in the bracket make up $\mathfrak{g}_{2(2)}$. Decomposing under $\SO(3)\times\Symp(6)\times\Symp(2)$ one can see that the non-compact generators of $\mathfrak{g}_{2(2)}$ match the $\rep{5}\oplus\rep{3}$ of $\SO(3)$ for the relevant exotic graviton, while the $\mathfrak{f}_{4(4)}$ term corresponds to the scalar coset. There is also a $\reppp{3'}{14}{1}$ for the vectors and $\reppp{3}{6}{2}$ for the self-dual two-forms. The final term is slightly harder to interpret, but the 14 non-compact generators could be matched to the three-form magnetic duals of the vectors. 

More generally, it appears that the special role played by $\SL(3,\bbR)$ for the exotic graviton for the $\cN=(4,0)$ multiplet could become $G_{2(2)}$ for the exotic graviton of the $\cN=(3,1)$ theory. For example, there is an $\cN=(1,0)$ supermultiplet (with $V=\reppp{3}{2}{1}$ in the notation of appendix~\ref{app:Ch6d}) with field content 
\begin{equation}
\label{eq:D-multiplet}
\begin{aligned}
		&\reppp{2}{2}{1} && \oplus  && \reppp{3}{2}{2} && \oplus & & \reppp{4}{2}{1} \\
		&A_\mu &&&& \psi^R_\mu &&&& D_{[\mu\nu]\lambda}
\end{aligned}
\end{equation}
This multiplet appears to match the decomposition of the group $\SO(4,3)$ by
\begin{equation}
\label{eq:SO43-G22}
		\mathfrak{so}(4,3)
		\ra \mathfrak{g}_{2(2)} \oplus \rep{7}
		\ra \Big( \mathfrak{sl}(3,\bbR) \oplus \repp{3}{1} \oplus \repp{3'}{1} \Big)
		\oplus \rep{3} \oplus \rep{3'} \oplus \rep{1}
\end{equation}
Again, the three terms in the bracket correspond to the field $D_{[\mu\nu]\lambda}$ while the $\rep{3}\oplus \rep{3'}$ of $\SL(3,\bbR)$ correspond to a vector field. Finally, the remaining non-compact singlet generator is the magnetic dual three-form to this vector. This pattern is repeated across other examples, with $\mathfrak{g}_{2(2)}$ playing the role of $D_{[\mu\nu]\lambda}$. 
It could thus be worth considering how the rest of our analysis would work out for these cases.\footnote{Many of the considerations of this paper could also be applied to exotic two-dimensional theories with fields built as product involving chiral bosons. We have not studied two-dimensional theories in this paper.}

The existence of the exotic six-dimensional multiplets could have been dismissed as a mere curiosity if not for the possible far reaching implications for gaining insights into strongly coupled gravitational theories \cite{Hull1, Hull2, Hull3}. Just like the gravity multiplet can be thought of the product of two YM multiplets, the $(4,0)$ multiplet is a product of two $(2,0)$ tensor multiplets~\cite{Chiodaroli:2011pp,Tensorsquared}. From other side while the circle reduction of a $(2,0)$ theory yields five-dimensional YM, it appears to be impossible to reconcile the nonlinear couplings of the five-dimensional maximal supergravity with the symmetries of the six dimensional $(4,0)$ and $(3,1)$ multiplet consistently with six-dimensional gauge or Lorentz invariance. 
We have not studied the possibility of couplings which manifestly break these properties in any detail.

Our results on anomalies in exotic six-dimensional multiplets are perhaps not too surprising -- after all they do not display full covariance. One should have in mind formal properties of elliptic operators on a six-manifold $M$, rather that anomalous box diagrams. While we express the result in terms of local curvatures on $M$, there can be no cancellation mechanism short of full automatic cancellation. Such cancellations are not happening for any of the exotic multiplets. In the $(2,0)$ case the SD Weyl multiplet is the only multiplet that does not have an anomaly polynomial proportional to $X_8$ \eqref{eq:X8}.  Both for $(2,0)$ and for  $(1,0)$ SD Weyl multiplets, trying to find a combination of matter that would lead to cancellation of the irreducible part of the anomaly is not useful, in spite of abundance of 2-form tensor fields. Due to absence of gravitons, one could not possibly compute counterterms that could lead to anomaly cancellation. On the other hand, for the  $(2,0)$ and $(1,0)$ SD Weyl multiplets one could contemplate coupling to respectively  $(2,0)$ and  $(1,0)$  gravity multiplets, together with appropriate matter, in order to cancel the irreducible part of the anomaly. We have neither studied if this can be done supersymmetrically  or  thought about any other aspects of such ``exotic bi-gravity" theories.

Finally, we have not discussed the quantisation of the exotic fields in detail. Some recent progress relevant to this direction includes~\cite{Action, Henneaux:2018rub, Lekeu:2018vhq, Henneaux:2019zod}. However, we have shown how to infer the information about at least the net degrees of freedom for the ghosts from anomaly calculations. As shown in section \ref{sec:ghost} comparing two ways of computing the anomaly of the exotic gravitino $\psi_{\mu \nu}$ leads to the conclusion that  its net ghost degrees of freedom are given by a chiral gravitino and a chiral spinor both of the same chirality as the exotic gravitino. It should be interesting to make a more direct and exhaustive study of this quantisation.


\section*{Acknowledgements} We would like to thank F. Bonetti, L. Borsten, J.J. Carrasco, D. Waldram for useful discussions. We owe special thanks to M. Duff, C. Hull and G. Bossard. The work of RM is supported in part by ERC Grant 787320 - QBH Structure and by ERC Grant 772408 - Stringlandscape.


\appendix

\section{Chiral 6d multiplets}
\label{app:Ch6d}

In this appendix, we briefly review the construction of massless multiplets of the chiral supersymmetry algebras in six dimensions. 

Consider the $\cN=(N,0)$ supersymmetry algebras, which have R-symmetry $\Symp(2N)$. Let $\alpha = 1, \dots, 4$ be an $\SU^*(4) \simeq \Spin(5,1)$ index and $A=1,\dots,2N$ be an $\Symp(2N)$ index. The supercharges $Q_{\alpha A}$ thus live in the $\repp{4}{2N}$ representation of $\Spin(5,1)\times\Symp(2N)$. Their anticommutator takes the form
\begin{equation}
\label{eq:6d-superalg}
	\{ Q_{\alpha A}, Q_{\beta B} \} = C_{AB} P_\mu {\gamma^\mu}_{[\alpha\beta]} 
		+ \dot{Z}_{[\alpha\beta][AB]} + Z_{(\alpha\beta)(AB)}
\end{equation}
where $C_{AB}$ is the $\Symp(2N)$ symplectic form, $P_\mu$ is the momentum and the quantities denoted with a $Z$ are central charges (with $\dot{Z}_{[\alpha\beta]A}{}^A = 0$).

As usual, to analyse the spin content of massless multiplets, we decompose under $\Spin(1,1)\times\Spin(4) \subset \Spin(5,1)$, writing the $\Cliff(5,1;\bbR)$ gamma matrices as the tensor products
\begin{equation}
	\gamma^{0} = \ii\sigma^{2} \otimes \id
	\qquad
	\gamma^{1} = \sigma^{1} \otimes \id
	\qquad
	\gamma^{m} = \sigma^{3} \otimes \gamma^m
\end{equation}
where $\sigma^i$ are the Pauli matrices and $\gamma^m$ are the generators of $\Cliff(4;\bbR)$. 
The transpose intertwiners $C_{5,1}$ for $\Cliff(5,1)$ and $C_{4}$ for $\Cliff(4)$, which we use to raise and lower spinor indices, are then related by $C_{5,1} = \sigma^1\otimes C_4$. 
Taking zero central charges and momentum $(P^\mu) = (k,k,0,\dots,0)$ for a massless representation, we see that
\begin{equation}
	\Big[ (P_\mu \gamma^\mu)_{[\alpha\beta]} \Big] = 
	2k \begin{pmatrix} 1 & 0 \\ 0 & 0 \end{pmatrix} \otimes C_4
\end{equation}
The supercharges with non-trivial algebra are thus those with positive chirality under the $\Spin(1,1)$.\footnote{The negative chirality supercharges are nilpotent and generate physically irrelevant zero-norm states, so we discard them at this point.} As $(\rep{4})_{\Spin(5,1)} \ra (\repp{2}{1}_+ + \repp{1}{2}_-)_{\SU(2)_\suone \times\SU(2)_\sutwo \times\Spin(1,1)}$ we have that these transform in the $\reppp{2}{1}{2N}$ representation of $\SU(2)_\suone \times\SU(2)_\sutwo \times\Symp(2N)$. 
Decomposing under $U(1)_\suone \subset \SU(2)_\suone$ we have $\rep{2} \ra \rep{1}_+ + \rep{1}_-$. Denote now by $Q_\pm$ the supercharges with $U(1)_\suone$ charge $\pm1$. In an appropriate complex basis we have that these satisfy the usual 
Clifford algebra of raising and lowering operators
\begin{equation}
	\{ Q_{+A}, Q_{+B} \} = 0
	\qquad
	\{ Q_{+A}, Q^{-B} \} = \delta_A{}^B
	\qquad
	\{ Q^{-A}, Q^{-B} \} = 0
\end{equation}
We can then build a multiplet by acting on a vacuum state $\vac$ with the raising operators $Q_{+A}$. 
The basic multiplet thus has the form 
\begin{equation}
\begin{array}{cccccccc}
	\vac
	\qquad
	&Q_{+A} \vac
	\qquad
	&Q_{+A} Q_{+B} \vac 
	\qquad 
	&Q_{+A} Q_{+B} Q_{+C} \vac 
	\qquad 
	&\dots
\end{array}
\end{equation}
With each term having one unit more $U(1)_\suone$ charge than the previous. These various terms can then be combined into $\SU(2)_\suone$ representations.


\subsection*{$\cN=(1,0)$}

Here, the R-symmetry is $\Symp(2)$ and the basic multiplet, in which the vacuum has only a $U(1)_\suone$ charge of $-1$, has the structure
\begin{equation}
\begin{array}{cccccccc}
	& &\vac
	&\qquad
	&Q_{+A} \vac
	&\qquad
	&Q_{+A} Q_{+B} \vac \\
	\text{$U(1)_\suone$ charge}
	& & -1 & & 0 & & +1 \\
	\text{$\Symp(2)$ irreps}
	& & \rep{1} & & \rep{2} & & \rep{1}
\end{array}
\end{equation}
This is the hyper-multiplet and by combining the $U(1)_\suone$ charges into $\SU(2)_\suone$ representations we can read-off its field content as
\begin{equation}
\begin{array}{cccccccc}
	\text{Spin} & & 0
	&\qquad
	& \frac12 \\
	\text{$\SU(2)_\suone \times \SU(2)_\sutwo \times \Symp(2)$ rep}
	& & \reppp{1}{1}{2} & & \reppp{2}{1}{1} &
\end{array}
\end{equation}

The other multiplets are then formed by taking tensor products of this multiplet with some representation $V$ of $G_{\text{little}} = \SU(2)_\suone \times \SU(2)_\sutwo \times \Symp(2)$. We have:
\begin{equation}
\begin{array}{cccccccccccc}
	V= \reppp{1}{1}{1} & & \text{(Hyper)} & \\
	\text{Field} & & \phi
	& & \lambda^R \\ 
	\text{$G_{\text{little}} $ rep}
	& & \reppp{1}{1}{2} & & \reppp{2}{1}{1} & \\ & \\
	V= \reppp{1}{2}{1} & & \text{(Vector)} & \\
	\text{Field} & & & & \lambda^L
	& & A_\mu \\ 
	\text{$G_{\text{little}} $ rep}
	& & & & \reppp{1}{2}{2} & & \reppp{2}{2}{1} \\ & \\
	V= \reppp{2}{1}{1} & & \text{(Tensor)} & \\
	\text{Field} & & \phi & & \lambda^R
	& & B_{\mu \nu}^- \\ 
	\text{$G_{\text{little}} $ rep}
	& & \reppp{1}{1}{1} & & \reppp{2}{1}{2} & & \reppp{3}{1}{1} \\ & \\
	V= \reppp{1}{3}{1} & & \text{(Gravitino$^L$)} & \\
	\text{Field} & & & & 
	& & B_{\mu \nu}^+ & & \psi_\mu^L \\ 
	\text{$G_{\text{little}} $ rep}
	& & & & & & \reppp{1}{3}{2} & & \reppp{2}{3}{1} \\ & \\
	V= \reppp{2}{2}{1} & & \text{(Gravitino$^R$)} & \\
	\text{Field} & & & & \lambda^L
	& & A_\mu & & \psi_\mu^R \\ 
	\text{$G_{\text{little}} $ rep}
	& & & & \reppp{1}{2}{1} & & \reppp{2}{2}{2} & & \reppp{3}{2}{1} \\ & \\
	V= \reppp{2}{3}{1} & & \text{(Gravity)} & \\
	\text{Field} & & & & 
	& & B_{\mu \nu}^+ & & \psi_\mu^L & & g_{\mu\nu} \\ 
	\text{$G_{\text{little}} $ rep}
	& & & & & & \reppp{1}{3}{1} & & \reppp{2}{3}{2} & & \reppp{3}{3}{1} \\ & \\
	V= \reppp{3}{1}{1} & & \text{(Exotic Gravitino)} & \\
	\text{Field} & & & & 
	\lambda^R & & B_{\mu \nu}^- & & \psi_{\mu\nu}^R \\ 
	\text{$G_{\text{little}} $ rep}
	& & & & \reppp{2}{1}{1}  & & \reppp{3}{1}{2} & & \reppp{4}{1}{1} \\ & \\
	V= \reppp{4}{1}{1} & & \text{(Exotic Gravity)} & \\
	\text{Field} & & & & 
	& & B_{\mu \nu}^- & & \psi_{\mu\nu}^R & & C_{[\mu\nu][\lambda\kappa]} \\ 
	\text{$G_{\text{little}} $ rep}
	& & & & & & \reppp{3}{1}{1} & & \reppp{4}{1}{2} & & \reppp{5}{1}{1} \\ & \\
\end{array}
\end{equation}
%


\subsection*{$\cN=(2,0)$}

Here, the R-symmetry is $\Symp(4)$ and the basic multiplet, in which the vacuum has only a $U(1)_\suone$ charge of $-2$, has the structure
\begin{equation}
\begin{array}{ccccccccccc}
	& &\vac
	&
	&Q_{+A} \vac
	&
	&Q_{+A} Q_{+B} \vac 
	&
	&Q_{+A} Q_{+B} Q_{+C} \vac 
	&
	&Q_{+A} Q_{+B} Q_{+C} Q_{+D} \vac \\
	\text{$U(1)_\suone$ charge}
	& & -2 & & -1 & & 0 & & +1 & & +2\\
	\text{$\Symp(4)$ irreps}
	& & \rep{1} & & \rep{4} & & \rep{1+5} & & \rep{4} & & \rep{1}
\end{array}
\end{equation}
This gives the tensor multiplet, whose field content is
\begin{equation}
\begin{array}{cccccccc}
	\text{Field} & & \phi
	&\quad & \lambda^R 
	&\quad & B_{\mu\nu}^- \\
	\text{$G_{\text{little}}$ rep}
	& & \reppp{1}{1}{5} & & \reppp{2}{1}{4} & & \reppp{3}{1}{1}
\end{array}
\end{equation}
The other multiplets are then formed by taking tensor products of this multiplet with some representation $V$ of $G_{\text{little}} = \SU(2)_\suone \times \SU(2)_\sutwo \times \Symp(4)$. We have:
\begin{equation}
\begin{array}{cccccccccccc}
	V= \reppp{1}{1}{1} & & \text{(Tensor)} & \\
	\text{Field} & & \phi
	& & \lambda^+ 
	& & B_{\mu\nu}^- \\ 
	\text{$G_{\text{little}} $ rep}
	& & \reppp{1}{1}{5} & & \reppp{2}{1}{4} & & \reppp{3}{1}{1} & \\ & \\
	V= \reppp{1}{2}{1} & & \text{(Gravitino$^+$)} & \\
	\text{Field} & & & & \lambda^-
	& & A_{\mu} & & \psi_\mu^+ \\ 
	\text{$G_{\text{little}} $ rep}
	& & & & \reppp{1}{2}{5} & & \reppp{2}{2}{4} & & \reppp{3}{2}{1} \\ & \\
	V= \reppp{1}{3}{1} & & \text{(Gravity)} & \\
	\text{Field} & & & & 
	& & B_{\mu \nu}^+ & & \psi_\mu^- & & g_{\mu\nu} \\ 
	\text{$G_{\text{little}} $ rep}
	& & & & & & \reppp{1}{3}{5} & & \reppp{2}{3}{4} & & \reppp{3}{3}{1} \\ & \\
	V= \reppp{2}{1}{1} & & \text{(Exotic Gravitino)} & \\
	\text{Field} & & \phi & & 
	\lambda^+ & & B_{\mu \nu}^- & & \psi_{\mu\nu}^+ \\ 
	\text{$G_{\text{little}} $ rep}
	& & \reppp{1}{1}{4} & & \reppp{2}{1}{5+1}  & & \reppp{3}{1}{4} & & \reppp{4}{1}{1} \\ & \\
	V= \reppp{3}{1}{1} & & \text{(Exotic Gravity)} & \\
	\text{Field} & & \phi & & \lambda^+
	& & B_{\mu \nu}^- & & \psi_{\mu\nu}^+ & & C_{[\mu\nu][\lambda\kappa]} \\ 
	\text{$G_{\text{little}} $ rep}
	& & \reppp{1}{1}{1} & & \reppp{2}{1}{4} & & \reppp{3}{1}{5+1} & & \reppp{4}{1}{4} & & \reppp{5}{1}{1} \\ & \\
\end{array}
\end{equation}
Note that there is no Gravitino$^L$ multiplet. This is consistent with the absence of a gravity mulitplet when $\cN=(3,0)$ or $\cN=(4,0)$. 

We can also decompose these multiplets into multiplets of the $\cN=(1,0)$ algebra. The resulting $\cN=(2,0) \ra \cN=(1,0)$ decompositions are given below.
\begin{equation}
\begin{array}{rcccl}
	\text{Tensor} & &\ra & & \text{Tensor} + 2 \times \text{Hyper} \\
	\text{Gravitino$^R$} & &\ra & & \text{Gravitino$^R$} + 2 \times \text{Vector} \\
	\text{Gravity} & &\ra & & \text{Gravity} + 2 \times \text{Gravitino$^L$} \\
	\text{Exotic Gravitino} & &\ra & & \text{Exotic Gravitino} + 2 \times \text{Tensor} 
		+ 2\times\text{Hyper} \\
	\text{Exotic Gravity} & &\ra & & \text{Exotic Gravity} + 2 \times \text{Exotic Gravitino} 
		+ \text{Tensor} \\
\end{array}
\end{equation}
%


\subsection*{$\cN=(4,0)$}

Here, the R-symmetry is $\Symp(8)$ and the basic multiplet, in which the vacuum has only a $U(1)_\suone$ charge of $-4$, is the exotic gravity multiplet and has the structure
\begin{equation}
\begin{array}{ccccccccccccccccccc}
	\text{$U(1)_\suone$ charge}
	& & -4 & & -3 & & -2 & & -1 \\ 
	\text{$\Symp(4)$ irreps}
	& & \rep{1} & & \rep{8} & & \rep{1+27} & & \rep{8+48} \\[8pt] 
	& & & & & & & & 0 \\
	& & & & & & & & \rep{1+27+42} \\[8pt]
	& & & & & & & & +1 & & +2 & & +3 & & +4 \\
	& & & & & & & &  \rep{8+48} & & \rep{1+27} & & \rep{8} & & \rep{1} 
\end{array}
\end{equation}
Thus the field content is
\begin{equation}
\begin{array}{cccccccccccccc}
	\text{Field} & & \phi & & \lambda^R
	& & B_{\mu \nu}^- & & \psi_{\mu\nu}^R & & C_{[\mu\nu][\lambda\kappa]} \\ 
	\text{$G_{\text{little}} $ rep}
	& & \reppp{1}{1}{42} & & \reppp{2}{1}{48} & & \reppp{3}{1}{27} 
		& & \reppp{4}{1}{8} & & \reppp{5}{1}{1}
\end{array}
\end{equation}
and we have the $\cN=(4,0) \ra \cN=(2,0)$ decomposition:
\begin{equation}
\begin{array}{rcccl}
	\text{Exotic Gravity} & &\ra & & \text{Exotic Gravity} + 4 \times \text{Exotic Gravitino} 
		+ 5\times\text{Tensor} \\
\end{array}
\end{equation}
%


\section{Conventions and useful formulae}
\label{app:Conv}

We start with a brief account of our conventions for the representations of the space-time Lorentz group $SO(5,1)$ and the orthogonal group $\SO(6)$. 
Their Lie algebras are different real forms of the complex Lie algebras of type $A_3 \sim D_3$ in the Cartan classification, with $\mathfrak{so}(6) \cong \mathfrak{su}(4)$ and $\mathfrak{so}(5,1) \cong \mathfrak{su}^*(4)$. 
There are then two common conventions for the ordering of the Dynkin labels, and we use both in places. 
In the ``D-type" conventions, the vector representation is $[1, 0, 0]$, the spinor with positive chirality is $[0, 1, 0]$ while the spinor with negative chirality is represented by $[0, 0, 1]$.
We then have, for instance, 
\begin{equation}
\label{tensorproduct1}
[1, 0, 0] \otimes [0, 1, 0] = [1, 1, 0] \oplus [0, 0, 1],
\end{equation}
which recovers the discussion below \eqref{RSCP}. 
In the ``A-type" conventions, we write the vector representation as $[0, 1, 0]$ and the spinor with positive chirality as $[1, 0, 0]$, while the spinor with negative chirality is represented by $[0, 0, 1]$. 
We use ``A-type" conventions whenever referring to the Lie algebra as $\mathfrak{su}(4)$ or $\mathfrak{su}^*(4)$.

The anomalies in $2n$ dimensions, $ I^{1}_{2n} $, can be related to the index of a Dirac operator in $2n+2$ dimensions via descent:
\begin{align*}
 I_{2n+2} &= \mathrm{d} I_{2n+1}  \\
 \delta I_{2n+1} &= \mathrm{d} I^{1}_{2n} 
\end{align*}
The Dirac index for a chiral spinor is given by  
\begin{equation}
I^{\mathrm{spin}\frac{1}{2}}_{2n+2} = [\hat{A}(M_{2n})\ch(F)]_{2n+2} \, ,
\end{equation}
where the roof-genus and the Chern character are defined as \cite{AGW}:
\begin{align}
  \Hat{A}(M_{2n}) &= 1- \frac{1}{24} p_1(TM) + \frac{1}{5760}(7p^2_1(TM)-4p_2(TM))  + ... \\
  \ch_{\bf R} (F) &\equiv \tr  \left( \frac{i}{2\pi} F \right) = \mbox{rk}({\bf R}) + \frac{i}{2\pi} \tr_{\bf R} F + ... + \frac{i^k}{k! (2\pi)^k} \tr_{\bf R} F^k + ...
\end{align}
$p_i(TM)$ are the Pontryagin classes of the tangent bundle which in conventions we use are given in terms of the curvature two-form as:
\begin{equation}
\det\left(1-\frac{R}{2\pi}\right) = 1 + p_1 + p_2+ p_3 +p_4+...
\end{equation}
The first two Pontryagin classes are sufficient for our purposes
\begin{equation}
\begin{aligned}
p_1 &=\frac{1}{(2\pi)^2}\left( -\frac{1}{2} \tr R^2\right)\\
p_2 &=\frac{1}{(2\pi)^4}\left(-\frac{1}{4} \tr R^4 +\frac{1}{8} (\tr R^2)^2\right).
\end{aligned}
\end{equation}
The spin $3/2$ fermion anomaly is computed using 
\begin{equation}
\begin{aligned} \label{app1}
\Hat{A}(M_{2n}) \left(\ch(R) -1\right)&=  \Hat{A}(M_{2n}) \left(\tr(e^{\frac{i}{2\pi}R}) -1\right)\\ 
&= \Hat{A}(M_{2n}) \left(\tr(e^{\frac{i}{2\pi}R}- \mathbb{I} ) + \text{dim}(T)- 1\right)\\
\end{aligned}
\end{equation}
where $\text{dim}(T)$ is the dimension of the tensor representation of $SO(2n)$ and $R$ is the curvature 2-form $R_{ab}$ with the orthogonal frame indices $a,b$ contracted with the generator $T^{ab}$ of $SO(2n)$.
Since $R_{ab}$ is anti-symmetric in $a$ and $b$, the matrix $\frac{1}{2 \pi}R$ can be brought in the skew-symmetric form
\begin{equation}
\begin{pmatrix}
 & x_1 &  &  &  &  &  & \\ 
 -x_1&  &  &  &  &  &  & \\ 
 &  &  &  x_2 &  &  &  & \\ 
 &  &  -x_2 &  &  &  &  & \\ 
 &  &  &  &  &  ..&  & \\ 
 &  &  &  &  &  & .. & \\ 
 &  &  &  &  &  &  &x_n  \\ 
 &  &  &  &  &  & -x_n & 
\end{pmatrix}
\end{equation}
where each $x_j$ is a $2$-form and the first two Pontryagin classes can also be expressed in power of $x_j$'s
\begin{equation} \label{eigenvalueR}
\begin{aligned}
p_1 &= \sum^{n}_{j=1} x_j^2\\
p_2 &=\sum^{n}_{i<j} x_i^2 x_j^2.
\end{aligned}
\end{equation}
We also make use of  the representation independent quantity
\begin{equation} \label{app2}
\Hat{A}(M_{2n}) \tr(e^{\frac{i}{2\pi}R}- \mathbb{I}) = \frac{1}{2^2}(4p_1)+\frac{1}{2^4}(\frac{2}{3} p^2_1-\frac{8}{3}p_2)
\end{equation}
and the Hirzebruch $L$-polynomial, expressed in terms of Pontryagin classes as
\begin{equation}
L(M_{2n}) = 1 + \frac{1}{3}p_1 + (-\frac{1}{45}p^2_1 + \frac{7}{45}p_2).
\end{equation}
The anomaly formulas for six-dimensional fields are given by \cite{AGW}
\begin{equation}
\begin{aligned}  \label{anomaly polynomial}
I^{\mathrm{spin}\frac{1}{2}} &= \frac{1}{5760} \left( 7 {p_1}^2 - 4p_2\right)\\
I^{\mathrm{spin}\frac{3}{2}} &=\frac{1}{5760} \left( 275 {p_1}^2 - 980 p_2\right) \\
I^{A}&=\frac{1}{5760} \left( 16 {p_1}^2 -  112p_2\right).
\end{aligned}
\end{equation}
The invariant polynomials in \eqref{anomaly polynomial} correspond to anomalies for the axial currents, i. e. the current of Dirac fermion  coupled to gauge field under the axial symmetry 
\begin{equation}
\psi \longrightarrow e^{i \alpha^{a}(x) T^{a} \gamma_5} \psi 
\end{equation}
with $T^a$ the Hermitian generator of the gauge group. Since we are interested in  the non-conservation of the currents of the Weyl fermion  coupled to gauge fields, $I^{\mathrm{spin}\frac{3}{2}}$ and $I^{\mathrm{spin}\frac{3}{2}}$ need to be divided by 2.


\subsection{Computational details for section \ref{sec:anomalies}}
\label{sec:calcul}

{\bf The Dirac operator for SD Weyl field: $\,\,\,$} In order to compute the relevant Dirac operator for the SD Weyl field one needs to extract the $[0,0,4]$ piece of the $\mathfrak{su}^*(4)$ representation in \eqref{2x2}:
\begin{equation}
\begin{aligned}
R \in &\mathcal{C}^{\infty}(S^-\otimes S^- \otimes S^-  \otimes S^-) - \mathcal{C}^{\infty}( T^*M \otimes T^*M) - \mathcal{C}^{\infty}(T^*M \otimes F_3^-) - \mathcal{C}^{\infty}(F_3^- \otimes T^*M) \\
&-\left( \mathcal{C}^{\infty}(T^*M \otimes F_3^-) - B \right) - g \\
= & \mathcal{C}^{\infty}(S^-\otimes S^- \otimes S^-  \otimes S^-) -\mathcal{C}^{\infty}( T^*M \otimes T^*M) - \mathcal{C}^{\infty}(T^*M \otimes (S^-\otimes S^- - T^*M)) 
\\&- \mathcal{C}^{\infty}((S^-\otimes S^- - T^*M) \otimes T^*M) - \mathcal{C}^{\infty}(T^*M \otimes (S^-\otimes S^- - T^*M)) + B - g\\
= & \mathcal{C}^{\infty}(S^-\otimes S^- \otimes S^-  \otimes S^-) -\mathcal{C}^{\infty}( T^*M \otimes T^*M) - \mathcal{C}^{\infty}(T^*M \otimes S^-\otimes S^- )   \\
& + \mathcal{C}^{\infty}(T^*M \otimes T^*M) - \mathcal{C}^{\infty}(S^-\otimes S^- \otimes T^*M) +  \mathcal{C}^{\infty}(T^*M \otimes T^*M)) \\
&- \mathcal{C}^{\infty}(T^*M \otimes S^-\otimes S^-) +  \mathcal{C}^{\infty}(T^*M \otimes T^*M)) + B - g \\
= & \mathcal{C}^{\infty}(S^-\otimes S^- \otimes S^-  \otimes S^-)  - \mathcal{C}^{\infty}(T^*M \otimes S^-\otimes S^- )   \\
& - \mathcal{C}^{\infty}(S^-\otimes S^- \otimes T^*M) - \mathcal{C}^{\infty}(T^*M \otimes S^-\otimes S^-)\\
&+  \mathcal{C}^{\infty}(T^*M \otimes T^*M))  +  \mathcal{C}^{\infty}(T^*M \otimes T^*M) + B - g\\
= & \mathcal{C}^{\infty}(S^-\otimes S^- \otimes S^-  \otimes S^-)  - \mathcal{C}^{\infty}(T^*M \otimes S^-\otimes S^- )   \\
& - \mathcal{C}^{\infty}(S^-\otimes S^- \otimes T^*M) - \mathcal{C}^{\infty}(T^*M \otimes S^-\otimes S^-)\\
&+  \mathcal{C}^{\infty}(T^*M \otimes T^*M)  +  \mathcal{C}^{\infty}(T^*M \otimes T^*M) + B - g \\
= & \mathcal{C}^{\infty}(S^-\otimes S^- \otimes S^-  \otimes S^-)  - \mathcal{C}^{\infty}(T^*M \otimes S^-\otimes S^- )   \\
& - \mathcal{C}^{\infty}(S^-\otimes S^- \otimes T^*M) - \mathcal{C}^{\infty}(T^*M \otimes S^-\otimes S^-)\\
&+  \mathcal{C}^{\infty}(S^- \otimes S^+ + g)  +  \mathcal{C}^{\infty}(S^- \otimes S^+ + g) + B - g\\
= & \mathcal{C}^{\infty}(S^-\otimes S^- \otimes S^-  \otimes S^-)  - \mathcal{C}^{\infty}(T^*M \otimes S^-\otimes S^- )   \\
& - \mathcal{C}^{\infty}(S^-\otimes S^- \otimes T^*M) - \mathcal{C}^{\infty}(T^*M \otimes S^-\otimes S^-)\\
&+  \mathcal{C}^{\infty}(S^- \otimes S^+) +  \mathcal{C}^{\infty}(S^- \otimes S^+) + B + g\\
=  & \mathcal{C}^{\infty}\left(S^-\otimes [S^- \otimes S^-  \otimes S^- - (S^- \otimes T^*M )^{\oplus 3} + (S^+)^{\oplus 2} ] \right)  + B + g
\end{aligned}
\end{equation}
\\
\noindent
{\bf The Dirac operator for exotic graviton in $(3,1)$ multiplet: $\,\,\,$} For  the $D$  field in the $(3,1)$ multiplet, we focus on its field strength $S$ in the $[1,0,3]$ of $\mathfrak{su}^*(4)$. From
\begin{equation}
[1,0,1] \otimes [0,0,2] = [1,0,3] \oplus [1,1,1] \oplus [0,0,2] \oplus [0,1,0]
\end{equation}
and 
\begin{equation}
[1,0,1] \otimes [0,1,0] = [1,1,1] \oplus [0,0,2] \oplus [0,1,0] \oplus [2,0,0]
\end{equation}
we get 
\begin{equation}
[1,0,3] = [1,0,1] \otimes [0,0,2] \ominus \left( [1,0,1] \otimes [0,1,0] \ominus [2,0,0] \right) 
\end{equation}
Thus
\begin{equation}
\begin{aligned}
S \in &\mathcal{C}^{\infty}(B\otimes F_3^-) - \mathcal{C}^{\infty}( B \otimes T^*M) +\mathcal{C}^{\infty}(F_3^+)\\
&=\mathcal{C}^{\infty}\left( [S^+ \otimes S^- - \phi] \otimes [S^- \otimes S^- - \phi]\right) - \mathcal{C}^{\infty}\left([S^+ \otimes S^- - \phi]\otimes T^*M \right)   +\mathcal{C}^{\infty}(S^+ \otimes S^+ -\phi)\\
&= \mathcal{C}^{\infty}\left(S^+ \otimes S^- \otimes S^- \otimes S^- \right) -  \mathcal{C}^{\infty}\left(S^- \otimes S^+ \otimes T^*M \right) - \mathcal{C}^{\infty}\left(S^- \otimes S^- \otimes \phi \right) + \mathcal{C}^{\infty}\left(T^*M \otimes \phi \right) \\
&-\mathcal{C}^{\infty}\left(S^- \otimes S^+ \otimes T^*M \right) +\mathcal{C}^{\infty}\left(T^*M \otimes \phi \right) + \mathcal{C}^{\infty}\left(S^+ \otimes S^+ \right) - \mathcal{C}^{\infty}\left(T^*M \right)\\
&= \mathcal{C}^{\infty}\left(S^- \otimes [S^- \otimes S^- \otimes S^+ - (S^+ \otimes T^*M )^{\oplus 2} - (S^-)^{\oplus 2}]\right).  \\
\end{aligned}
\end{equation}
\\
\noindent
{\bf The $SO(6)$ generator in the rank 2-tensor representation: $\,\,\,$} In order to find the index of the exotic gravitino (see subsection \ref{sec:ghost}), 
\begin{equation}
\tr e^{i M} =  n(2n-1) - \frac{1}{2}\tr M^2 + \frac{1}{4!} \tr M^4 + \ldots
\end{equation}
for a $ n(2n-1) \times n(2n-1)$ matrix M given by  
\begin{equation}
M_{cd,ef} \equiv \frac{1}{2\pi} \left(R_{ce} \delta_{df} +R_{df} \delta_{ce} -R_{cf} \delta_{de}-R_{de} \delta_{cf}  \right) \, 
\end{equation}
has to be computed. $\tr M^2$ and $\tr M^4$ are evaluated as follows.
\begin{equation*}
\begin{aligned}
(M^2)_{ab, ef} &= \frac{1}{2} M_{ab, cd}  M_{cd, ef}\\
&=\frac{1}{2(2\pi)^2}\left( R_{ac}\delta_{bd}+R_{bd}\delta_{ac}-R_{ad}\delta_{bc}-R_{bc}\delta_{ad}\right) \left( R_{ce}\delta_{df}+R_{df}\delta_{ce}-R_{cf}\delta_{de}-R_{de}\delta_{cf}\right)\\
&=   \frac{1}{(2\pi)^2} (R^2_{ae}\delta_{bf}+R^2_{bf}\delta_{ae}-R^2_{af}\delta_{be}-R^2_{be}\delta_{af}+2R_{ae}R_{bf}-2R_{af} R_{be} )\\
\end{aligned}
\end{equation*}
\begin{equation}
\begin{aligned}
\Rightarrow \tr M^2 &= \frac{1}{2} \sum_{a,b} (M^2)_{ab,ab}\\
&=\frac{1}{2(2\pi)^2} \sum_{a,b} ( R^2_{aa}\delta_{bb} +R^2_{bb}\delta_{aa}-R^2_{ab}\delta_{ba}-R^2_{ba}\delta_{ab}+2R_{aa}R_{bb}-2R_{ab}R_{ba})\\
&=\frac{1}{2(2\pi)^2} ( 2n \tr R^2 +2n \tr R^2 - \tr R^2- \tr R^2+0 - 2  \tr R^2 )\\
&= \frac{1}{(2\pi)^2} (2n-2) \tr R^2
\end{aligned}
\end{equation}

\begin{equation*}
\begin{aligned}
(M^4)_{ab, ef} &= \frac{1}{2} (M^2)_{ab, cd}  (M^2)_{cd, ef}\\
&=\frac{1}{(2\pi)^4} ( R^4_{ae}\delta_{bf}+6R^2_{ae}R^2_{bf}-6 R^2_{be}R^2_{af}-R^4_{be}\delta_{af}+4 R^3_{ae}R_{bf}-4 R^3_{be}R_{af} \\
&\;\;\;\;+R^4_{bf}\delta_{ae}- R^4_{af}\delta_{be}+4 R^3_{bf}R_{ae}-4 R^3_{af}R_{be} )\\
\end{aligned}
\end{equation*}
\begin{equation}
\begin{aligned}
\Rightarrow \tr M^4 &= \frac{1}{2} \sum_{a,b} (M^4)_{ab,ab}\\
&=\frac{1}{2(2\pi)^4} \sum_{a,b} ( R^4_{aa}\delta_{bb} + 6 R^2_{aa}R^2_{bb}-6 R^2_{ba}R^2_{ab}-R^4_{ba}\delta_{ab}+4R^3_{aa}R_{bb}\\
& \;\;\;\;-4R^3_{ba}R_{ab}+  R^4_{bb}\delta_{aa} -  R^4_{ab}\delta_{ba} +  4R^3_{bb}R_{aa} - 4 R^3_{ab}R_{ba} )\\
&=\frac{1}{(2\pi)^4} ( (2n - 8) \tr R^4 +3 (\tr R^2 )^2)
\end{aligned}
\end{equation}


\subsection{Independent components of SD Weyl field strength}

Deducing which components of the field strength $G_{MNP,QRS}$ of the SD Weyl field on $T^3$ are independent is a cumbersome task due to the double self-duality of the field strength. Here we present a brief group-theoretical account which enables us to be sure that we have not missed parts of the equations of motion in equation~\eqref{eq:Ceom*}.

The components of the SD Weyl field form a representation of $\SO(6)$, whose Lie algebra coincides with that of $\SU(4)$. Using Dynkin label conventions in which the six-dimensional vector representation is $[1,0,0]$, while the positive chirality spinor representation is $[0,1,0]$, the SD Weyl field strength $G_{MNP,QRS}$ transforms in the reducible  representation $[0,4,0] + [2,0,0]$. Under the relevant $\SO(3)\times\SO(3)$ subgroup we have the decompositions
\begin{equation}
\begin{aligned}{}
	[1,0,0] &\longrightarrow [2,0] + [0,2] \\[0pt]
	[0,1,0] &\longrightarrow [1,1] \\[0pt]
	[0,4,0] &\longrightarrow [0,0] + [2,2] + [4,4] \\[0pt]
	[2,0,0] &\longrightarrow [0,0] + [2,2] + [4,0] + [0,4] 
\end{aligned}
\end{equation}
Splitting the index $M = (\alpha, i)$ as in section~\ref{sec:SDW-T3}, 
the corresponding parts of the field $G$ can be identified as follows (the symbol $\sim$ here is taken to mean ``represents the same independent components of $G$"):
\begin{equation}
\begin{aligned}{}
	G_{ijk,i'j'k'} \sim G_{ijk,\alpha\beta\gamma} \sim G_{\alpha\beta\gamma,ijk} 
		\sim G_{\alpha\beta\gamma,\alpha'\beta'\gamma'} \sim [0,0] \\
	G^{ij\alpha}{}_{ij\alpha} \sim G^{i\alpha\beta}{}_{i\alpha\beta} \sim [0,0] \\
	G^{i}{}_{\alpha\beta,ijk} \sim G^{\alpha\beta}{}_{i,\alpha\beta\gamma} \sim [2,2] \\
	G^{ij}{}_{\alpha,ij\beta} \sim G^{\alpha}{}_{ij,\alpha\beta\gamma} \sim [2,2] \\
	\text{$\alpha\beta$-traceless part of } G^{ij}{}_{\alpha, ij\beta } \sim [4,0] \\
	\text{$ij$-traceless part of } G^{\alpha\beta}{}_{i,\alpha\beta j} \sim [0,4]\\
	\text{$ij$- and $\alpha\beta$-traceless parts of } G^{k}{}_{\alpha i, k\beta j} \sim G^{\gamma}{}_{\alpha i, \gamma \beta j} \sim [4,4] \\
\end{aligned}
\end{equation}
Imposing that $\der_i C_{MNPQ} = 0$ as in section~\ref{sec:SDW-T3},  we see that the first $[0,0]$ parts and the first [2,2] parts in this list vanish. All of the remaining components are then related to $G^{\gamma}{}_{\alpha i, \gamma \beta j}$ and its traces. Thus we conclude that the equation of motion $G^{M}{}_{NP,MRS} = 0$ indeed reduces to $G^{\gamma}{}_{\alpha i, \gamma \beta j} = 0$.


%
%



\begin{thebibliography}{99}

\bibitem{Hull1} 
  C.~M.~Hull,
``Strongly coupled gravity and duality,"
  Nucl.\ Phys.\ B {\bf 583} (2000) 237
  [hep-th/0004195].
 
\bibitem{Hull2}
  C.~M.~Hull,
  ``Symmetries and compactifications of (4,0) conformal gravity,''
  JHEP {\bf 0012} (2000) 007
  [hep-th/0011215].
 
 
\bibitem{Hull3}
  C.~M.~Hull,
  ``BPS supermultiplets in five-dimensions,''
  JHEP {\bf 0006} (2000) 019
  [hep-th/0004086].
 
 \bibitem{Chiodaroli:2011pp}
M.~Chiodaroli, M.~Gunaydin and R.~Roiban,
``Superconformal symmetry and maximal supergravity in various dimensions,''
JHEP \textbf{03}, 093 (2012)
[arXiv:1108.3085 [hep-th]].


\bibitem{magicsquare}
L.~Borsten, M.~J.~Duff, L.~J.~Hughes and S.~Nagy,
``Magic Square from Yang-Mills Squared,''
Phys. Rev. Lett. \textbf{112}, no.13, 131601 (2014)
[arXiv:1301.4176 [hep-th]].

\bibitem{magicpyramid}
  A.~Anastasiou, L.~Borsten, M.~J.~Duff, L.~J.~Hughes and S.~Nagy,
  ``A magic pyramid of supergravities,''
  JHEP {\bf 1404} (2014) 178
  [arXiv:1312.6523 [hep-th]].

\bibitem{Anastasiou:2014qba}
A.~Anastasiou, L.~Borsten, M.~J.~Duff, L.~J.~Hughes and S.~Nagy,
``Yang-Mills origin of gravitational symmetries,''
Phys. Rev. Lett. \textbf{113}, no.23, 231606 (2014)
[arXiv:1408.4434 [hep-th]].

\bibitem{YangMillssquared} 
  A.~Anastasiou, L.~Borsten, M.~J.~Hughes and S.~Nagy,
``Global symmetries of Yang-Mills squared in various dimensions,"
  JHEP {\bf 1601} (2016) 148
  [arXiv:1502.05359 [hep-th]].

  \bibitem{Tensorsquared} 
  L.~Borsten,
``$D=6$, $\mathcal{N}=(2,0)$ and $\mathcal{N}=(4,0)$ theories,"
  Phys.\ Rev.\ D {\bf 97} (2018) no.6, 066014
  [arXiv:1708.02573 [hep-th]].

\bibitem{Action} 
  M.~Henneaux, V.~Lekeu and A.~Leonard,
``The action of the (free) (4, 0)-theory,"
  JHEP {\bf 1801} (2018) 114
   Erratum: [JHEP {\bf 1805} (2018) 105]
  [arXiv:1711.07448 [hep-th]].

\bibitem{Cachazo:2018hqa}
  F.~Cachazo, A.~Guevara, M.~Heydeman, S.~Mizera, J.~H.~Schwarz and C.~Wen,
  ``The S Matrix of 6D Super Yang-Mills and Maximal Supergravity from Rational Maps,''
  JHEP {\bf 1809} (2018) 125
  [arXiv:1805.11111 [hep-th]].




  \bibitem{Strathdee} 
  J.~A.~Strathdee,
``Extended Poincare Supersymmetry,"
  Int.\ J.\ Mod.\ Phys.\ A {\bf 2} (1987) 273.

\bibitem{CK1} 
  Z.~Bern, J.~J.~M.~Carrasco and H.~Johansson,
  ``New Relations for Gauge-Theory Amplitudes,''
  Phys.\ Rev.\ D {\bf 78}, 085011 (2008)
  [arXiv:0805.3993 [hep-ph]].

\bibitem{CK2} 
  Z.~Bern, J.~J.~M.~Carrasco and H.~Johansson,
  ``Perturbative Quantum Gravity as a Double Copy of Gauge Theory,''
  Phys.\ Rev.\ Lett.\  {\bf 105}, 061602 (2010)
  [arXiv:1004.0476 [hep-th]].

\bibitem{CK3} 
  Z.~Bern, T.~Dennen, Y.~t.~Huang and M.~Kiermaier,
  ``Gravity as the Square of Gauge Theory,''
  Phys.\ Rev.\ D {\bf 82}, 065003 (2010)
  [arXiv:1004.0693 [hep-th]].

\bibitem{Bern:2019prr} 
  Z.~Bern, J.~J.~Carrasco, M.~Chiodaroli, H.~Johansson and R.~Roiban,
  ``The Duality Between Color and Kinematics and its Applications,''
  arXiv:1909.01358 [hep-th].


\bibitem{Nagy:2014jza}
S.~Nagy,
``Chiral Squaring,''
JHEP \textbf{07}, 142 (2016)
[arXiv:1412.4750 [hep-th]].

\bibitem{Borsten:2015pla}
L.~Borsten and M.~J.~Duff,
``Gravity as the square of YangÐMills?,''
Phys. Scripta \textbf{90}, 108012 (2015)
[arXiv:1602.08267 [hep-th]].

\bibitem{Anastasiou:2017nsz}
A.~Anastasiou, L.~Borsten, M.~J.~Duff, A.~Marrani, S.~Nagy and M.~Zoccali,
``Are all supergravity theories YangÐMills squared?,''
Nucl. Phys. B \textbf{934}, 606-633 (2018)
[arXiv:1707.03234 [hep-th]].

\bibitem{ClassicalDCopy1} 
  R.~Monteiro, D.~O'Connell and C.~D.~White,
  ``Black holes and the double copy,''
  JHEP {\bf 1412}, 056 (2014)
  [arXiv:1410.0239 [hep-th]].

\bibitem{ClassicalDCopy2} 
  A.~Luna, R.~Monteiro, D.~O'Connell and C.~D.~White,
  ``The classical double copy for Taub–NUT spacetime,''
  Phys.\ Lett.\ B {\bf 750}, 272 (2015)
  [arXiv:1507.01869 [hep-th]].

\bibitem{Cardoso:2016ngt}
G.~L.~Cardoso, S.~Nagy and S.~Nampuri,
``A double copy for $ \mathcal{N}=2 $ supergravity: a linearised tale told on-shell,''
JHEP \textbf{10}, 127 (2016)
[arXiv:1609.05022 [hep-th]].

\bibitem{Cardoso:2016amd}
G.~Cardoso, S.~Nagy and S.~Nampuri,
JHEP \textbf{04}, 037 (2017)
doi:10.1007/JHEP04(2017)037
[arXiv:1611.04409 [hep-th]].

\bibitem{ClassicalDCopy3} 
  D.~S.~Berman, E.~Chacón, A.~Luna and C.~D.~White,
  ``The self-dual classical double copy, and the Eguchi-Hanson instanton,''
  JHEP {\bf 1901}, 107 (2019)
  [arXiv:1809.04063 [hep-th]].




\bibitem{CSW2} 
  A.~Coimbra, C.~Strickland-Constable and D.~Waldram,
  ``$E_{d(d)} \times \mathbb{R}^+$ generalised geometry, connections and M theory,''
  JHEP {\bf 1402}, 054 (2014)
  [arXiv:1112.3989 [hep-th]].
  
\bibitem{CSW3} 
  A.~Coimbra, C.~Strickland-Constable and D.~Waldram,
  ``Supergravity as Generalised Geometry II: $E_{d(d)} \times \mathbb{R}^+$ and M theory,''
  JHEP {\bf 1403}, 019 (2014)
  [arXiv:1212.1586 [hep-th]].


\bibitem{CSC} 
  C.~Strickland-Constable,
  ``Subsectors, Dynkin Diagrams and New Generalised Geometries,''
  JHEP {\bf 1708}, 144 (2017)
  [arXiv:1310.4196 [hep-th]].

\bibitem{HS} 
  O.~Hohm and H.~Samtleben,
  ``Exceptional Form of D=11 Supergravity,''
  Phys.\ Rev.\ Lett.\  {\bf 111}, 231601 (2013)
  [arXiv:1308.1673 [hep-th]].

\bibitem{E77EFT} 
  H.~Godazgar, M.~Godazgar, O.~Hohm, H.~Nicolai and H.~Samtleben,
  ``Supersymmetric E$_{7(7)}$ Exceptional Field Theory,''
  JHEP {\bf 1409}, 044 (2014)
  [arXiv:1406.3235 [hep-th]].

\bibitem{HS-E88} 
  O.~Hohm and H.~Samtleben,
  ``Exceptional field theory. III. E$_{8(8)}$,''
  Phys.\ Rev.\ D {\bf 90}, 066002 (2014)
  [arXiv:1406.3348 [hep-th]].

\bibitem{Schnakenburg:2001he} 
  I.~Schnakenburg and P.~C.~West,
  ``Kac-Moody symmetries of 2B supergravity,''
  Phys.\ Lett.\ B {\bf 517}, 421 (2001)
  [hep-th/0107181].

\bibitem{Ciceri:2016hup} 
  F.~Ciceri, G.~Dibitetto, J.~J.~Fernandez-Melgarejo, A.~Guarino and G.~Inverso,
  ``Double Field Theory at SL(2) angles,''
  JHEP {\bf 1705}, 028 (2017)
  [arXiv:1612.05230 [hep-th]].

\bibitem{Vafa:1996xn}
  C.~Vafa,
  ``Evidence for F theory,''
  Nucl.\ Phys.\ B {\bf 469} (1996) 403
  [hep-th/9602022].

\bibitem{Liu:1997mb}
  J.~T.~Liu and R.~Minasian,
``U-branes and T**3 fibrations,"
  Nucl.\ Phys.\ B {\bf 510} (1998) 538
  [hep-th/9707125].

\bibitem{Cremmer1}
  E.~Cremmer,
  ``Supergravities in 5 Dimensions,''
  In *Salam, A. (ed.), Sezgin, E. (ed.): Supergravities in diverse dimensions, vol. 1* 422-437. (In *Cambridge 1980, Proceedings, Superspace and supergravity* 267-282) and Paris Ec. Norm. Sup. - LPTENS 80-17 (80,rec.Sep.) 17 p. (see Book Index)
  
\bibitem{Bonetti1} 
  F.~Bonetti, T.~W.~Grimm and S.~Hohenegger,
  ``One-loop Chern-Simons terms in five dimensions,''
  JHEP {\bf 1307} (2013) 043
  [arXiv:1302.2918 [hep-th]].
 
 
\bibitem{Bonetti2}
  F.~Bonetti, T.~W.~Grimm and S.~Hohenegger,
  ``Exploring 6D origins of 5D supergravities with Chern-Simons terms,''
  JHEP {\bf 1305} (2013) 124
  [arXiv:1303.2661 [hep-th]].


 \bibitem{Bonetti3}   
  F.~Bonetti, T.~W.~Grimm and S.~Hohenegger,
  ``A Kaluza-Klein inspired action for chiral p-forms and their anomalies,''
  Phys.\ Lett.\ B {\bf 720} (2013) 424
  [arXiv:1206.1600 [hep-th]].


\bibitem{Ohmori:2014kda}
  K.~Ohmori, H.~Shimizu, Y.~Tachikawa and K.~Yonekura,
  ``Anomaly polynomial of general 6d SCFTs,''
  PTEP {\bf 2014} (2014) no.10,  103B07
  [arXiv:1408.5572 [hep-th]].


\bibitem{Eguchi:1980jx} 
  T.~Eguchi, P.~B.~Gilkey and A.~J.~Hanson,
  ``Gravitation, Gauge Theories and Differential Geometry,''
  Phys.\ Rept.\  {\bf 66}, 213 (1980).





\bibitem{deWit:2002vz}
  B.~de Wit,
  ``Supergravity,''
  hep-th/0212245.

\bibitem{deWitNicolai} 
  B.~de Wit and H.~Nicolai,
  ``$d=11$ Supergravity With Local SU(8) Invariance,''
  Nucl.\ Phys.\ B {\bf 274}, 363 (1986).

\bibitem{CremmerJulia} 
  E.~Cremmer and B.~Julia,
  ``The SO(8) Supergravity,''
  Nucl.\ Phys.\ B {\bf 159}, 141 (1979).

\bibitem{West} 
  P.~C.~West,
  ``E(11) and M theory,''
  Class.\ Quant.\ Grav.\  {\bf 18}, 4443 (2001)
  [hep-th/0104081].

\bibitem{DHN} 
  T.~Damour, M.~Henneaux and H.~Nicolai,
  ``E(10) and a 'small tension expansion' of M theory,''
  Phys.\ Rev.\ Lett.\  {\bf 89}, 221601 (2002)
  [hep-th/0207267].

\bibitem{Hull-EGG}
C.~M.~Hull,
``Generalised Geometry for M-Theory,''
JHEP \textbf{07}, 079 (2007)
[arXiv:hep-th/0701203 [hep-th]].

\bibitem{PW}
P.~Pires Pacheco and D.~Waldram,
``M-theory, exceptional generalised geometry and superpotentials,''
JHEP \textbf{09}, 123 (2008)
[arXiv:0804.1362 [hep-th]].

\bibitem{Kleinschmidt:2003mf} 
  A.~Kleinschmidt, I.~Schnakenburg and P.~C.~West,
  ``Very extended Kac-Moody algebras and their interpretation at low levels,''
  Class.\ Quant.\ Grav.\  {\bf 21}, 2493 (2004)
  [hep-th/0309198].

\bibitem{Bossard:2019ksx}
G.~Bossard, A.~Kleinschmidt and E.~Sezgin,
``On supersymmetric E$_{11}$ exceptional field theory,''
JHEP \textbf{10}, 165 (2019)
[arXiv:1907.02080 [hep-th]].

\bibitem{Hull:1997kt} 
  C.~M.~Hull,
  ``Gravitational duality, branes and charges,''
  Nucl.\ Phys.\ B {\bf 509}, 216 (1998)
  [hep-th/9705162].

\bibitem{Blair:2013gqa}
C.~D.~Blair, E.~Malek and J.~H.~Park,
``M-theory and Type IIB from a Duality Manifest Action,''
JHEP \textbf{01}, 172 (2014)
[arXiv:1311.5109 [hep-th]].

\bibitem{Schwarz:1995dk}
J.~H.~Schwarz,
``An SL(2,Z) multiplet of type IIB superstrings,''
Phys. Lett. B \textbf{360}, 13-18 (1995)
[arXiv:hep-th/9508143 [hep-th]].

\bibitem{Aspinwall:1995fw}
P.~S.~Aspinwall,
``Some relationships between dualities in string theory,''
Nucl. Phys. B Proc. Suppl. \textbf{46}, 30-38 (1996)
[arXiv:hep-th/9508154 [hep-th]].

\bibitem{Schwarz:1995jq}
J.~H.~Schwarz,
``The power of M theory,''
Phys. Lett. B \textbf{367}, 97-103 (1996)
[arXiv:hep-th/9510086 [hep-th]].

\bibitem{Guillaume}
G.~Bossard, 
Private communication.

\bibitem{Godazgar:2013pfa}
H.~Godazgar, M.~Godazgar and H.~Nicolai,
``Nonlinear Kaluza-Klein theory for dual fields,''
Phys. Rev. D \textbf{88}, no.12, 125002 (2013)
[arXiv:1309.0266 [hep-th]].

\bibitem{LSW2}
K.~Lee, C.~Strickland-Constable and D.~Waldram,
``New Gaugings and Non-Geometry,''
Fortsch. Phys. \textbf{65}, no.10-11, 1700049 (2017)
[arXiv:1506.03457 [hep-th]].


\bibitem{SO8-gaugings}
G.~Dall'Agata, G.~Inverso and M.~Trigiante,
``Evidence for a family of SO(8) gauged supergravity theories,''
Phys. Rev. Lett. \textbf{109}, 201301 (2012)
[arXiv:1209.0760 [hep-th]].


\bibitem{VanProeyen} 
  A.~Van Proeyen,
  ``Special geometries, from real to quaternionic,''
  hep-th/0110263.

\bibitem{AGW}
  L.~Alvarez-Gaume and E.~Witten,
 ``Gravitational Anomalies,"
  Nucl.\ Phys.\ B {\bf 234} (1984) 269.
  
 
\bibitem{ASZ}   
  O.~Alvarez, I.~M.~Singer and B.~Zumino,
``Gravitational Anomalies and the Family's Index Theorem,"
  Commun.\ Math.\ Phys.\  {\bf 96} (1984) 409.

\bibitem{AGG1} 
  L.~Alvarez-Gaume and P.~H.~Ginsparg,
``The Structure of Gauge and Gravitational Anomalies,"
  Annals Phys.\  {\bf 161} (1985) 423
   Erratum: [Annals Phys.\  {\bf 171} (1986) 233].
  
  
\bibitem{AGG2}
  L.~Alvarez-Gaume and P.~H.~Ginsparg,
``The Topological Meaning of Nonabelian Anomalies,"
  Nucl.\ Phys.\ B {\bf 243} (1984) 449.


 \bibitem{Bilal} 
A.~Bilal,
``Lectures on Anomalies,''
[arXiv:0802.0634 [hep-th]].


\bibitem{ATS}
  M.~F.~Atiyah and I.~M.~Singer,
``The Index of elliptic operators. 3.,"
  Annals Math.\  {\bf 87} (1968) 546.

\bibitem{BNV}
N.~Berline, E.~Getzler and M.~Vergne,
``Heat kernels and Dirac operators,"
Springer-Verlag, New York, (1992) 369 p.


\bibitem{Corvilain:2017luj}
  P.~Corvilain, T.~W.~Grimm and D.~Regalado,
  ``Chiral anomalies on a circle and their cancellation in F-theory,''
  JHEP {\bf 1804} (2018) 020
  [arXiv:1710.07626 [hep-th]].
 


  
\bibitem{Corvilain:2020tfb}
P.~Corvilain,
``6d $\mathcal N=(1,0)$ anomalies on $S^1$ and F-theory implications,''
[arXiv:2005.12935 [hep-th]].

\bibitem{BOS}
A.~Baguet, O.~Hohm and H.~Samtleben,
``Consistent Type IIB Reductions to Maximal 5D Supergravity,''
Phys. Rev. D \textbf{92}, no.6, 065004 (2015)
[arXiv:1506.01385 [hep-th]].


\bibitem{FMS}
S.~Ferrara, R.~Minasian and A.~Sagnotti,
``Low-energy analysis of M and F theories on Calabi-Yau threefolds,''
Nucl. Phys. B \textbf{474} (1996), 323-342
[arXiv:hep-th/9604097 [hep-th]].

  
 \bibitem{Henneaux:2018rub}
   M.~Henneaux, V.~Lekeu, J.~Matulich and S.~Prohazka,
   ``The Action of the (Free) $\mathcal{N} = (3,1)$ Theory in Six Spacetime Dimensions,''
   JHEP \textbf{06}, 057 (2018)
   [arXiv:1804.10125 [hep-th]].


\bibitem{Lekeu:2018vhq}
  V.~Lekeu,
  ``Aspects of electric-magnetic dualities in maximal supergravity,''
  arXiv:1807.01077 [hep-th].

\bibitem{Henneaux:2019zod}
M.~Henneaux, V.~Lekeu and A.~Leonard,
``A note on the double dual graviton,''
J. Phys. A \textbf{53} (2020) no.1, 014002
[arXiv:1909.12706 [hep-th]].



\end{thebibliography}
\end{document}